\documentclass[prb,preprintnumbers,superscriptaddress,showpacs,twocolumn]{revtex4-1}
\usepackage{graphicx}
\usepackage{rotating}
\usepackage{dcolumn}
\usepackage{bm}
\usepackage{amsmath,amssymb,amsfonts,textcomp}
\usepackage{multirow}
\usepackage{color}
\usepackage{array}
\usepackage{hhline}
\usepackage[normalem]{ulem}

\usepackage{amsmath}    
\usepackage{verbatim}   
\usepackage{subfigure}  
\begin{document}

\title{The dispersion relation of Landau elementary excitations\\ and the thermodynamic properties of superfluid $^4$He}

\author{H. Godfrin}
 \email[Corresponding author: ]{henri.godfrin@neel.cnrs.fr}
  \affiliation{Univ. Grenoble Alpes, CNRS, Grenoble INP{\,$^\dagger$}, Institut N\'eel, 38000 Grenoble, France}
 \thanks{Institute of Engineering Univ. Grenoble Alpes}
\author{K. Beauvois} 
  \affiliation{Univ. Grenoble Alpes, CNRS, Grenoble INP{\,$^\dagger$}, Institut N\'eel, 38000 Grenoble, France}
	\thanks{Institute of Engineering Univ. Grenoble Alpes}
	  \affiliation{Institut Laue-Langevin, CS 20156, 38042 Grenoble Cedex 9, France}
\author{A. Sultan}	
  \affiliation{Univ. Grenoble Alpes, CNRS, Grenoble INP{\,$^\dagger$}, Institut N\'eel, 38000 Grenoble, France}
	\thanks{Institute of Engineering Univ. Grenoble Alpes}
	  \affiliation{Institut Laue-Langevin, CS 20156, 38042 Grenoble Cedex 9, France}
\author{E.~Krotscheck}
   \affiliation{Department of Physics, University at Buffalo, SUNY Buffalo NY 14260, USA}
   \affiliation{Institute for Theoretical Physics, Johannes Kepler University, A 4040 Linz, Austria}
\author{J. Dawidowski}
   \affiliation{Comisi\'on Nacional de Energ\'ia At\'omica and CONICET, Centro At\'omico Bariloche, (8400) San Carlos de Bariloche,
R\'io Negro, Argentina}
\author{B. F\aa k} 
   \affiliation{Institut Laue-Langevin, CS 20156, 38042 Grenoble Cedex 9, France}
\author{J. Ollivier}
   \affiliation{Institut Laue-Langevin, CS 20156, 38042 Grenoble Cedex 9, France}

\date{\today}

\begin{abstract}

The dispersion relation $\epsilon(k)$ of the elementary excitations of superfluid $^4$He has been 
measured at very low temperatures, from saturated vapor pressure up to solidification, 
using a high flux time-of-flight neutron scattering spectrometer equipped with a high spatial resolution  detector (10$^5$ `pixels').  
A complete determination of $\epsilon(k)$ is achieved, from very low wave-vectors up to the end of Pitaeskii's plateau. The results compare favorably in the whole the wave-vector range with the predictions of the dynamic many-body theory (DMBT). 
At low wave-vectors, bridging the gap between ultrasonic data and former neutron measurements, the evolution with the pressure from anomalous to normal dispersion, as well as the peculiar wave-vector dependence of the phase and group velocities, are accurately characterized.         
The thermodynamic properties have been calculated analytically, developing Landau's model, using the measured dispersion curve. A good agreement is found  below 0.85\,K between direct heat capacity measurements and the calculated specific heat, if thermodynamically consistent power series expansions are used. The thermodynamic  properties have also been calculated numerically; in this case, the results are applicable with excellent accuracy up to 1.3\,K, a temperature above which the dispersion relation itself becomes temperature dependent.      

\end{abstract}


\maketitle 

\section{Introduction}
One of the most fundamental properties of a many-body system is the dispersion relation $\epsilon$(k) of its elementary excitations \cite{NozieresPines2018,FetterWalecka,ThoulessBook}, i.e., the dependence of their energy on their wave-vector. The prediction by Landau \cite{Landauroton,Landauroton2} of the phonon-roton spectrum of the excitations in superfluid $^4$He, the canonical example of correlated bosons, has paved the way for the development of several areas of modern physics, like Bose-Einstein condensation, superfluidity, phase transitions, quantum field theory, cold atoms, cosmology and astrophysics. 

In the first version of his theory, published in 1941, Landau assumed that phonons and rotons had two separate dispersion relations; he  corrected this idea in the 1947 paper, where he reached the conclusion that helium was described by a single dispersion curve.  
The evolution of the dispersion relation from the quadratic law of independent atoms to the sophisticated form proposed by Landau is a spectacular example of emergent physics. 

At low wave-vectors, the phonon linear dispersion progressively builds up as the interactions are switched on, as shown by Bogoliubov \cite{Bogoljubov}. At atomic-like wave-vectors, a roton minimum appears, which is the signature of the hard core and strong interactions, as solidification is approached \cite{NozieresPines2018,SmithSolid,NozSolid}. Excitations created  from the superfluid condensate, let's name it `the Vacuum', have the characteristics of waves and identical particles \cite{volovik2003,wen2004}.  

The dispersion relation $\epsilon$(k) has been directly observed in $^4$He by measurements of 
the dynamic structure factor $S(Q,\omega)$ using inelastic neutron scattering
techniques \cite{GlydeBook,Glyde2017}, fully confirming Landau's prediction. 
Substantial theoretical work has been devoted to the description of the single-excitations dispersion of superfluid $^4$He, a simple Bose system where the atomic interaction potential is well known. Variational, Monte-Carlo, and phenomenological approaches have brought valuable contributions to the present understanding of helium physics \cite{Landauroton,Landauroton2,Bogoljubov,Aldrich,LeeLee,FeenbergBook,EKthree,1994-BoronatQMC,CeperleyRMP,BoronatRoton,Pistolesi,KroTriesteBook,KroNavarroBook,Vitali2010,2016FerreBoronat,GriffinBook}, but many important questions remain open. 

Superfluid helium also has important applications in experimental physics. In particular, the properties of phonons and rotons are exploited in quantum measurements at the nanoscale level \cite{Guenault2019,2020-PRB-nano-osc,Guenault2019,2020-PRB-nano-osc} and in detectors for particle physics \cite{Guo2013,PRL2016Zurek,PRL2017MarisDetector,Zurek2017}. 

In two previous articles \cite{Beauvois2016,Beauvois2018} we investigated in detail the \emph{multi-excitations} of superfluid $^4$He. 
Here we provide experimental results on \emph{single-excitations} in the whole dynamic range where they are well defined, and we compare them to the predictions of recent dynamic many-body theory (DMBT) calculations \cite{eomIII,phonons}. In the second part of the manuscript, starting from the measured dispersion curve at  saturated vapor pressure, we calculate the thermodynamic properties analytically and numerically. The results are compared to high accuracy thermodynamic data. Tabulated values are provided for different usual parameters (see also Supplemental Material at [URL will be inserted by publisher] for additional tables).

\section{Previous works}
\label{sec:prevwork}
\begin{figure}[h]
	\begin{center}	
	\resizebox{1.0\columnwidth}{!}{\includegraphics{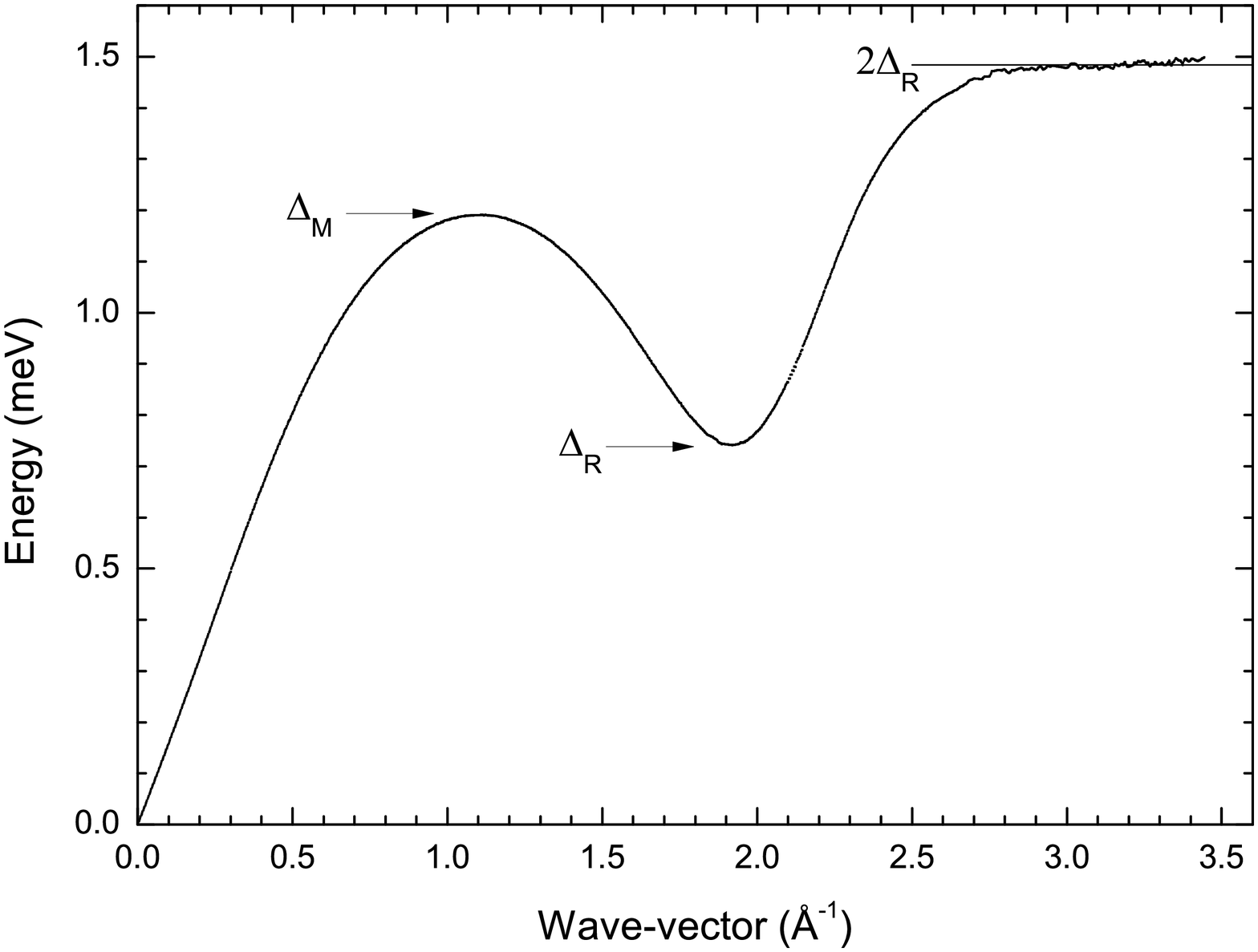}}
	\caption{The dispersion relation of $^4$He at P=0 and T$<$0.1\,K, determined by neutron scattering for wave-vectors k$>$0.15\,\AA$^{-1}$ (this work) . Below this value: extrapolation of ultrasonic data \cite{Junker1977,Rugar1984}. Error bars are not visible at this scale in most of the range (see Table \ref{tab:dispersionP0}). $\Delta_M$ and $\Delta_R$ are the maxon and roton energies.}
	\label{fig:dispersion}
	\end{center}
\end{figure}

The measured phonon dispersion relation of $^4$He is shown in Fig. \ref{fig:dispersion}; it closely resembles the curve predicted 
by Landau \cite{Landauroton,Landauroton2}: the linear (`phonon') part at low 
wave-vectors is followed by a broad maximum (`maxon') at wave-vectors k$\sim$1\AA$^{-1}$ and a deep (`roton') minimum (`roton gap') at  k$\sim$2\AA$^{-1}$. The dispersion curve becomes flat for k$\geq$2.8\,\AA$^{-1}$ as the energy reaches twice the roton gap, forming 'Pitaevskii's plateau'\cite{GlydeBook,Glyde2017}. 

In the long wavelength limit, explored by ultrasonic techniques, the deviations from linearity are described by the expression 

\begin{equation}
	\epsilon(k) \approx \hbar c k (1-\gamma k^2)
\end{equation}

where $c$ is the speed of sound, and $\gamma$ is the {\em phonon dispersion coefficient\/}. 
The general shape of the Landau spectrum suggests that $\gamma>0$, but this was found to be inconsistent with experiments. The latter found an explanation with the  suggestion made by Maris and Massey \cite{Maris1970,MarisRMP} that the dispersion is anomalous ($\gamma<0$) at low pressures. This effect attracted considerable attention both from the theoretical and experimental points of view. Phonon damping due to 3-phonon processes, for instance, is then allowed up to a critical wave-vector $k_c$. The dispersion becomes `normal' at high pressures, near solidification. Details can be found in a critical review by Sridhar \cite{Sridhar87}. 

Thermal phonons at usual temperatures involve much higher wave-vectors. Going from macroscopic to atomic wavelengths is obviously a challenge, which has been taken up by neutron scattering.   

\subsection{The long wavelength limit}
\label{sec:sound}
Deviations from the linear dispersion relation are often described by a polynomial expansion of the excitation energy in powers of the wave-vector modulus $k$: 
\begin{equation}
  \epsilon(k) = \hbar c k\left(1+ \alpha_2 k^2 + \alpha_3 k^3 + \alpha_4 k^4 + ...\right)
\label{eq:dispfit}
\end{equation}
where $\alpha_2$ = -$\gamma$ and $\alpha_1$ is assumed to be zero. 

A different type of expression, frequently used in the analyis of  experimental data, is the Pad\'e approximant \cite{MarisRMP,Maris1980}
\begin{equation}
  \epsilon(k) = \hbar c k\left(1 -
  \gamma k^2\frac{1-k^2/Q_a^2}{1+k^2/Q_b^2}\right)\,.
\label{eq:dispfitMaris}
\end{equation} 

Its series expansion does not contain the term $\alpha_3$. 

Microscopic theory, in fact, suggests a different description of the low-$k$ regime.
Starting from Bogoliubov's formula (see the discussion in Ref. \onlinecite{phonons}), 
we derive the simple expression 
\begin{equation}
 \epsilon(k) = \hbar c k \sqrt{1+ d_2 k^2 + d_3 k^3+...}
\label{eq:dispfitBogoliubov}
\end{equation} 
which is physically correct in the Feynman limit. 

Comparing its power series expansion with Eq. \ref{eq:dispfit} shows that $\alpha_1$=0, $\gamma$=-$d_2$/2, and $\alpha_3$=$d_3$/2. Since higher order terms are generated in the expansion, it is interesting to see if Eq. \ref{eq:dispfitBogoliubov} can describe the experimental data, eventually with a smaller number of parameters.

The term $\alpha_3$ has been calculated analytically \cite{Pitaevskii1970,Feenberg1971} 
from the asymptotic form of the microscopic two-body
interaction. For $V(r) = C_6 r^{-6}$, 
\begin{equation}
  \alpha_3 = \frac{\pi^2}{24}\frac{\rho}{m_4 c^2}C_6\,
  \label{eq:alpha3}
\end{equation}
where m$_4$ is the mass of a $^4$He atom, and $\rho$ the density of the liquid.
At saturated vapor pressure, this estimate gives $\alpha_3$=-3.34\,\AA$^3$ (see Refs. \onlinecite{Pitaevskii1970,Feenberg1971,Rugar1984,Sridhar87}). 
The pseudo-potential theory of Aldrich and Pines \cite{Aldrich,aldrich1976phonon,Sridhar87} provides a similar estimate, $\alpha_3$=-3.7\,\AA$^3$ with a value for the dispersion coefficient, $\gamma$ $\approx$-1.5\,\AA$^2$, consistent with experiments.

The experimental determination of the dispersion relation at long wavelengths has been attempted by different techniques.  
The speed of sound is known in the whole pressure range: ultrasound measurements at very low temperatures yield c=238.3 $\pm$ 0.1\,m/s at saturated vapor pressure. It increases rapidly with pressure (see Refs. \onlinecite{Abraham,DonnellyBarenghi}, and references therein), exceeding 366\,m/s at the melting pressure. 

Since $c=\sqrt{\frac{1}{\rho \kappa}}$, where $\kappa$ is the isothermal compressibility and $\rho$ the density, one can obtain the pressure dependence of the density by measuring the sound velocity as a function of pressure.
Abraham \textit{et al.} \cite{Abraham} found that the expressions
\begin{equation}
P = A_1 (\rho - \rho_0) + A_2 (\rho - \rho_0)^2 + A_3 (\rho - \rho_0)^3
  \label{eq:PofRhoAbraham}
\end{equation}
and 
\begin{equation}
c = \sqrt{A_1 + 2 A_2 (\rho - \rho_0) + 3 A_3 (\rho - \rho_0)^2}.
  \label{eq:CofRhoAbraham}
\end{equation}
accurately describe their results. A fit of their data yields the coefficients
A$_1$=5.679 10$^4$ bar cm$^3$ g$^{-1}$ (corresponding to c$_0$= 238.3\,m/s),  
A$_2$=1.1115 10$^6$ bar cm$^6$ g$^{-2}$, and 
A$_3$=7.43 10$^6$ bar cm$^9$ g$^{-3}$.
Here $\rho_0$=0.14513 g/cm$^3$ is the density at P=0 (see Ref. \onlinecite{Abraham} and references therein). 
The sound velocity is almost linear as a function of density, and one can use the expression
\begin{equation}
c = c_0 + c_1(\rho - \rho_0) + c_2 (\rho - \rho_0)^2.
  \label{eq:clinear}
\end{equation} 
where $c_0$=238.3$\pm$0.1, $c_1$=4671.0$\pm$1.3 and $c_2$=496$\pm$45 for  velocities in m/s and densities in g/cm$^3$.

Ultrasonic measurements are accurate in the determination of the pressure dependence of the sound velocity, but there are some uncertainties in the way the reference velocity c$_0$=238.3 $\pm$ 0.1 m/s (the value at zero pressure) has been determined \cite{Abraham} . 
It is therefore interesting to compare the ultrasonic values with those obtained by other techniques.

Tanaka \textit{et al.} \cite{Tanaka2000} measured the molar volume of pure liquid $^4$He at very low temperatures as a function of pressure. We obtain from their data the velocity of sound, either by derivation of the polynomial of order 9 given by Tanaka \textit{et al.} or by derivation of their data in a small range around the relevant pressures, using the compressibility:  
$c^2 = {V_m}^2/(m_4 \partial{V_m}/\partial{P})$,                       
where m$_4$ is the atomic mass of $^4$He (4.0026032 g/mol).
The molar volume V$_0$ at P=0 is 27.5793 cm$^3$/mol, and the number density 0.021836 atoms/\AA$^3$.

In the pressure range from 0 to 15 bar, the sound velocities determined from the compressibility are systematically below the ultrasonic values, but they agree with the latter within 0.7\,m/s. At higher pressures (partial data are given in Ref. \onlinecite{Tanaka2000} up to the melting pressure), we find a strong deviation of the sound velocity (up to 3\,m/s) from an almost linear density dependence, resulting from a small systematic error in the molar volume data above 15 bar, as can be seen by comparing them to the results of Abraham \textit{et al.} \cite{Abraham} (a useful formula for V$_m$(P) is given by Greywall \cite{Greywall78}). 
  
Anomalous dispersion in superfluid $^4$He was observed by Phillips \cite{Phillips1970} using heat capacity techniques. The results have been extended to lower temperatures by Greywall \cite{Greywall78,Greywall79ERRATUM}, motivated by discrepancies observed between heat capacity and neutron scattering data.  Since the heat capacity is obtained as an integral of the dispersion curve over a substantial range of wave-vectors, extracting the dispersion curve from data is not a unique procedure (this point will be discussed in detail in section \ref{sec:Cv}). 
The velocity of sound determined by Greywall at low temperatures from the coefficient of the T$^3$ term in the heat capacity, in addition, is affected by uncertainties in the thermometry \cite{Greywall78,Greywall79ERRATUM}. The values of $c$ from heat capacity are lower than the ultrasonic ones, and less accurate, but their density dependence is similar. Thermometry calibration improvement reduced the values of the sound velocities by about 3 to 6\,m/s for increasing pressures, which is an indication of  the typical uncertainties in heat capacity data. Paradoxically, uncorrected data were closer to the ultrasonic results.

Accurate measurement of the deviations from linearity of the phonon dispersion have been made by Rugar and Foster \cite{Rugar1984}. Ultrasonic measurements at two fundamental frequencies showed that $\alpha_1$$<$10$^{-3}$\AA \, at P=0 and 6.3\,bar, and $\alpha_2$ = (1.56$\pm$0.06)\AA$^2$ at SVP. If $\alpha_1$ is assumed to be zero, then $\alpha_2$ =(1.55$\pm$0.01)\AA$^2$ at SVP. 
The excitation spectrum is probed for k$<$0.011\AA$^{-1}$. Their analysis is insensitive to assumed values of $\alpha_4$, $\alpha_5$, etc., and it is only slightly sensitive to the value of $\alpha_3$, which is taken from theory \cite{Pitaevskii1970,Feenberg1971}. 
The density dependence of $\alpha_2$, measured from SVP to 10\,bar, is almost linear.
The data agree well with former results of Junker and Elbaum \cite{Junker1977} on the temperature dependence of the ultrasonic velocity, 
 which reach higher pressures (about 15 bar). 

We shall use the ultrasonic values in the following, since they are the most accurate, and confirmed (but only within about 0.7\,m/s) by other measurements.
The values obtained in the present work will be compared to these data in Section \ref{sec:results}.

The experiments described above provided a good description of the dispersion relation for small wave-vectors, and convincing evidence of anomalous dispersion for pressures below about 20 bar was progressively gathered. 
To achieve a direct observation of the dispersion curve and explore the dynamics at atomic wave-vectors, the privileged tool is inelastic neutron scattering.

\subsection{Previous neutron scattering results}
\label{sec:PreviousNeutron}

Previous neutron scattering data have been described in detail by Glyde in a book \cite{GlydeBook} and a recent review article \cite{Glyde2017}.
Tables of the properties of liquid helium have been published by Brooks and Donnelly \cite{BrooksDonnelly} and by Donnelly and Barenghi \cite{DonnellyBarenghi}; neutron scattering data from a variety of sources and smoothed values are provided. Original references should be consulted, however, for error bars.  

The quantitative knowledge of the dispersion relation is based on measurements by Cowley and Woods \cite{CowleyWoods}, Woods \textit{et al.} \cite{WoodsHilton}, Svensson \textit{et al.} \cite{Svensson1975,SvenssonMartel,SvenssonScherm}, Stirling \textit{et al.} \cite{stirling-83,StirlingPhonon,StirlingExeter} and others \cite{Talbot-Glyde,Glyde2017} mainly performed on triple-axis spectrometers. The different data sets are not totally compatible, and the dispersion relation which emerges from these studies is therefore not fully satisfactory. 
 
The main advantage of triple-axis spectrometers is their good accuracy in the determination of energies and wave-vectors. This point-by-point measuring technique is time consuming, and therefore not appropriate to investigate the whole wave-vector range. 

The small-k phonon region was studied, pushing the technique to its limits, to explore a possible anomalous dispersion. The first results \cite{CowleyWoods,Svensson1975,SvenssonMartel} were highly speculative and at best qualitative, since error bars growing at low wave-vectors precluded a thorough comparison to ultrasonic sound velocity measurements. Higher accuracy measurements performed by Stirling \textit{et al.} \cite{stirling-83,StirlingPhonon,StirlingExeter} finally confirmed the anomalous character of the dispersion at low pressures. However, these measurements showed a systematic disagreement with ultrasound measurements, which will be discussed further below. 

Time-of-flight spectrometers (TOF) with large detector arrays allow measurements of dispersion curves over a large range of energies and wave vectors simultaneously. Early experiments by Dietrich \textit{et al.} \cite{Dietrich72} and Stirling \textit{et al.} \cite{StirlingCopley} were followed by more recent measurements on IN6 at the ILL by Stirling, Andersen, and coworkers \cite{AndersenThesis,Fak-Andersen-91,Andersen92,Andersen94a,Andersen94b,GibbsThesis,AndersenRoton}. Two new data sets with an energy resolution of about 100\,$\mu$eV  were obtained through the latter works, referred to as `Andersen' \cite{AndersenThesis,Andersen94a} and `Gibbs' \cite{GibbsThesis,AndersenRoton}.

A good agreement with triple-axis data was found in the roton and the maxon regions, but strong deviations were observed both at low and high wave-vectors. IN6 shares with triple axis spectrometers the use of graphite monochromators (3 focusing ones), thus complicating the resolution function shape, and significant corrections for sample absorption or off-center sample position were needed in the data analysis. 

Additional measurements were performed at ISIS on the IRIS time of flight inverted-geometry crystal analyzer spectrometer, with an excellent energy resolution of 15\,$\mu$eV, but a coarse wave-vector resolution \cite{Glyde-Gibbs-98}. An important result was obtained at high wave-vectors, showing that the single-excitation dispersion curve is slightly below twice the roton energy \cite{Glyde-Gibbs-98}. In this case, where the dispersion is flat, the resolution characteristics of IRIS constituted a major advantage.

Measurements by Pearce \textit{et al.} \cite{Pearce-Azuah-Stirling} on the same instrument, mainly around the roton energy, showed discrepancies with former works, in particular in the magnitude of the temperature dependence of the roton parameters determined at ILL's IN10 backscattering spectrometer \cite{GrenobleRotonsT} with an energy resolution better that 1\,$\mu$eV.   

It was difficult to decide which set, among these partly conflicting TOF data, was correct. 
The potential of the TOF technique motivated the present studies on IN5.

\section{Experimental details}
\label{sec:cryogenics}

The cylindrical sample cell was made out of 5083 aluminum alloy, selected because of its good mechanical and neutron scattering properties. The minority chemical constituents (4.4\% Mg, 0.7\% Mn, 0.15\% Cr, etc.) have a modest effect on the neutron scattering and absorption cross-sections compared to the values for pure aluminum, with an increase of less than 15\% of the total cross-section. The gain in mechanical properties allows reducing the thickness in a much larger proportion, by a factor of 3. High pressure studies could be made using a thin cell, of  1\,mm wall thickness, for pressures up to 24\,bar. 

The cell had a 15\,mm inner diameter, which is small compared to the 30 to 50\,mm  diameters used in other works.  
Cadmium disks of 0.5\,mm thickness were placed inside the cell every 10\,mm, to reduce multiple scattering. This was not needed for the present studies, and it even had an undesirable effect, reducing the signal on some neutron detectors placed far from the sample horizontal plane. We did not place Cd masks on the sides of the cell; preserving the cylindrical geometry turned out to be favorable for the data analysis. 

High purity (99.999\,\%) helium gas was condensed in the cell at 
temperatures on the order of 1\,K. 
The stainless steel gas-handling system consisted of a set of high quality valves, tubes and calibrated volumes.  The gas was admitted through a ``dipstick'', placed  in a helium storage dewar, which was used to purify, condense and pressurize the helium sample.  

The dispersion relation of helium is very sensitive to the applied pressure. For this reason,  pressures in the system were measured with a high accuracy 0-60\,bar 
Digiquartz gauge, located at the top of the cryostat. This gauge has a precision of 6\,mbar, but the pressures inside the cell are known only within 
20\,mbar, due to helium hydrostatic-head corrections. 
The corrected pressures in the cell are given in Table \ref{table:Pressures}. 

\begin{table}[h]
\begin{ruledtabular}
\begin{tabular}{lccccccc} 
Helium samples \\ 
\hline \\
Nominal P (bar) & 0 & 0.5 & 1 & 2 & 5 & 10 & 24   \\ 
Corrected P (bar) & 0 & 0.51 & 1.02 & 2.01 & 5.01 & 10.01 & 24.08   \\
\end{tabular}
\end{ruledtabular}
	\caption{Nominal and corrected values for the pressures investigated in the present  work. The estimated uncertainty is $<$0.02\,bar. }
	\label{table:Pressures}
\end{table}

The cell was carefully centered in a dilution refrigerator providing temperatures well below 100\,mK. The thermal connection to the 
mixing chamber was achieved by using massive OFHC-copper pieces.  Sintered
silver powder heat exchangers placed at the top of the cell provided a good 
thermal contact between the cell and the helium sample. 
Two long, small diameter, Cu-Ni filling capillaries were used in parallel, for safety. 
They were thermally anchored along the dilution unit, insuring a negligible heat leak to the cell.
Thermometry was provided by calibrated carbon and RuO$_2$ resistors.  

Measurements were made for a vanadium sample (a rolled foil, mass 9.81\,g, external diameter 12\,mm, height 60\,mm, used for the detectors efficiency calibration), for the empty cell, and then for the cell filled with $^4$He at several pressures (see Table \ref{table:Pressures}). The helium measurements were performed at temperatures below 100\,mK. 
The data acquisition consists in several runs of one hour duration. The longest measurements were made at P=0 (9h) and P=24\,bar (6h). Two hours runs were made  at all other pressures. The empty cell signal, measured for 10h, was used as background  and subtracted from all the helium measurements. 

\section{Inelastic neutron scattering}
\label{sec:INS}

\subsection{Inelastic neutron scattering equations}
\label{sec:TOF}

The quantity measured by a neutron spectrometer \cite{Lovesey,schober2014introduction} is the double  differential 
scattering cross section per target atom, which is proportional to the dynamic structure factor:  
\begin{equation}
\label{eq:neutrons}
\frac{\partial^{2}\sigma} {\partial\Omega~\partial{E_f}} = \frac{b^{2}_{c}} {\hbar} \frac{k_f} {k_i} S(Q,\omega)  
\end{equation}
where $b_{c}$ is the bound atom coherent scattering length. 
The incident neutron has an initial energy $E_i$ and a wave-vector $\vec{k_i}$, leaving the sample with a final energy $E_f$ and a wave-vector $\vec{k_f}$; the wave-vector transfer is     
$\vec{Q}=\vec{k_i}-\vec{k_f}$, and the energy transfer $\hbar\,\omega = E_i-E_f$.

The wave-vector transfer is written in terms of the scattering angle $\varphi$  between $\vec{k_i}$ and $\vec{k_f}$:
\begin{eqnarray}
   Q^2={k_i}^2+{k_f}^2-2{k_i}{k_f}{\cos\varphi} \\
	 Q^2=\frac{2m_n}{\hbar^2} \left[2E_i-\hbar\omega\ -2\sqrt{E_i(E_i-\hbar\,\omega)} \cos\varphi\right]
	\label{QAngle}
\end{eqnarray}

The number of neutrons detected as a function of the scattering angle $\varphi$ and the energy transfer yields $S(Q,\omega)$ through Eq. \ref{eq:neutrons}.

At zero temperature there are no thermal excitations, and the only allowed process is the creation of excitations. 
When a single-excitation of energy $\epsilon$ and wave-vector $\vec{k}$ is created, conservation of energy and wave-vector leads to $\epsilon = \hbar\,\omega$ and $\vec{k}=\vec{Q}$. Single-excitations on the dispersion curve $\epsilon(k)$ are observed in the dynamic structure factor $S(Q,\omega)$ as sharp peaks.  

\subsection{The time of flight spectrometer IN5}
		\label{sec:IN5}

\begin{figure}[h]
	\begin{center}
		\resizebox{0.95\columnwidth}{!}{\includegraphics{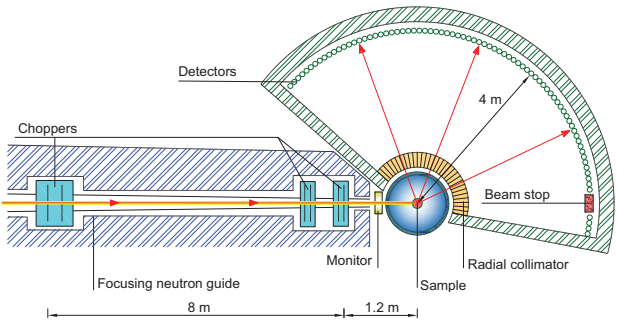}}
		\caption{Disk chopper time of flight spectrometer IN5.
		}
		\label{fig:IN5}
	\end{center}
\end{figure}

The measurements were performed on the IN5 time of flight spectrometer \cite{IN5_2010,IN5_2011} at the Institut Laue Langevin (see Fig. \ref{fig:IN5}).

A pulsed monochromatic beam is provided by three groups of two choppers.  
A key feature is that the resolution is well represented by a Gaussian function. 

\begin{figure}[t]
	\begin{center}
		\resizebox{0.8\columnwidth}{!}{\includegraphics{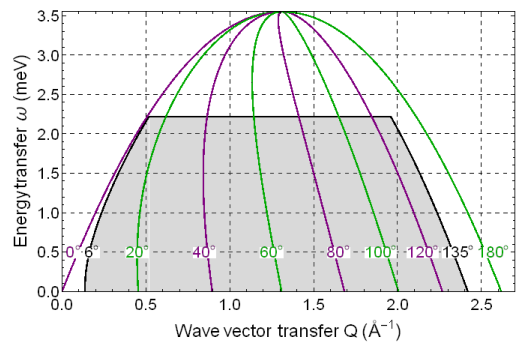}}		
		\caption{The ($Q$, $\omega$) space accessible for an incident neutron energy $E_i$ = 3.52\,meV, calculated from equation (\ref{QAngle}). Constant angle lines are shown for selected values between 0$^\circ$ et 180$^\circ$. The gray area indicates the region actually used in the present measurements.}
		\label{FigQWspace}
	\end{center}
\end{figure}

The neutron energy was $E_i$ = 3.52\,mev for the low wave-vector range, which gives a convenient access to excitations of wave-vectors $k$ from 0.15 to 2.3 \AA$^{-1}$ and covers the energy range between 0 and 2.22\,meV as shown in Fig. \ref{FigQWspace}. In the conditions of the experiment, IN5 has a very large neutron flux of $\phi_n=2\times10^5$ neutrons/(cm$^2$s) at the sample position. 
The uncertainty in the incident energy is $\approx$1\%, a point which will be further discussed below. 
The complete wave-vector range was explored using different incident neutron energies $E_i$ = 3.520, 5.071, 7.990, and 20.45\,meV, with energy resolutions (FWHM) at elastic energy transfer of  0.07, 0.12, 0.23 and 0.92\,meV, respectively, for a chopper speed of 16900\,rpm. 

\begin{figure}[t]
	\begin{center}	
			\resizebox{1.0\columnwidth}{!}{\includegraphics{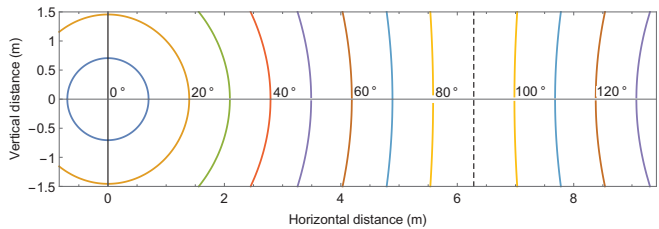}}
	\caption{IN5 detectors set-up (drawn to scale): constant scattering angle curves on the detector surface of $\approx($3$\times$10)\,m$^2$. The horizontal positions of the pixels are characterized by the azimutal angle $\varphi_0$ of their detector tube (indicated on the curves); vertical positions are given by the height $z$ measured along the tube.  }
		\label{Constant-angle-curves}
	\end{center}
\end{figure}

A large array of $^3$He+CF$_4$ position sensitive neutron detectors (PSD) is located in a vacuum chamber which surrounds the sample space. Key features of the detection system are its large angular coverage and resolution. 
The 384 detector tubes are placed at a distance of 4\,m from the axis of the instrument. The angular position of the tubes with respect to the direction of the neutron beam is given by the `detector angle' $\varphi_0$, covering the range from -12$^\circ$ to 135$^\circ$.  The tubes are straight and long, their vertical range goes from -1.47 to +1.47\,m. The PSD system provides 241 `pixels' of small size (26$\times$11.24 mm$^2$) per tube, characterized by their position (angle $\varphi_0$, height $z$) in the detector surface. The pixels corresponding to the same Debye-Scherrer cones, i.e., at the same scattering angle $\varphi$  (Fig. \ref{Constant-angle-curves}), are grouped by software \cite{lamp}, resulting in 346 different scattering angles in the interval 6$^\circ$ to 135$^\circ$.   
The detection process is more efficient than with triple-axis spectrometers, where a single detector has to be moved over the whole angular range. 

The distance from each pixel to the center of the instrument varies substantially due to the tall, vertical tube geometry. The Debye-Scherrer cones `standard procedure' \cite{lamp} groups the individual detector pixels into equivalent units located at the in-plane angles $\varphi_0$ and in-plane nominal distance D=4\,m.  

\begin{figure}[h]
	\begin{center}
		\resizebox{1.0\columnwidth}{!}{\includegraphics{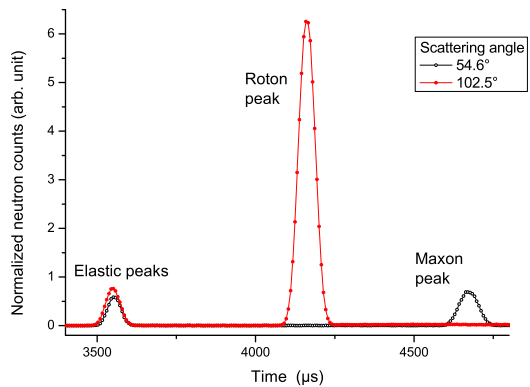}}
		\caption{Time of flight measurement at an incident neutron energy E$_i$=3.52\,meV, for scattering angles of 54.6$^\circ$ (near the maxon), and 102.5$^\circ$ (near the roton). Note that the elastic peaks of these signals are slightly shifted.}
		\label{fig:TOF}
	\end{center}
\end{figure}

\begin{figure}[h]
	\begin{center}
		\resizebox{1.0\columnwidth}{!}{\includegraphics{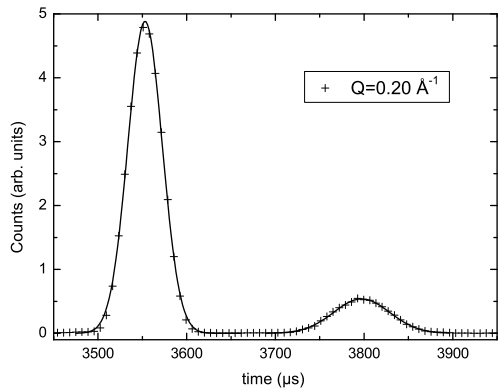}}
		\caption{Time of flight measurement at an incident neutron energy E$_i$=3.52\,meV, for a scattering angle of 26$^\circ$ (Q=0.20\,\AA$^{-1}$) corresponding to the phonon region.}
		\label{fig:TOFphonon}
	\end{center}
\end{figure}

The neutron arrival signal from each pixel of the PSD is read into a data acquisition system of 1024 time channels of 6.9084 $\mu$s duration (`time frame'). Since the neutron velocity is on the order of 820 m/s, the time of flight over the 4\,m instrumental distance is on the order of 4.9\,ms, or 700 channels. 
The `time origin' of the data acquisition is set in such a way that both the elastic peak and the helium excitation peak are measured within the same time frame, as shown in Figs. \ref{fig:TOF} and \ref{fig:TOFphonon}. 

\section{Standard data reduction}
\label{sec:data-acquisition}

Standard data-reduction \cite{lamp} was initially    
used \cite{Beauvois2016,Beauvois2018} to calculate from the raw data the dynamic structure factor $S(Q,\omega)$ of broad multi-excitations. 
The `standard analysis' data consist of time of flight spectra (matrix of the number of counts for the 1024 time channels for 346 angles $\varphi$): the raw data from the detectors pixels have been grouped by scattering angle $\varphi$, as described above. In this section \ref{sec:data-acquisition}, therefore, $\varphi$ represents the scattering angle of an effective detector located in the horizontal plane, at $z$=0 and $\varphi$=$\varphi_0$ (`in-plane effective description'). 
The very narrow single-excitations, however, require a more sophisticated `pixel-by-pixel analysis', described in section \ref{sec:pixelanalysis}, where the same raw data are processed, but the TOF data (1024 time channels) of the 384$\times$241 detector pixels are analyzed individually.  

\subsection{Time of flight (TOF) equations}
\label{sec:TOFequations}

We first proceed to fit the spectra to determine very accurately the time of arrival at the detectors of elastically scattered neutrons, $t_{elast}$ measured, as described above, using the system clock times.
The `elastic peaks' (see Fig. \ref{fig:TOF}) can be approximated by simple Gaussians. Since superfluid helium does not scatter elastically, we use the signal of the aluminum cell, and compare it to that of the vanadium sample. 
 
The equation 
$\tau_0=t_{elast}-t_s=D/v_i$
for the neutron flight over the distance $D$ separating the sample from the detectors, determines the important parameter $t_s$, the time of scattering at the sample, according to the system clock. 
This supposes that the detectors are located at the same distance of the sample, which is often a good approximation.

The energy of the excitations is determined from the measurement of the time of flight $\tau$ of \emph{inelastically} scattered neutrons, over the same  distance $D$.  
A neutron creating an excitation of energy $\epsilon$ reaches a detector located at an angle $\varphi$ at a time $t_{inel}(\varphi)$, obtained from the gaussian fits of the `helium peaks' (see Fig \ref{fig:TOF}).
The time of flight is now $\tau(\varphi)=t_{inel}(\varphi)-t_s$.  
The final velocity of the neutron $v_f=D/\tau$ provides the neutron final energy $E_f$. 
The energy of the excitations is $\epsilon = E_i-E_f$, where the initial energy of the neutrons $E_i$ is known from the mechanical characteristics of the choppers system. 
The excitation wave-vector $k$ is obtained from equ. \ref{QAngle}.

In this simple scheme, there are only two independent instrumental parameters, selected among the initial neutron energy $E_i$, the average sample-detector distance $D_{av}$, and the average time of arrival of the neutrons at the sample position, $t^{av}_s$. 
The energy of the excitations is  obtained from the equation:

\begin{equation}
	\label{equ:standard}
  \epsilon(\varphi) = \frac{1} {2} {m_n} {D_{av}^2} {\left[ \frac{1} {(t_{elast}(\varphi) - t^{av}_s)^2} - \frac{1} {(t_{inel}(\varphi) - t^{av}_s)^2}  \right]}
\end{equation}
where $m_n$ is the neutron mass, and $	t^{av}_s=t^{av}_{elast}-D_{av}/v_i$.\\

A more convenient form can be used when the sample-detector distances differ by a significant amount:    

\begin{equation}
	\label{Erefined}
  \epsilon(\varphi) = E_i {\left[ {1} - \left(\frac{t_{elast}(\varphi) - t_s} {t_{inel}(\varphi) - t_s}\right)^2  \right]}
\end{equation}

The dependence on the initial energy $E_i$ is made explicit. The distances $D(\varphi$) to all individual detectors do not appear. Instead, we find the inelastic and elastic times for each angle $\varphi$, which are the measured parameters. Last but not least, one has to determine $t_s$. In experiments using a small diameter cylindrical sample with a small absorption, like in the present case, this time of arrival at the sample is very well defined, and unique: it does not depend on $\varphi$.      

The analysis yielding the energies $\epsilon(\varphi)$ \textsl{involves only two instrumental parameters}: the initial neutron energy $E_i$, and the neutron arrival time $t_s$ at the sample, central to the present discussion, which can be estimated using the nominal  sample-detector distance of IN5, 4.00\,m. The actual flight distances in the sample plane should be close to this value, but they can be significantly affected by other effects. For instance, the analysis assumes that the elastically scattered neutrons follow the same flight path as the inelastically scattered ones reaching a given detector. This is not true if absorption plays an important role; it introduces, in addition, undesired angular shifts. Correcting for systematic errors, fortunately, can be done as shown in the next section.
 
\subsection{Distance and angle corrections}
\label{sec:Corrections}  
Distance and angular corrections arise from imperfections in the instrument and sample geometries and from the finite size of the components. Neutron beam, sample and detectors have typical dimensions on the order of centimeters, the instrument lengths are on the order of meters, thus requiring  finite size optics analytical calculations or computer simulations if uncertainties on the order of 10$^{-3}$ are desirable.  
We have used both techniques to evaluate possible effects, and retained the corresponding corrections, listed below, when their influence on the excitations energies was larger than 1\,$\mu$eV. 

\subsubsection{Sample off-center}
\label{sec:off-center}
Corrections may be necessary if the sample is not placed exactly at the geometrical center of the instrument. 
Large sample off-set effects were observed by Andersen \textit{et al.} \cite{AndersenThesis,Andersen94a}. Their characteristic symptom, essentially a parabolic angular dependence of the elastic times of flight, is also observed here and ascribed, however, to a very different cause, namely, a rigid-plate distortion of the detectors plane. Both effects are discussed below. 
 
\begin{figure}[h]
	\begin{center}
		\resizebox{0.75\columnwidth}{!}{\includegraphics{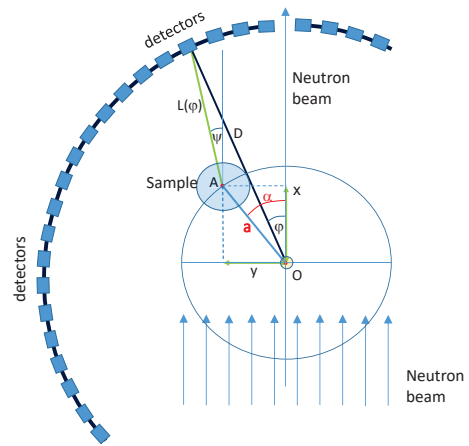}}
		\caption{Instrument parameters when the sample is not centered. Reference points: instrument center $O$, sample center $A$; $x$ and $y$ are the coordinates of $\vec{OA}$ and $\alpha$ the angle between $\vec{OA}$ and the neutron beam. Distances: instrument center to sample $a$, instrument center-detectors $D$, sample-detectors $L(\varphi)$; $\varphi$ is the nominal scattering angle, $\psi$ the physical scattering angle. }
		\label{Sample-off-Center}
	\end{center}
\end{figure}

A description of the sample off-center geometry, not to scale, is given in Fig. \ref{Sample-off-Center}. 
The detectors are placed on a rigid frame, forming a circle around the `instrument center' $O$.  Their distances and angles have been carefully characterized using theodolites.  
In principle the sample is centered with respect to the cryostat, which is centered with respect to the cylindrical experimental space, aligned with the detectors bank. 

If the `sample position' $A$ is shifted from the geometrical `instrument center' $O$ (Fig. \ref{Sample-off-Center}), the TOF distance $D$ is replaced by $L$ :

\begin{equation}
L(\varphi)=\sqrt{D^2+a^2-2 a D \cos (\alpha -\varphi )}
\label{eq:lengthL}
\end{equation}

while the physical scattering angle $\psi$ is related to the detector angle $\varphi$ by the expression  
\begin{equation}
\cos(\psi)=(D \cos (\varphi )-a \cos (\alpha ))/L
\label{eq:cospsi}
\end{equation}

\begin{figure}[t]
	\begin{center}	
			\resizebox{0.9\columnwidth}{!}{\includegraphics{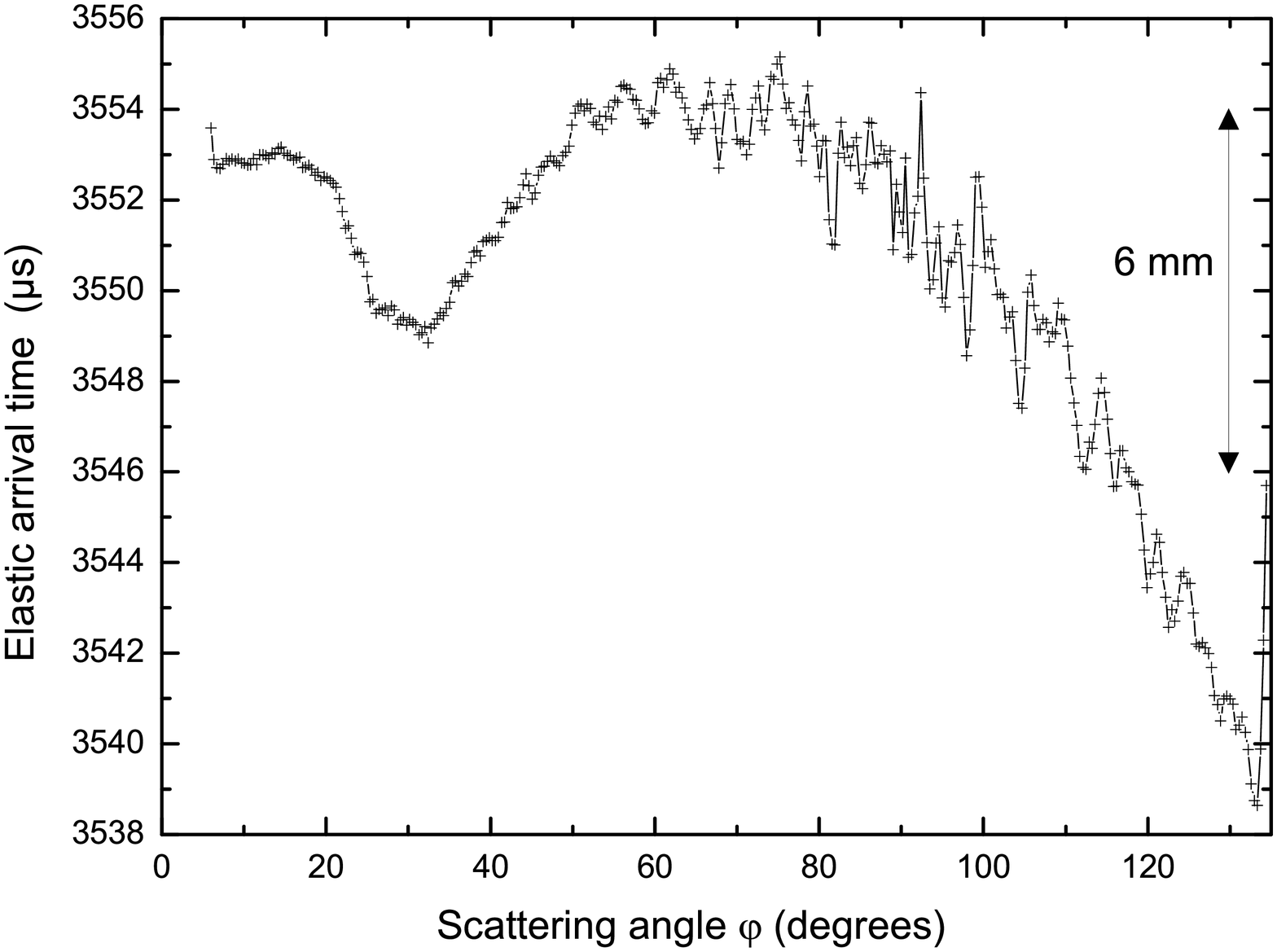}}
	\caption{Arrival time of neutrons elastically scattered by the aluminum cell, measured in the standard analysis as a function of the scattering angle (the time origin t$_s$ is -1323 $\mu$s), at an incident neutron energy E$_i$=3.52\,meV. The arrow on the right hand side indicates the corresponding variation of the  flight distance.} 
		\label{elasticTOF}
	\end{center}
\end{figure}

The elastic TOFs of the vanadium sample and the aluminum cell both display a visible angular dependence, indicating a significant variation of the $L(\varphi)$. 
Fig. \ref{elasticTOF} shows the time of arrival of neutron elastically scattered by the aluminum cell. 
The apparent dispersion observed on the data points, reproducible in different scans, corresponds to small differences in sample-detector distances. 
The arrival times at large angles display the characteristic parabolic shape of a sample position shifted with respect to the detectors center.  
A fit to the data in Fig. \ref{elasticTOF} using the sample-offset model  

\begin{equation}
t_{elast}(\varphi) = \sqrt{D^2+a^2-2 a D \cos (\alpha -\varphi)}/v_i + t_s
\label{eq:TOFfit}
\end{equation}

with $D$=4.00\,m and $v_i$=824.17\,m/s (these values are not critical), would yield as off-set parameters a=19.0(5)\,mm and $\alpha$=245.8(7)$^\circ$. The fit also yields $t_s$, but this parameter, strongly correlated to the initial energy $E_i$ and the distance $D$, will be determined consistently later on.    
The off-set distance and angle are surprisingly similar to those calculated by Andersen \textit{et al.} \cite{AndersenThesis,Andersen94a}.
In the present case, however, we can show that such a large off-set is incompatible with the complete calculation of the elastic TOF for our 3-dimensional detector array. In particular, the detectors covering positive and negative low angles are highly sensitive to sideways displacements. We found that sample off-set corrections are on the order of 2\,mm or less. 
The TOF results determined with the vanadium sample are almost identical to those  described above, the corresponding differences in flight distances are again less than 2\,mm. This indicates that the sample is very well centered inside the cryostat, and the latter within the instrument. Some small differences between the vanadium and  the aluminum cell data can be ascribed to their different geometry, in particular the effect of the cadmium disks inside the cell.

\begin{figure}[h]
	\begin{center}	
			\resizebox{0.95\columnwidth}{!}{\includegraphics{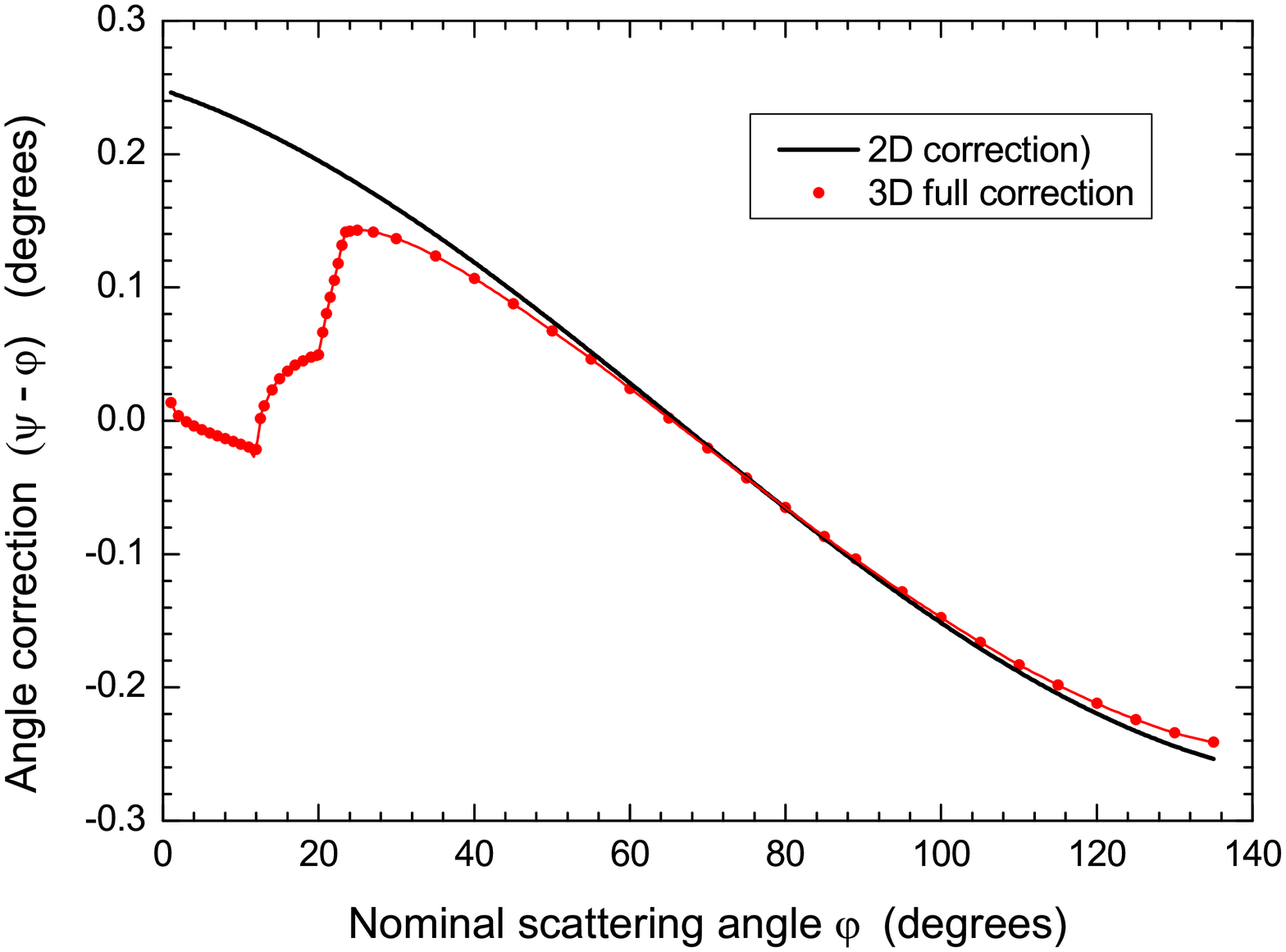}}
	\caption{Angular correction to the scattering angle $\varphi$, calculated for a supposed sample position off-set (a=19.0(5)\,mm and $\alpha$=245.8(7)$^\circ$). $\psi$ is the corrected scattering angle. The black line corresponds to a simple in-plane-only detectors scheme, while the red line describes the values calculated for the actual 3-dimensional IN5 detectors geometry. Such a correction could be discarded here, but may have affected earlier works (see text). }
		\label{fig:PsiMinusPhi3D}
	\end{center}
\end{figure}

Scattering \emph{angle} corrections have also been considered for a case where the sample would be off-center. They were calculated for the present geometry, taking into account the 3-dimensional positions of the detector pixels, shown in Fig. \ref{Constant-angle-curves}. Debye-Scherrer rings grouping  would produce a peculiar angular correction, shown in Fig. \ref{fig:PsiMinusPhi3D}, if the sample had been off-center. Such a correction is not compatible with our measured data for the dispersion curve: it would have given visible accidents. We have therefore concluded that distance and angular corrections due to a sample center off-set are very small in the present work. 

\subsubsection{Effect of strong scattering and absorption}
\label{sec:diffcorr}
A different correction may be caused by strong scattering and/or absorption in large samples. 
Essentially, the sample regions which are both closer to the reactor and to the detectors provide a larger contribution to the scattered neutrons flux, than those located further away. Both TOF distances and scattering angles are affected.  

\begin{figure}[t]
	\begin{center}	
			\resizebox{0.8\columnwidth}{!}{\includegraphics{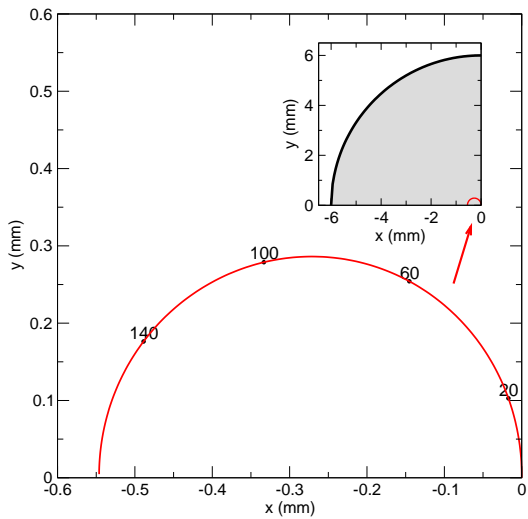}}
	\caption{Displacement of the effective position of the sample center for the vanadium sample (used in the calibration procedure) calculated for neutron absorption. The coordinates axes are defined in Fig. \ref{Sample-off-Center}, and the corresponding angles are indicated along the parametric curve. Inset: magnitude of the effect compared to the sample size. }
		\label{fig:vanaOffset}
	\end{center}
\end{figure}

We have calculated the effective sample center position for the vanadium  cylindrical sample used for calibration purposes, and for the thin wall aluminum alloy sample cell.
The relevant parameter is the ratio of the sample dimensions to the penetration depth $\lambda_{scatt}=(\sum{n_i}\sigma_i)^{-1}$. The sum runs over the scattering and absorption cross-sections (at the incident energy E$_i$) of the different elements present in the sample with number densities $n_i$.  
The apparent sample center position depends now on the scattering angle $\varphi$. 
 
For the vanadium sample, $\lambda_{scatt}$=32\,mm, significantly larger that the vanadium radius of 6\,mm. In this case, the calculation (Fig. \ref{fig:vanaOffset})  yields a maximum shift of less that 1\,mm, a small effect on the distances of flight. 

\begin{figure}[t]
	\centering
		\includegraphics[width=0.8\columnwidth]{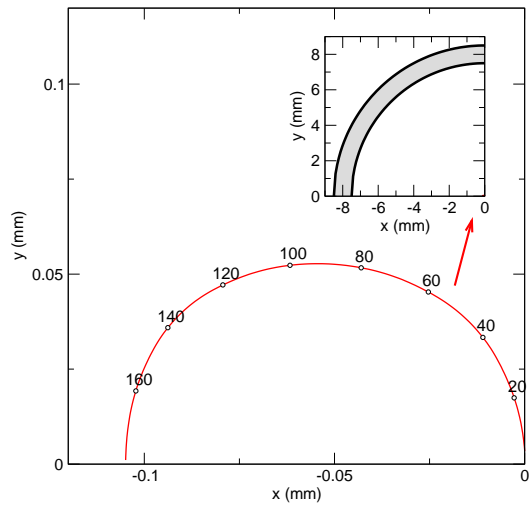}
	\caption{Displacement of the effective sample center position calculated for neutron scattering and absorption in the thin wall cylindrical aluminum alloy sample cell. The coordinates axes are defined in Fig. \ref{Sample-off-Center}, and the corresponding angles are indicated along the parametric curve. The magnitude of the effect is much smaller than the sample size (inset).}
	\label{fig:alumOffset}
\end{figure}

The aluminum cell may also display a displacement of its effective center due to scattering and absorption. The shift, however, is even smaller. For our aluminum alloy, $\lambda_{scatt}$=65\,mm,  considerably larger that the can dimensions (7.5\,mm internal radius, wall thickness 1\,mm) and the corresponding neutron paths. The effective sample center position calculated for this hollow cylinder geometry is given in Fig. \ref{fig:alumOffset}.    
\vspace{5mm}

Corrections ascribed to strong absorption have been applied by Gibbs \cite{GibbsThesis} to his TOF data.  A much larger and thicker aluminum cell was used in this work; nevertheless, the present work suggest that other causes  are more probable. The accuracy of these data may thus be slightly lower than initially believed.     

The effective sample center displacement due to strong scattering and absorption in the sample also affects the scattering angles.
The calculated angular shift for the vanadium sample is very small, in particular at low angles, where accuracy is needed. The same remark is valid for the measurements with the experimental cell. We can therefore conclude that angular corrections due to strong scattering and/or absorption are small in the conditions of the present experiment (small diameter, thin aluminum cell).  

\subsubsection{Distance corrections in the detectors}
\label{sec:corrdet}
Neutrons are not detected, in average, at the center of the detectors. 
Due to the strong neutron absorption of $^3$He, the detection process takes place with a short characteristic distance $\lambda_{abs}$, which depends on the density of the $^3$He gas in the detector, and on the neutron energy. 
For the parameters of the present experiment, the penetration length is $\approx$5\,mm. 
The distance between the instrument center and the position where neutrons are detected is in fact somewhat shorter than the nominal distance $D$=4.00\,m between the center of the instrument to the center of the detector tubes. 
The latter have an internal diameter $D_{det}$=24.4\,mm. 
The neutron detection process occurs at an average distance $\bar{y}$ from the plane of the detector centers. 
At an incident neutron energy E$_i$=3.52\,meV, $\bar{y}\approx$5\,mm.
This value depends on the final neutron energy, and hence on the scattering angle $\varphi$ for neutrons scattered from the sharp excitations on the dispersion curve of $^4$He.
Corrections for this effect have been calculated for the detectors geometry, and applied to the data. 

\subsubsection{Other corrections}
\label{sec:othercorr}
A related effect is the apparent displacement of the sample center, when using cadmium shields or windows on the sides of the cell \cite{AndersenThesis,Andersen94a}, slightly masking the helium sample or the aluminum cell for some scattering angles. The effect is absent in the present experiment, where the cylindrical symmetry has been preserved, thus ensuring an excellent angular average.

The finite diameter of the sample can lead to distance and angular corrections in small instruments, in particular for the 30 to 50\,mm diameter cells used in previous works. 
These corrections are negligible for the present work on IN5 with a 15\,mm inner diameter cell.

\section{Detailed pixel-by-pixel analysis}
\label{sec:pixelanalysis}
   
The single-excitation dispersion curve is intense and extremely sharp, and we are therefore interested in the best resolution and accuracy both in energy and wave-vector. For this reason, we proceed now with a refined analysis using a `high resolution configuration'. \\

Analysis step 1: The `high resolution configuration' consists of a pixel-by-pixel treatment of the multidetectors signals (see section \ref{sec:INS} for hardware details). Neutrons are collected in 1024 time channels for each PSD detector pixel.
Thanks to the very high flux of IN5, elastic and inelastic times of flight can be determined by means of  gaussian fits, for \textsl{each} of the 241x384 pixels in the detector matrix (see section \ref{sec:IN5}). The software LAMP\cite{lamp} is used to read and fit the nxs raw data files of IN5. 
We obtain about 9$\times$10$^4$ values of pairs (t$_{elast}$($\varphi_0$,$z$),t$_{inel}$($\varphi_0$,$z$)), where $\varphi_0$ is the angle of a detector tube and $z$ the height of a pixel within the corresponding tube. 

\begin{figure}[h]
	\centering
		\includegraphics[width=1.0\columnwidth]{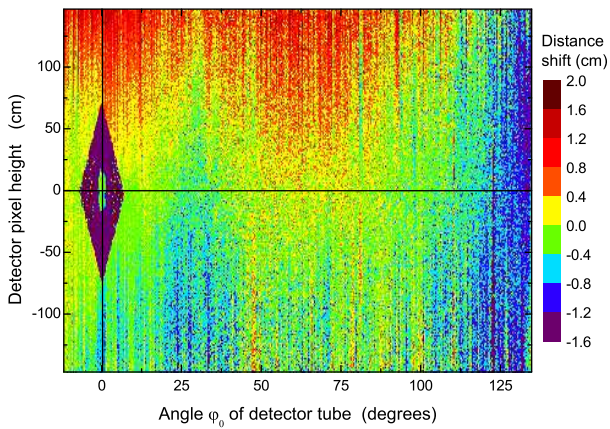}
	\caption{Distortion of the detectors plane: difference between the radial distance determined from the time of flight and the nominal IN5 radius (4.00\,m), for all detector pixels (angle $\varphi_0$, height $z$).  }
	\label{fig:DistanceAllDetectors}
\end{figure}

The elastic times yield, using at this early stage the value of t$_s$ obtained from the standard analysis (see Section \ref{sec:Corrections}), the distances of flight for each pixel. These are visualized by representing their projections on the horizontal plane (i.e., the radial distances), as a 2D-array shown in Fig. \ref{fig:DistanceAllDetectors}, where the nominal 4.00\,m have been subtracted. 
The data reveal a systematic distortion of the detectors plane along the angular direction, confirming the previous observation based on the standard analysis (see Fig. \ref{elasticTOF}). It is now shown, in addition, that the distortion is also present in the vertical direction: an undulation of the detector plane, similar to that of a distorted incompressible plate, is observed. 
A Debye-Scherrer average along the lines depicted in Fig. \ref{Constant-angle-curves} depends now on the detailed shape of the detector plane distortion, and the standard procedure is clearly inaccurate. 

\begin{figure}[h]
	\centering
		\includegraphics[width=0.8\columnwidth]{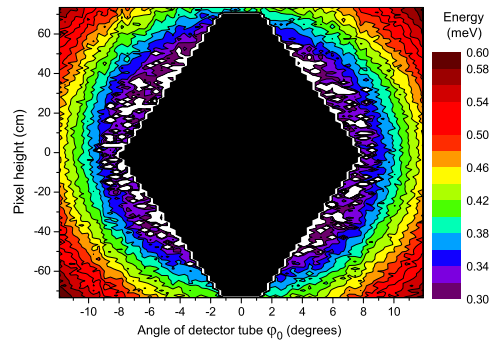}
	\caption{Debye-Scherrer rings seen on the low-angle pixels of the detector plane, in the phonon region (see energy scale on the right hand side). The black diamond is the beam-stop shadow.}
	\label{fig:Phonon-Debye-Scherrer-Ring}
\end{figure}

Another important information can be obtained from the pixelized analysis. The energy of the excitations and their scattering angle can now be calculated and visualized is a two-dimensional array, as shown in Fig. \ref{fig:Phonon-Debye-Scherrer-Ring} for the smallest angles.  
Contour fits made along the phonon Debye-Scherrer rings show that large sample off-centering (see Section \ref{sec:off-center})  can be excluded to a very good accuracy (a few mm). 
We have calculated the possible distortions of the detector assembly, which is by construction a rather rigid cylindrical wall, fixed at the floor level, rigidly held at the level of the middle plane, and rather free to move at the top. 
As suggested by Fig. \ref{fig:DistanceAllDetectors}, the detectors plane simply undulates. As a result, the distances to the center vary substantially, but angular corrections, a second order effect, are small. 
For our data, the angular correction is easily calculated by iteration, adding the successive angular deviations calculated for the measured distances corresponding to each detector tube. The correction, which reaches its maximum (0.07$^\circ$) at the largest angles (135$^\circ$), is very small.\\ 

Analysis step 2: we determine the value of E$_i$,the initial neutron energy, and  t$_s$, the time of arrival of the neutrons at the sample. 
As was explained in Section \ref{sec:TOFequations}, the nominal values of these parameters are not  accurate enough, and the dispersion relation calculated with these values is systematically too high by about 9\,$\mu$eV. We thus calibrate our energy scale \emph{at a single point} using the roton energy determined by Stirling \cite{stirling-83,StirlingExeter} on IN12, a high resolution triple-axis spectrometer: $\Delta_R$=0.7418$\pm$0.001\,meV.
The inelastic times we have measured, plotted as a function of the scattering angle, have a maximum value of t$_{inel}^{rot}$=(602.68$\pm$0.02)\,$\tau_{ch}$ at the roton; the corresponding elastic time is t$_{elast}^{rot}$=(514.02$\pm$0.05)\,$\tau_{ch}$ (a time channel is $\tau_{ch}$=6.9084\,$\mu$s). 

The data of Fig. \ref{fig:DistanceAllDetectors} show that the detector distances are close to the nominal value of 4.00\,m in the lower part of the plane and that they increase near the top. Taking into account the reduction of the effective flight distance due to the average penetration length in the $^3$He detectors (see Section \ref{sec:corrdet}), we estimate the average sample-detector distance in the roton region, L$_{rot}$$\approx$4.000$\pm$0.005\,m. 
Knowing the distance and the arrival time gives a relation between the initial neutron velocity v$_i$ (and hence the energy E$_i$) and  t$_s$, the time of arrival of the neutrons at the sample:
$L_{rot}=(t_{elast}^{rot}-t_s) v_i$, 
which can be solved together with Eq. \ref{Erefined} expressed at the roton: 

$\Delta_R= E_i {\left[ {1} - \left(\frac{t_{elast}^{rot}  - t_s} {t_{inel}^{rot} - t_s}\right)^2  \right]}$

We obtain E$_i$=3.520$\pm$0.003\,meV, v$_i$=820.62$\pm$0.3\,m/s, and t$_s$=(-191.55$\pm$0.4)\,$\tau_{ch}$.
As expected, the corrected neutron energy is slightly lower (by 0.85\%) than the nominal value.\\  

\begin{figure}[h]
	\centering
		\includegraphics[width=0.85\columnwidth]{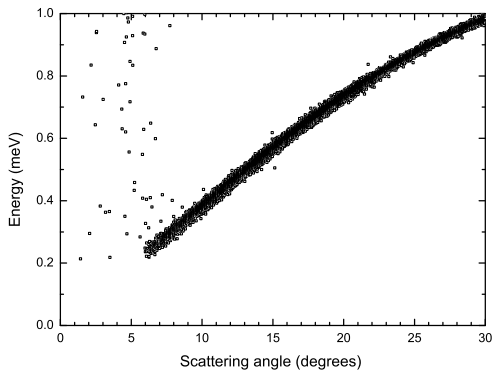}
	\caption{$\epsilon(\varphi)$ (analysis step 3) in the phonon region. Most of the data display the usual statistical distribution, but spurious data points are also present at very low angles, systematically above the main curve. The corresponding neutrons travel through indirect paths, they must be identified and eliminated from the analysis. }
	\label{fig:rawDataPhonon}
\end{figure}

Analysis step 3: with these parameters, we analyze with a Mathematica program the set of data pairs t$_{elast}$($\varphi_0$,$z$),t$_{inel}$($\varphi_0$,$z$). For each pixel, we calculate the excitation energy using Eq. \ref{Erefined}, and the corresponding Debye-Scherrer angle $\varphi$. The result is a curve $\epsilon(\varphi)$ with a very large number (9$\times$10$^4$) of independent data points. 
Fig. \ref{fig:rawDataPhonon} shows the results in the most delicate region, at low angles. 

There are obvious spurious data points, corresponding to neutrons reaching the detectors in an indirect way, as commonly observed in neutron scattering experiments.        
Also, a simple inspection of Fig. \ref{fig:DistanceAllDetectors} shows the presence of a few bad tubes and  bad pixels. In addition, some detectors located behind the beam-stop (diamond-like shadow at zero angle) or in its vicinity, cannot be exploited. Removing these spurious points leaves typically 88000 independent points of good quality. It is clearly desirable to average over several data points in order to improve the statistical uncertainty in the energy, as suggested by the dispersion seen in Fig. \ref{fig:rawDataPhonon}, at the expense of a reduced wave-vector resolution.  \\ 
 
\begin{figure}[h]
	\centering
		\includegraphics[width=0.8\columnwidth]{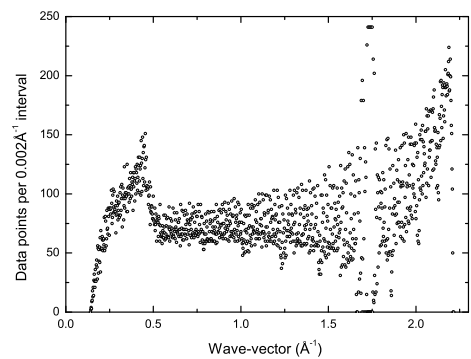}
	\caption{Number of data points per wave-vector interval of width 0.002\,\AA$^{-1}$.  }
	\label{fig:DataPointsPerQbin}
\end{figure}

Analysis step 4: the wave-vectors k corresponding to the E($\varphi$) data points are calculated using Eq. \ref{QAngle}. The resulting $\epsilon(k)$ data sets are averaged within 0.002\,\AA$^{-1}$ bins. There are about 10$^{3}$ bins on the dispersion relation at each pressure in the wave-vector range 0.14$<$k$<$2.25\,\AA$^{-1}$. 
The number of points per bin varies, as shown in Fig. \ref{fig:DataPointsPerQbin}, as a function of wave-vector. This is mainly due to the detectors geometrical layout: there are gaps between different groups of detector tubes, as described in Section \ref{sec:IN5}. Empty bins are also found around Q=1.729\,\AA$^{-1}$, which corresponds to angles near 90$^\circ$, where the Debye Scherrer cone is essentially a vertical plane.  

At the lowest wave-vectors, typically between 0.15 and 0.2\,\AA$^{-1}$, the number of data points per bin is small. Binning carries no benefit, and error bars in this region are dominated by statistical errors. Except for this small region, binning is done over a substantial number of points, typically more than 70. By trying different bin sizes, it becomes clear that going beyond about 50 points/bin does not improve the resulting dispersion curve: statistical errors become negligible compared to systematic errors. 

 \begin{figure}[t]
	\centering
		\includegraphics[width=0.9\columnwidth]{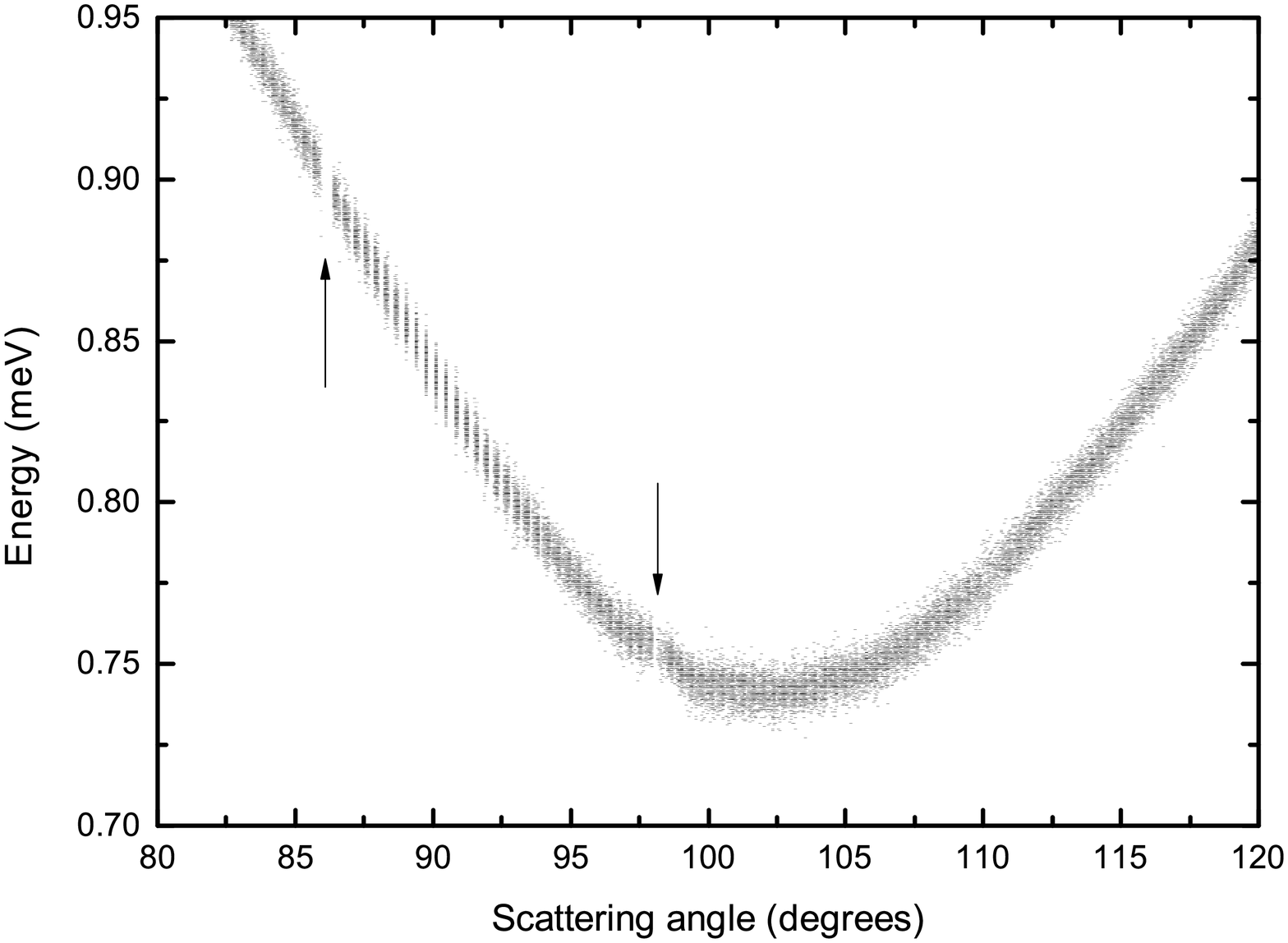}
	\caption{Raw data in the roton region: a point is the measurement from one pixel. The effect of the gaps present by construction between different groups of detector tubes (marked by arrows) are visible. Systematic errors can be seen in one of these regions, probably due to a local defect of the detector groups at their junction. Individual tubes are visible around 90$^\circ$.}
	\label{fig:MissingTubesAroundRoton}
\end{figure}

Some small oscillations can be seen in the data. An example is given in Fig. \ref{fig:MissingTubesAroundRoton}.
They are due to several factors, essentially deviations from the assumed parameters (instrument geometry, sample environment characteristics, detector properties, electronic delays, etc.). Correcting for these cannot be achieved by averaging neighboring points. These deviations correspond well to the error bars, calculated using the uncertainties in all these parameters, for data above 0.2\,\AA$^{-1}$. 
The uncertainty in Q, due to the uncertainty of the instrument angles (0.07$^\circ$, about 2/10 of a detector tube angular range) and to the uncertainty of the initial energy (0.003\,meV) (see Eq. \ref{QAngle}), can be represented by the expression $\Delta$Q=10$^{-4}$(7+7.2Q)\,\AA$^{-1}$. This corresponds essentially to a fraction of a bin. 
The uncertainty in the energy $\epsilon(k)$ has been determined by varying the parameters E$_i$, t$_s$, $D_{rot}$, $\Delta_R$ in the allowable parameter range. This is needed due to the non-linear character of the equations, and the strong correlation between E$_i$ and t$_s$, a problem already noted by Andersen \textit{et al.} \cite{AndersenThesis,Andersen94a}. The calculated relative uncertainty is essentially constant, $\Delta$E/E$\approx$ 2.1$\times10$$^{-3}$.

\section{The dispersion relation in the whole range}
\label{sec:results}
\begin{figure}[t]
	\centering
		\includegraphics[width=0.95\columnwidth]{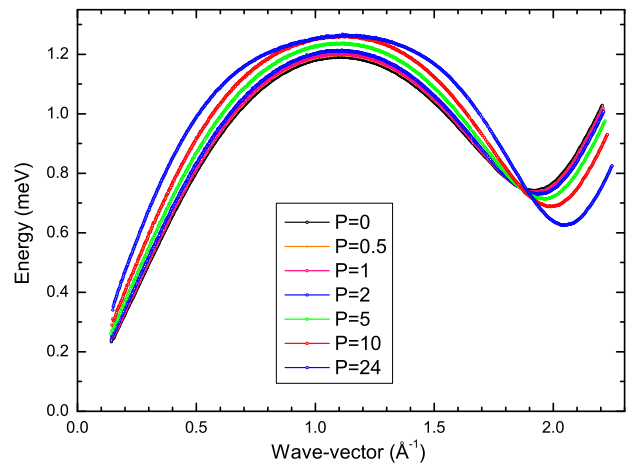}
	\caption{Dispersion curves $\epsilon(k)$ measured for several pressures in the 0 to 24\,bar range. The individual data points are represented by small circles (best seen on-line). Accurate values for the pressures are given in Table \ref{table:Pressures}. }
	\label{fig:allDispCurves}
\end{figure}

The dispersion relation at saturated vapor pressure in the whole wave-vector range is shown in Fig. \ref{fig:dispersion}. In this section, we first present high accuracy measurements of the pressure dependence in the particularly interesting wave-vector range below 2.3\,\AA$^{-1}$, shown in Fig. \ref{fig:allDispCurves}. Error bars are comparable to the size of the data points. 
The effects of pressure are clearly seen: the phonon sound velocity and the maxon energy increase, while the roton minimum decreases and shifts towards higher wave-vectors. A spectacular flattening of the maxon is observed at high pressures. In the following paragraphs, we provide a quantitative analysis of the experimental dispersion curves. 

\subsection{Phonons}
\label{sec:lowQ}

\begin{figure}[t]
	\centering
		\includegraphics[width=1.0\columnwidth]{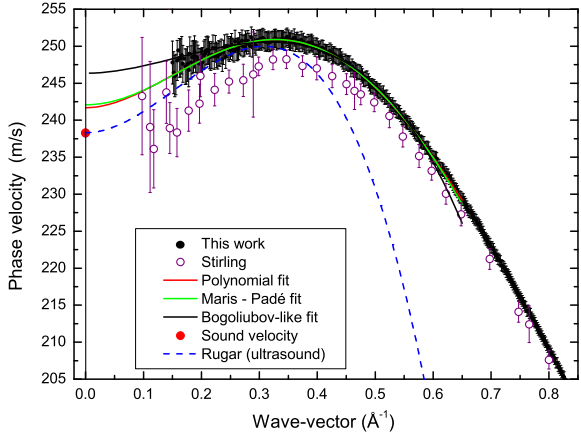}
	\caption{Phase velocity $\epsilon(k)$/$\hbar$k at P=0. Black dots with error bars: present results. 
	  Different fits made in the wave-vector range 0.2$<$k$<$0.6\,\AA$^{-1}$ are shown by thick lines (see legend and text). 
	Red dot at k=0: sound velocity \cite{Abraham,DonnellyBarenghi}. Extrapolated non-linear ultrasound data \cite{Rugar1984} are shown as a dashed blue line.
	Stirling's neutron data \cite{stirling-83,GlydeBook,StirlingExeter} are represented by purple circles.}
	\label{fig:phaseVelocity}
\end{figure}

\begin{figure}[htbp]
	\centering
		\includegraphics[width=1.0\columnwidth]{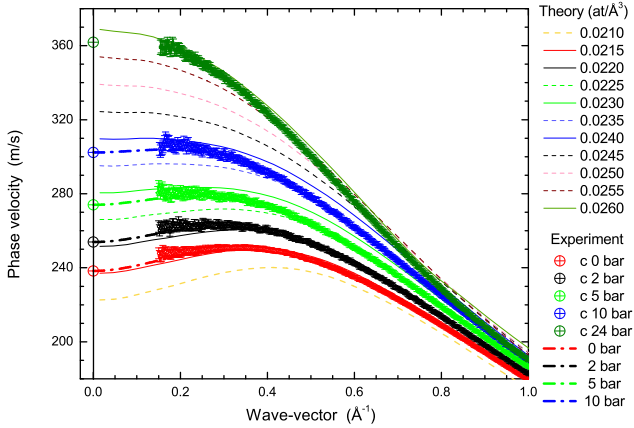}
	\caption{Phase velocity $\epsilon(k)$/$\hbar$k at P=0, 2, 5, 10, and 24 bar from bottom to top, dots with error bars (this work). The corresponding sound velocities \cite{Abraham,DonnellyBarenghi} are indicated at k=0 by crossed circles. 
	Non-linear ultrasound data \cite{Junker1977,Rugar1984} (available only between 0 and 15 bar), interpolated to P=0, 2, 5, 10 bar, and extrapolated to large wave-vectors: thick dash-dotted lines. 
	Theory: DMBT \cite{eomIII,phonons} curves are shown by thin lines for several atomic densities (see graph legend and Table \ref{table:soundveloc}). }
	\label{fig:phasevelocExpTheo}
\end{figure}

The behavior at low wave-vectors is shown in Fig. \ref{fig:phaseVelocity}, where the phase velocity $\epsilon(k)$/$\hbar$k 
is represented as a function of wave-vector k at P=0. 
The k$\sim$0 value of the ultrasonic data for the sound velocity (238.3$\pm$0.1\,m/s) and the curve calculated 
using Rugar and Foster non-linear ultrasonic data \cite{Rugar1984}, strongly extrapolated from low wave-vectors, are also shown.
It is already rewarding to observe that ultrasound and neutron data, in spite of their non-overlapping validity region, are perfectly compatible  
and smoothly merge around 0.2-0.25\,\AA$^{-1}$. 
For k$<$0.2\AA$^{-1}$, however, the neutron data are slightly too high in energy.
This is not surprising: spurious data points proliferate at the lower end of the wave-vector range, as discussed above 
(see Fig. \ref{fig:rawDataPhonon}), leading to systematic errors that increase the energies.
For wave-vectors as low as 0.15$<$k$<$0.2\AA$^{-1}$, Rugar and Foster's curve is still in good agreement with the neutron data within error bars (at their lowest limit). 
A similar behavior is observed at all pressures (Fig. \ref{fig:phasevelocExpTheo}). DMBT calculations, to be discussed in  detail below, are clearly in good quantitative agreement with the experiments. 

We also show in Fig. \ref{fig:phaseVelocity} the neutron scattering data obtained at saturated vapor pressure by 
Stirling \cite{stirling-83,GlydeBook,StirlingExeter}. 
The comparison is of particular importance, because they have been measured on a   
triple-axis spectrometer (IN12), i.e., using a very different neutron technique.
It is obvious that the IN12 phonon energies are too low: the difference with our data as well as with Rugar and Foster's curve exceeds error bars for k$<$0.3\,\AA$^{-1}$, and matching the ultrasound velocity is clearly impossible   
unless the error bars of IN12 data are significantly increased. 
Systematic errors at low wave-vectors are indeed expected, given the large size of Stirling's helium sample and the short length of the IN12 instrument, 
as well as other particular features of the resolution function of triple-axis (TAS) spectrometers. 
At higher wave-vectors, where both techniques are relatively free from systematic errors, 
a very good agreement between our TOF data and Stirling's TAS data is observed, which constitutes an important experimental test.

Several functional forms describing the low wave-vector sector of the dispersion curve have been proposed (see Section \ref{sec:sound}). 
Figure \ref{fig:phaseVelocity} shows fits  made in the range 0.2$<$k$<$0.6\,\AA$^{-1}$. 
The first fit uses the polynomial expansion of Eq.\,\ref{eq:dispfit}, written now in practical units as 
 
\begin{equation}
\epsilon(k)=0.0065821 c k (1+ \alpha_2 k^2 + \alpha_3 k^3 + \alpha_4 k^4) 	
\label{eq:anomdisppol}
\end{equation}

where  $\alpha_2=-\gamma$, $\epsilon(k)$ is expressed in meV, 
k in \AA$^{-1}$, c in m/s, and the $\alpha_i$ coefficients in \AA$^{i}$.
The sound velocities obtained from neutron data using the polynomial fit are given in Table\,\ref{table:soundveloc}. 
At 24 bar the dispersion is normal, and a simple quadratic fit ($\alpha_3=\alpha_4=0$) is sufficient to describe the data very well, 
changing the speed of sound by a small amount, within error bars, with respect to the result obtained with the full expression. 
The Pad\'e approximant (Eq. \ref{eq:dispfitMaris}) is very sensitive to the upper limit of k used in the fit. 
It tends to overestimate the sound velocity, and  
the same conclusion applies to the expression derived from Bogoliubov's formula (Eq. \ref{eq:dispfitBogoliubov}). 

\begin{table}[h]
\begin{ruledtabular}
\begin{tabular}{llclllll}
P     & n          & Neutron  & err  & Sound  & err   & V$_m$(P)  & C$_V$     \\ 
bar   & at/\AA$^3$ & m/s      & m/s  & m/s    & m/s   & m/s    & m/s    \\ \hline
0     & 0.021836   & 241.7    & 2.9  & 238.3  & 0.1   & 237.76 & 236.8  \\ 
0.51  & 0.021968   & 246.4    & 7.2  & 242.6  & 0.1   & 241.98 & 240.7  \\ 
1.02  & 0.022096   & 251.0    & 6.6  & 246.5  & 0.1   & 246.04 & 244.3  \\ 
2.01  & 0.022334   & 257.9    & 6.9  & 253.9  & 0.1   & 253.53 & 251.2  \\ 
5.01  & 0.022983   & 278.1    & 7.1  & 274.0  & 0.1   & 273.73 & 270.5  \\ 
10.01 & 0.023889   & 308.6    & 7.8  & 302.3  & 0.1   & 301.75 & 298.6  \\ 
24.08 & 0.025804   & 364.7    & 1.9  & 361.9  & 0.1   & 359.28 & 357.9  \\ 
24.08 & 0.025804   & \textit{367.1}    & \textit{4.6}  &        &       &     &        \\ 
\end{tabular}
\end{ruledtabular}
\caption{Sound velocities obtained from neutron scattering (this work), ultrasound \cite{Abraham,DonnellyBarenghi}, molar volume pressure dependence \cite{Tanaka2000}, and heat capacity (Greywall, analysis 4)\cite{Greywall78,Greywall79ERRATUM}, at different pressures (n is the atomic density). At 24 bar, the upper line corresponds to a quadratic fit, while the last  line gives the result of the full fit, for comparison (see text).}
\label{table:soundveloc}
\end{table}

The sound velocities deduced from the present neutron scattering measurements are compared in Table\,\ref{table:soundveloc} to the ultrasonic sound velocities \cite{Abraham,DonnellyBarenghi}, to those obtained (at interpolated densities) from Greywall's heat capacity (C$_V$) measurements \cite{Greywall78,Greywall79ERRATUM}, and to the values we calculate from the compressibility (molar volume pressure dependence) determined by Tanaka \textit{et al.} \cite{Tanaka2000} (see Section \ref{sec:sound}. As noted for the P=0 data, the neutron scattering values are higher than the ultrasonic ones, but the difference is within error bars. 
The speed of sound we calculate from the compressibility agrees very well with the ultrasonic data, except at the highest pressure, where either the ultrasonic data or, most likely, the compressibility data become somewhat inaccurate. 
Heat capacity data for the speed of sound are systematically lower than the ultrasonic values. Error bars are not quoted, but their sensitivity to different methods of data analysis \cite{Greywall78,Greywall79ERRATUM} suggests that the uncertainties are comparable to our estimated errors for the neutron data.

\begin{figure}[h]
	\centering
		\includegraphics[width=1.0\columnwidth]{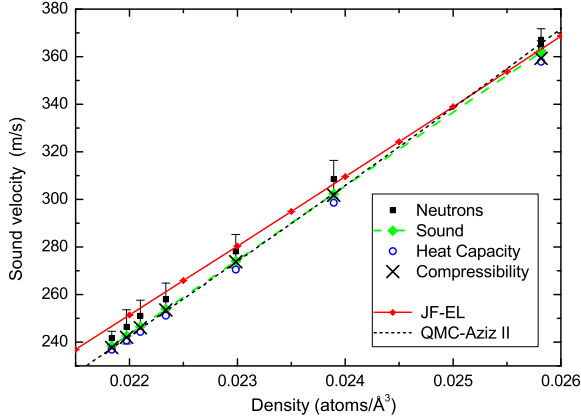}
	\caption{Sound velocities from ultrasound \cite{Abraham,DonnellyBarenghi}, heat capacity \cite{Greywall78,Greywall79ERRATUM}, our analysis of compressibility \cite{Tanaka2000} data, and the present neutron scattering measurements, as a function of density (see Table \ref{table:soundveloc}). Red line: Jastrow-Feenberg Euler-Lagrange calculation \cite{eomIII,phonons}. Short-dash line: Quantum Monte-Carlo calculation \cite{1994-BoronatQMC} with Aziz II potential.}
	\label{fig:SoundVeloc}
\end{figure}

The values of the speed of sound and their density dependence determined using different techniques are in excellent agreement, they only display a small overall shift within error bars, as can be seen in Fig. \ref{fig:SoundVeloc}. The same observation applies to the polynomial fits of the dispersion curve made in different ranges (0.015$<$k$<$0.5, 0.015$<$k$<$0.6, and 0.18$<$k$<$0.6\,\AA$^{-1}$) with the Jastrow-Feenberg Euler-Lagrange microscopic theory \cite{eomIII,phonons}. The resulting sound velocities display a small dependence on the selected k-range, not visible at the scale of Fig. \ref{fig:SoundVeloc}. The figure also shows the results of Quantum Monte-Carlo calculations \cite{1994-BoronatQMC} performed with the Aziz II potential, in excellent agreement with the experiments.
 
We focus now our attention on the determination of the anomalous dispersion parameter $\gamma$ as a function of density. 
Fits were made with equation \ref{eq:anomdisppol} using the ultrasonic sound velocities $c$ to reduce the free parameters to $\alpha_2$=$-\gamma$, $\alpha_3$ and $\alpha_4$. 
As seen from Fig.\,\ref{fig:phaseVelocity} and its accompanying discussion, the choice of the wave-vector fit range of the experimental data has to be made with care. The chosen lower limit was k=0.25\,\AA$^{-1}$ to reduce systematic errors, and k=0.5\,\AA$^{-1}$ was used as the upper limit; we also checked with fits extended to k=0.6\,\AA$^{-1}$ the effect of the fit range on the accuracy. With 125 data points in this range, statistical error bars were small.     

\begin{figure}[h]
	\centering
		\includegraphics[width=1.0\columnwidth]{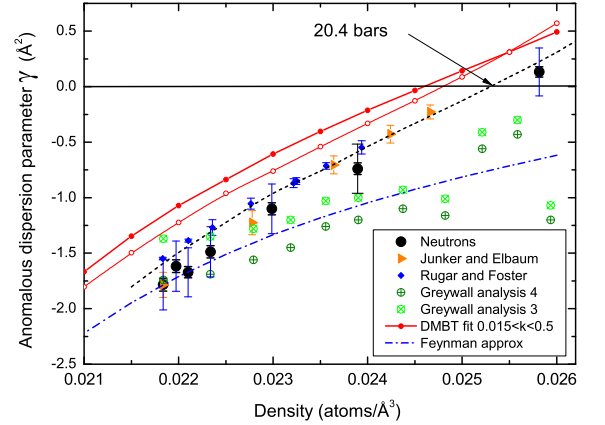}
	\caption{The anomalous dispersion parameter $\gamma$ as a function of density. Black dots: neutrons, this work (small error bars: statistical uncertainty, larger bars: systematic uncertainty). Ultrasound: blue lozenges \cite{Rugar1984} and orange triangles \cite{Junker1977}. Green circles: heat capacity  \cite{Greywall78,Greywall79ERRATUM}. Blue dash-dotted line: Feynman approximation. Red lines: DMBT results \cite{eomIII,phonons} (dots and circles correspond to polynomial fits in different k-ranges. Black dashed line: DMBT \cite{eomIII,phonons} fit rescaled in density by 0.0005\,\AA$^{-3}$. See legend and text for details. 
}
	\label{fig:gamma}
\end{figure}

The anomalous dispersion parameter $\gamma$ obtained from this analysis is shown in Fig.\,\ref{fig:gamma}. Two types of error bars are given for each data point; the smaller bars indicate the statistical uncertainty, while the larger ones give the estimated systematic errors, associated to the uncertainty in the global parameters of the analysis, described in Section \ref{sec:data-acquisition}.
The statistical error bars are small, and therefore, {\em{only a global shift of the whole experimental curve within the systematic error bars is allowed}}. 
The results are compared in Fig. \ref{fig:gamma} to data from two different types of ultrasonic measurements, performed respectively by  
 Junker and Elbaum\cite{Junker1977} (temperature dependence), and Rugar and Foster \cite{Rugar1984} (non-linear measurements)
(see Ref. \onlinecite{Sridhar87} for a critical review and references to former data). 
The neutron data agree well in magnitude with these, and their density dependence, in particular, agrees extremely well. A small global systematic shift, as described above, is observed.  
At the highest densities, where ultrasound data do not exist, the neutron scattering result confirms the evolution towards positive values of $\gamma$ extrapolated from the ultrasonic data. 

The density dependence of neutron scattering and ultrasonic data for $\gamma$ can be described quite remarkably by the DMBT theory \cite{eomIII,phonons}. 
Polynomial fits are a simple method to compare theory to experiments, but the results depend on the choice of the wave-vector range. 
Several fits were made with Eq. \ref{eq:anomdisppol}, starting with the range (0.015-0.5\,\AA$^{-1}$). We then made fits in a higher range, 
0.18$<$k$<$0.6\,\AA$^{-1}$, comparable to the experimentally accessible range, in order to estimate a possible correction on the experimental $\gamma$. 
The correction suggested by theory places the neutron data exactly on top of the ultrasonic results. Normal dispersion (i.e., $\gamma$=0) is recovered for pressures larger than 20.4\,bars.  
A perfect quantitative theoretical fit of the experimental data is obtained in the whole density range if  the theoretical densities are globally increased by 0.0005\,\AA$^ {-3}$, a very small correction  which is compatible with the uncertainties in the theoretical equation of state \cite{eomIII}.

On the other hand, the curve calculated using the Feynman approximation is clearly inadequate at high densities, where correlations are strongest. 
The phenomenological theory of Pines and coworkers \cite{Aldrich,aldrich1976phonon} estimated $\gamma$$\sim$-1.5\,\AA$^2$ at the saturated vapor pressure, a value in good agreement with the present experiment and the DMBT microscopic theory. 

For completeness, we also show in Fig.\,\ref{fig:gamma} Greywall's heat capacity results \cite{Greywall78,Greywall79ERRATUM}. Error bars, not provided, can be roughly estimated from the change in $\gamma$ observed in different types of analysis. Heat capacity data agree reasonably well at low densities with the results discussed above. However, a very large systematic discrepancy is seen at high densities; in particular, the transition to normal dispersion is not observed.     

\begin{figure}[h]
	\centering
		\includegraphics[width=1.0\columnwidth]{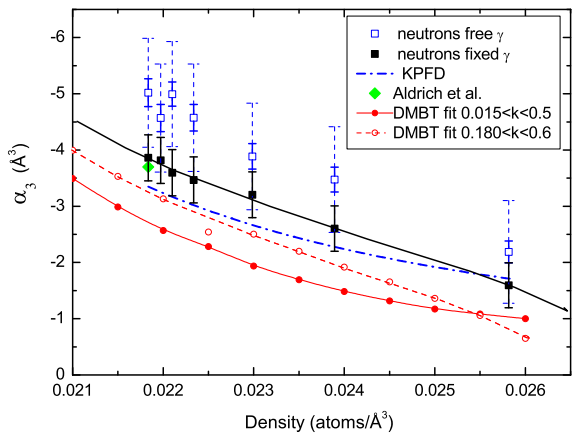}
	\caption{The parameter $\alpha_3$ as a function of density. Open blue squares: present work, fit parameters =($\gamma$,$\alpha_3$,$\alpha_4$) (small error bars: statistical uncertainty, larger bars: systematic uncertainty, see text). Black squares: present work, fit parameters =($\alpha_3$,$\alpha_4$); error bars include both types of uncertainties. Dashed blue line: KPFD expression \cite{Pitaevskii1970,Feenberg1971,Davison_1966}. Red dots: DMBT results \cite{eomIII,phonons} (fits in indicated k-ranges). Diamond: Aldrich, Pethick and Pines \cite{Aldrich,aldrich1976phonon}. }
	\label{fig:alpha3}
\end{figure}

The parameter $\alpha_3$, which originates in the long-range part of the van der Waals interaction between helium atoms, is of theoretical interest.  
Kemoklidze and Pitaevskii \cite{Pitaevskii1970} and Feenberg \cite{Feenberg1971} calculated $\alpha_3=-3.34$\,\AA$^{3}$ at saturated vapor pressure (see Eq. \ref{eq:alpha3}). The density dependence was obtained from Davison's formula \cite{Davison_1966,Rugar1984,Greywall78,Greywall79ERRATUM}. 
The results are shown in Fig. \ref{fig:alpha3}. This parameter is negative in the whole density range, its magnitude is about -3\,\AA$^{3}$, with a slow variation as a function of density. 
Fits were made using $\gamma$ as a free parameter, and also using our `best estimate' for $\gamma$ (black dashed line in Fig. \ref{fig:gamma}) discussed above. The statistical uncertainties are rather small in both cases, and systematic uncertainties dominate. Having checked that the two sets of results are consistent,  we use in the following the values of $\alpha_3$ calculated with the ultrasound values of  $\gamma$. 
The magnitude of $\alpha_3$ and its density dependence are in good agreement with the Kemoklidze-Pitaevskii-Feenberg-Davison (KPFD) expression \cite{Pitaevskii1970,Feenberg1971,Davison_1966}. 
The pseudo-potentials phenomenological calculation by Aldrich, Pethick and Pines \cite{Aldrich,aldrich1976phonon} yields $\gamma$=-1.5\,\AA$^{2}$ and $\alpha_3$=-3.7\,\AA$^{3}$.
Our fits of their published curves show that these results depend on the wave-vector range.  For 0$<$k$<$0.4\,\AA$^{-1}$ we find 
$\gamma$=-(1.57$\pm$0.02\,\AA$^{2}$ and $\alpha_3$=-(4.0$\pm$0.1)\,\AA$^{3}$. 
The original values \cite{Aldrich,aldrich1976phonon} are found only if we extend the fit range beyond k=0.6. 

The results of the microscopic DMBT calculation \cite{eomIII,phonons} are shown in Fig. \ref{fig:alpha3}. Again, the results depend on the wave-vector range selected for the fits. 
We note that the density dependence of $\alpha_3$ calculated using a low wave-vector range (0.015$<$k$<$0.5\,\AA$^{-1}$) is remarkably similar to that predicted by KPFD. The fit done at higher wave-vectors (0.18$<$k$<$0.6\,\AA$^{-1}$) agrees particularly well with the neutron data. 
We conclude that the neutron scattering measurement and the microscopic DMBT calculation agree reasonably well with the $\alpha_3$ values predicted by KPFD. 
We also note that the present work is the only source of experimental data on $\alpha_3$ up to now. 

\begin{figure}[h]
	\centering
		\includegraphics[width=0.9\columnwidth]{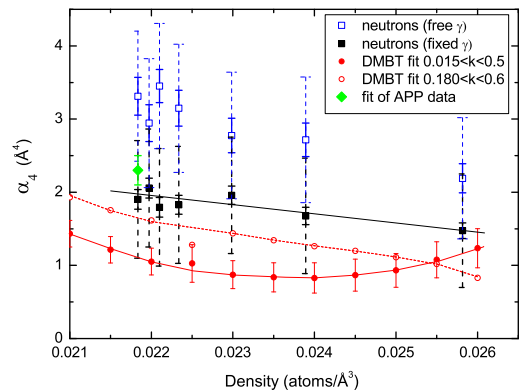}
	\caption{The parameter $\alpha_4$ as a function of density. Open blue squares: present work, fit parameters =($\gamma$,$\alpha_3$,$\alpha_4$) (small error bars: statistical uncertainty, larger bars: systematic uncertainty, see text). Black squares: present work, fit parameters =($\alpha_3$,$\alpha_4$); error bars include both types of uncertainties. Red dots: DMBT results \cite{eomIII,phonons} (fits in indicated k-ranges). Diamond: our fit of data by Aldrich, Pethick and Pines \cite{Aldrich,aldrich1976phonon}.} 
	\label{fig:alpha4}
\end{figure}

The results for $\alpha_4$, the next term in the series expansion obtained with the fits described above, are shown in Fig. \ref{fig:alpha4}. This parameter, determined here experimentally for the first time, is positive in the whole density range, its magnitude is about 2\,\AA$^{4}$, with a slow variation as a function of density. The results can be discussed in a very similar way as done above for $\alpha_3$. The values depend on the fit range. Our fits to the pseudo-potential theory published curves \cite{Aldrich,aldrich1976phonon} give $\alpha_4$$\sim$2.3$\pm$0.2, which agrees well with our neutron data. The values calculated with the microscopic theory (DMBT) are in good agreement with the neutron data, when comparable wave-vector ranges are used for the fits. 

\begin{figure}[h]
	\centering
		\includegraphics[width=1.0\columnwidth]{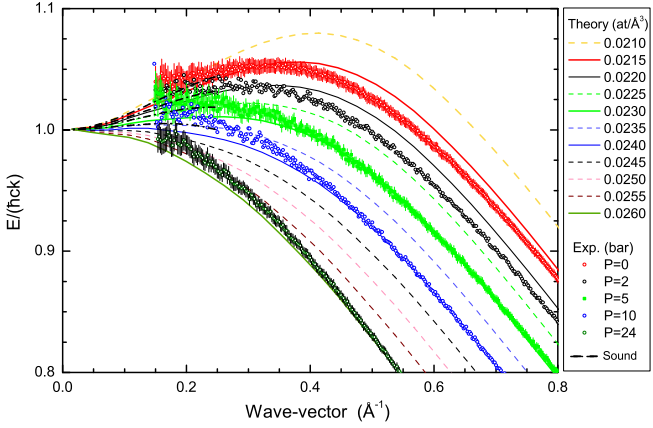}
	\caption{Normalized phase velocity $\epsilon(k)$/$\hbar$ck as a function of wave-vector k. Circles with error bars: present results at P=0, 2, 5, 10, and 24 bar (from top to bottom). Non-linear ultrasound data \cite{Junker1977,Rugar1984} (available only between 0 and 15 bar) have been interpolated to P=0, 2, 5, 10 bar, and extrapolated to wave-vectors 0$<$k$<$0.25\,\AA$^{-1}$ (black dashed lines). Theory: DMBT \cite{eomIII,phonons} curves for several atomic densities (see Table \ref{table:soundveloc}). } 
	\label{fig:AnomDispNorm}
\end{figure}

The results are often presented in terms of the normalized phase velocity, $\epsilon$/$\hbar$ck, in order to emphasize the transition from anomalous to normal dispersion as a function of pressure. The experimental data (neutrons and ultrasound) are shown in Fig. \ref{fig:AnomDispNorm}, together with the set of curves calculated by the DMBT \cite{eomIII,phonons}. 

\begin{figure}[h]
	\centering
		\includegraphics[width=1.0\columnwidth]{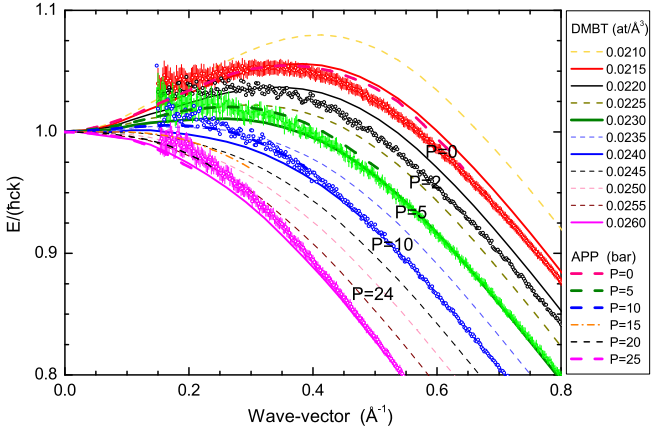}
	\caption{Normalized phase velocity $\epsilon(k)$/$\hbar$ck as a function of wave-vector k. Experimental data at at P=0, 2, 5, 10, and 24 bar (circles with error bars) are compared to the pseudo-potential calculation of Aldrich, Pethick and Pines \cite{Aldrich,aldrich1976phonon} (dashed curves) and to the DMBT results \cite{eomIII,phonons} for several atomic densities (see Table \ref{table:soundveloc}).  } 
	\label{fig:AnomExpPinesDMBT}
\end{figure}

It is interesting to compare in this `anomalous dispersion' representation, the results obtained from two very different theoretical approaches, pseudo-potentials and DMBT. Even at the substantially expanded vertical scale of Fig. \ref{fig:AnomExpPinesDMBT}, a remarkable agreement is observed, within the uncertainties of comparable magnitude estimated for experiments and theory.  

\subsection{Phase and group velocities}
\label{sec:highQ}

\begin{figure}[h]
	\centering
		\includegraphics[width=1.0\columnwidth]{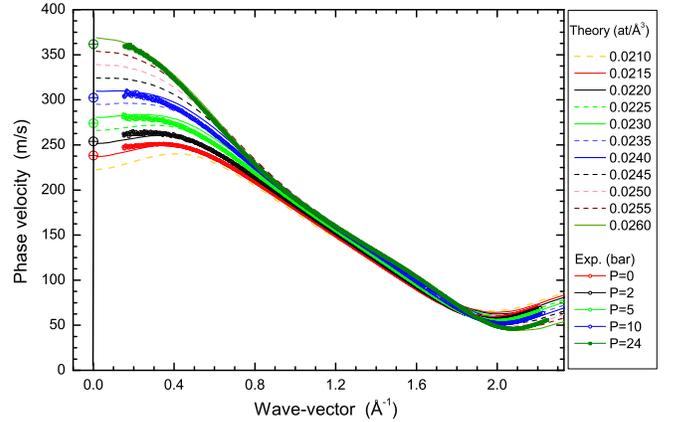}
	\caption{Phase velocity $\epsilon(k)$/$\hbar$k at P=0, 2, 5, 10, and 24 bar (from bottom to top, circles with error bars). DMBT curves \cite{eomIII,phonons} are shown for several atomic densities (see Table \ref{table:soundveloc}). A linear dependence on the wave-vector is observed in a large range (see text).}
	\label{fig:phasevelocLargeRange}
\end{figure}

We have shown in the previous section that the polynomial expansion (Eq. \ref{eq:anomdisppol}) becomes inaccurate as one considers wave-vectors in the atomic range. As seen in Fig. \ref{fig:phasevelocLargeRange}, the phase velocity curves display a very peculiar behavior for 0.5$<$k$<$1.8\,\AA$^{-1}$: they become linear to a high degree of accuracy. This is observed both in the experiment and in the DMBT curves, at all densities (with some changes at the highest pressures, where the maxon is strongly damped). It is obvious that adding higher order terms in a series expansion around k=0 in order to describe this type of high-k dispersion requires a strong compensation of successive terms, and is inadequate. 
One can easily check (series expansion of the phase velocity around the maxon wave-vector) that this linear term is a consequence of the maxon parabolic dispersion relation, combined with the fact that the maxon energy at low pressures is numerically very close to $-\hbar^2 k_M^2/2 m_M$, where $k_M$ and $m_M$ are the maxon wave-vector and its (negative) mass, respectively. 

\begin{figure}[h]
	\centering
		\includegraphics[width=1.0\columnwidth]{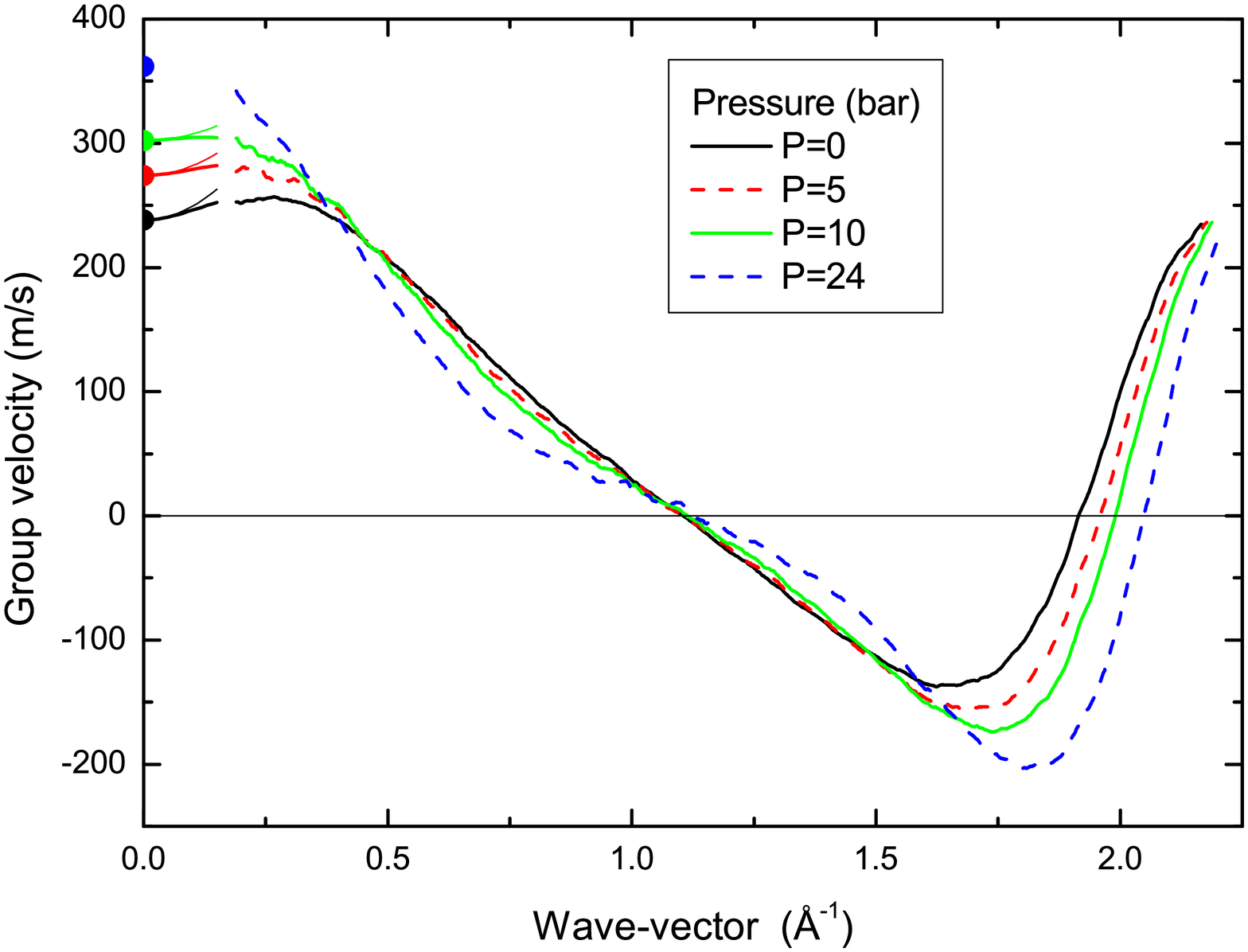}
	\caption{Group velocity $\partial$$\epsilon(k)$/$\hbar$$\partial$k at P=0, 5, 10, and 24 bar. Sound velocities are indicated by dots at k=0. At low wave-vectors, thin lines show the parabolic dependence calculated from ultrasound data \cite{Junker1977,Rugar1984} (available only between 0 and 15 bar), and thick lines the polynomial expansion including the calculated $\alpha_3$ term \cite{Pitaevskii1970,Feenberg1971,Davison_1966}.}
	\label{fig:Group velocity}
\end{figure}

\begin{figure}[h]
	\centering
		\includegraphics[width=1.0\columnwidth]{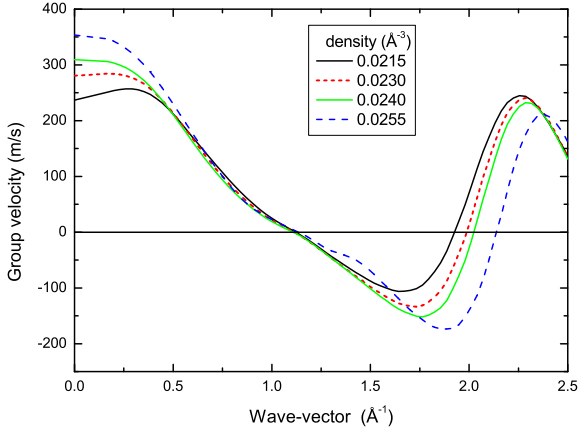}
	\caption{Group velocity $\partial$$\epsilon(k)$/$\hbar$$\partial$k calculated by DMBT \cite{eomIII} at selected atomic densities (compare to experiment, Fig. \ref{fig:Group velocity}).}
	\label{fig:GroupVelocityDMBT}
\end{figure}

Our dense data-set allows to calculate the group velocity by numerical differentiation with a good accuracy. The result is shown in Fig. \ref{fig:Group velocity}, where a 40-points average is used for clarity. The graph emphasizes the behavior at low wave-vectors (the importance of the $\alpha_3$ term), and the behavior around the maxon and roton wave-vectors, where the group velocity vanishes. The corresponding DMBT results are shown in Fig. \ref{fig:GroupVelocityDMBT}. The selected densities are extremely close to the values corresponding to the experimental pressures (P=0, 5, 10, and 24 bar), see Table \ref{table:soundveloc}. The agreement between theory and experiment is remarkable.  

\subsection{Rotons}
\label{sec:roton}

\begin{figure}[h]
	\centering
		\includegraphics[width=1.0\columnwidth]{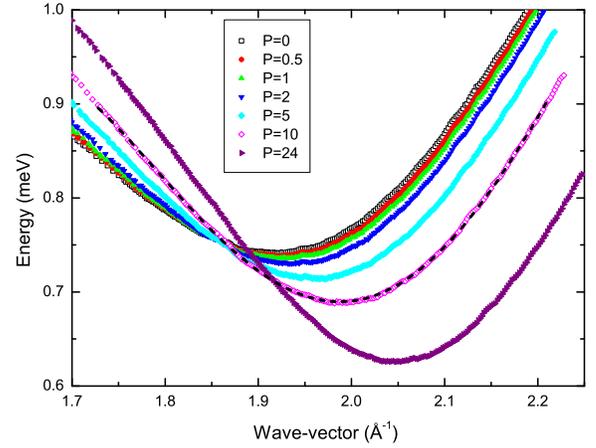}
	\caption{The dispersion relation in the vicinity of the roton minima at different pressures (accurate values of P are given in Table \ref{table:Pressures}). The dashed line on the 10 bar curve shows a quartic fit using the largest acceptable wave-vector range (see text). }
	\label{fig:rotonfit}
\end{figure}

We now concentrate on the properties of the dispersion curve around the roton. 
Given the large number of data points with small error bars, it is possible to calculate the main parameters of these excitations (energy, wave-vector, and mass) for the different pressures using a quartic polynomial expression:
\begin{equation}
	\epsilon_R(k)=\Delta_R+\frac{\hbar^2}{2m_4\mu_R}(k-k_R)^2+B_R(k-k_R)^3+C_R(k-k_R)^4
	\label{Eq:fitroton}
\end{equation}

where $\Delta_R$, $k_R$, $\mu_R$, $B_R$, and $C_R$ are adjustable parameters.
$\Delta_R$ is the roton gap defined before, $k_R$ the roton wave-vector, and $\mu_R$ the effective roton mass. The best fits are obtained using an asymmetric wave-vector range, from $k_R$-0.2\AA$^{-1}$ to $k_R$+0.3\AA$^{-1}$. Different ranges were tested, with a number of data points in the  100 to 200 points range. Under these conditions, the parameters $\Delta_R$, $k_R$, and $\mu_R$ do not depend on the wave-vector range selected for the fits. Quadratic fits in a small range ($\leq$0.1\AA$^{-1}$) give essentially the same results: statistical errors on the resulting parameters are very small, and the dominating uncertainties essentially originate from systematic errors. For example, the very small wiggles seen in the dispersion curves are due to imperfections of the detectors, and not to statistics.
A typical fit is shown in Fig. \ref{fig:rotonfit}, and the results for all pressures are given in Table \ref{tab:roton}.

The present data at saturated vapor pressure are compared to the results of previous works in Table \ref{tab:rotonCompare}. 
The roton gap $\Delta_R$ is known with an accuracy of 1\,$\mu$eV. Early measurements indicated values close to 0.743\,meV or higher, but a slightly lower value ($\Delta_R$=0.7418(10)) was  obtained by Stirling using a high resolution spectrometer. As seen above, we have used this value in order to calibrate the IN5 spectrometer in energy, and a remarkable agreement with the ultrasonic sound velocities was obtained at very low wave-vectors. 

\begin{table}[h]
\begin{ruledtabular}
\begin{tabular}{lllllll}
$P$ (bar) & & $\Delta_R$ (meV)& &  $k_R$ (\AA$^{-1}$) & & $\mu_R$   \\ \hline
0    & & \textit{0.7418(10)}& &  1.918(2)   & & 0.141(2)      \\ 
0.51 & & 0.7388(10)         & &  1.923(2)   & & 0.139(2)      \\ 
1.02 & & 0.7360(10)         & &  1.927(2)   & & 0.137(2)      \\ 
2.01 & & 0.7307(10)         & &  1.935(2)   & & 0.135(2)      \\ 
5.01 & & 0.7148(10)         & &  1.957(2)   & & 0.125(2)      \\ 
10.01& & 0.6895(10)         & &  1.988(2)   & & 0.114(2)      \\ 
24.08& & 0.6261(10)         & &  2.048(2)   & & 0.091(2)      \\ 
\end{tabular}
\end{ruledtabular}
\caption{Roton parameters deduced from fits using Eq. \ref{Eq:fitroton} around the roton minimum. The roton energy at P=0 is taken from Refs. \onlinecite{stirling-83,GlydeBook,StirlingExeter} in our instrument calibration procedure. }
\label{tab:roton}
\end{table}

\begin{table}[h]
\begin{ruledtabular}
\begin{tabular}{llll}
		& $\Delta_R$ (meV)  & $k_R$ (\AA$^{-1}$) &  $\mu_R$ \\ \hline
		This work     & \textit{0.7418(10)}    & 1.918(2)  & 0.141(2)\\
		Woods 1977    & 0.7426(10) & 1.926(5)  & 0.126(30)    \\
		Stirling 1991 & 0.7418(10) & 1.920(2)  & 0.136(5)    \\
		Andersen 1992-1994 & 0.743(1) & 1.931(3)  & 0.144(3)    \\
		Gibbs 1999    & 0.7426(21) & 1.929(2)  & 0.161(4)    \\
		Pearce 2001   & 0.7440(20) & 1.926(-)  & 0.166(10)    \\
\end{tabular}
\end{ruledtabular}
\caption{Zero pressure roton parameters (this work) compared to previous results:  Woods \textit{et al.} \cite{WoodsHilton}, Stirling \cite{stirling-83,GlydeBook,StirlingExeter}, Andersen \textit{et al.} \cite{AndersenThesis,Fak-Andersen-91,Andersen92,Andersen94a,Andersen94b},  Gibbs \textit{et al.} \cite{GibbsThesis,AndersenRoton}, and Pearce \textit{et al.} \cite{Pearce-Azuah-Stirling}. $\Delta_R$ at P=0 is taken from Ref. \onlinecite{stirling-83,StirlingExeter} in our instrument calibration procedure. }
	\label{tab:rotonCompare}
\end{table}

We find a value of $k_R$ slightly lower than Stirling's (the most accurate available so far), within small and comparable error bars. Higher values, outside error bars, are found in the literature (Table \ref{tab:rotonCompare}). A similar remark applies to the roton effective mass:  
we obtain a value that agrees well with that found by Stirling  \cite{stirling-83,GlydeBook,StirlingExeter} and Andersen \textit{et al.} \cite{AndersenThesis,Fak-Andersen-91,Andersen92,Andersen94a,Andersen94b}, but disagrees with those found by Gibbs \textit{et al.} \cite{GibbsThesis,AndersenRoton} and Pearce \textit{et al.}\cite{Pearce-Azuah-Stirling}.
The position and curvature of the roton minimum can be determined in our case with a better accuracy, simply because of our much larger number of data points in the small wave-vector range of interest. 

\begin{figure}[t]
	\centering
		\includegraphics[width=1.0\columnwidth]{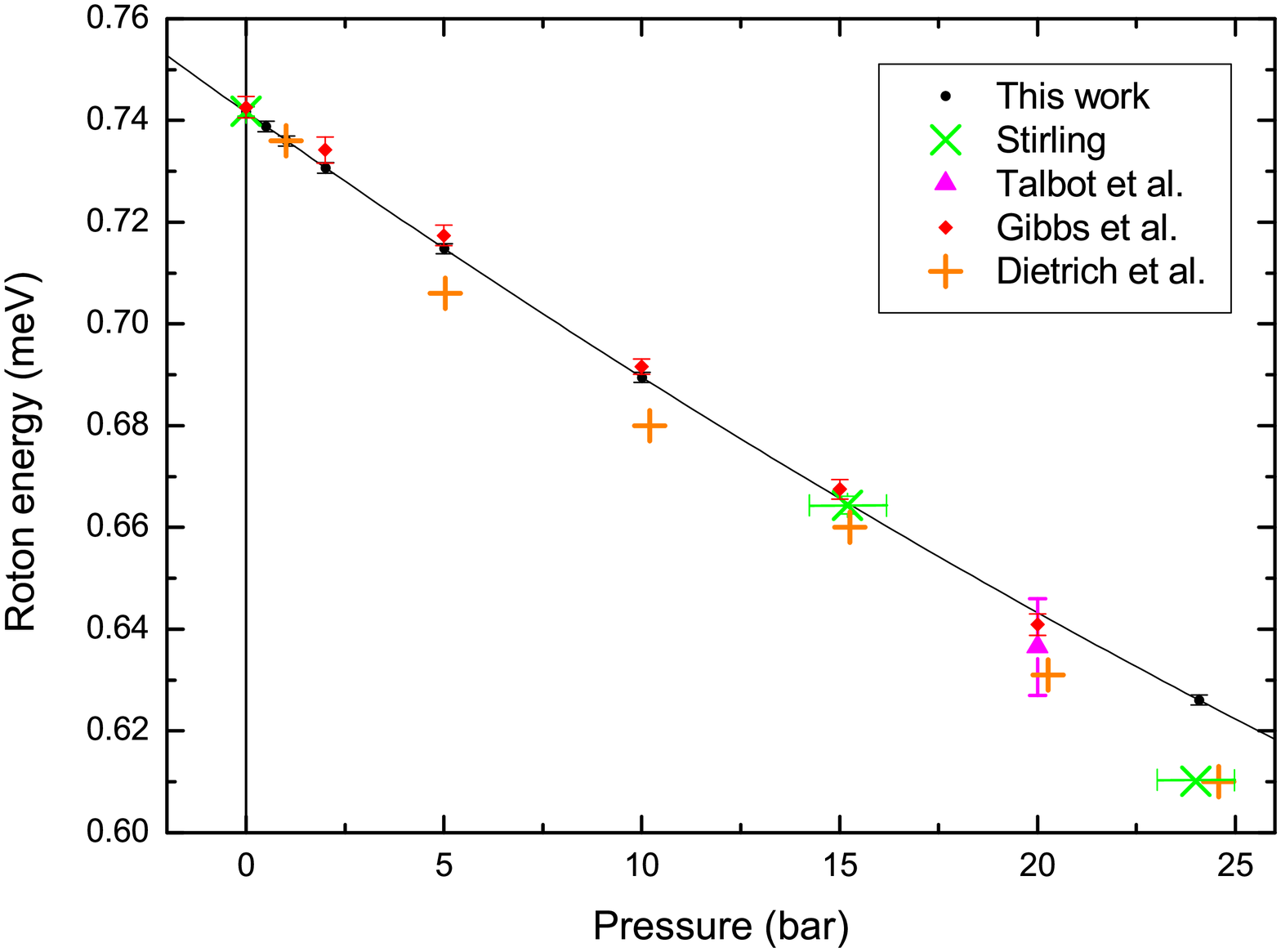}
	\caption{Pressure dependence of the roton energy at very low temperatures. Small black dots: this work; the solid line is a 2$^d$ order polynomial fit. Energy statistical error bars, and uncertainties in the pressures, are not visible at this scale; the plotted energy error bars correspond to the systematic uncertainty in the P=0 roton gap (see text). Green $\times$: Stirling \cite{stirling-83,GlydeBook,StirlingExeter}. Red diamonds: Gibbs \textit{et al.} \cite{GibbsThesis,AndersenRoton} (pressure error bars unknown). Orange +: Dietrich \textit{et al.} \cite{Dietrich72} (at T=1.3\,K). }
	\label{fig:RotonVsPressure}
\end{figure}

This becomes obvious in the representation of the pressure dependence of the roton gap, shown in Fig. \ref{fig:RotonVsPressure}. Our data (see Table \ref{tab:roton}) have a smooth dependence on pressure, easily fitted by a second order polynomial through the statistical error bars. 
We remind that an overall shift in energy is allowed, within the 1\,$\mu$eV systematic uncertainty in $\Delta_R$(P), if the accuracy of $\Delta_R$(P=0) is improved in future measurements. 

There are few results in the literature on the pressure dependence of the dispersion relation. Results for the roton gap are shown  in Fig. \ref{fig:RotonVsPressure}. 
Early data of Dietrich \textit{et al.} \cite{Dietrich72} cover a large pressure range, with large uncertainties in energy (about 5\,$\mu$eV) and wave-vector (about 0.005\,\AA$^{-1}$), and a reasonable accuracy on the pressure ($\pm$0.14 bar). The temperature, on the order of 1.3\,K is unfortunately too high, and a finite temperature correction, estimated using the roton gap temperature dependence measured more recently by Gibbs \textit{et al.} \cite{GibbsThesis,AndersenRoton}, would shift Dietrich's data upwards in energy by 5 to 10\,$\mu$eV. This would bring them in good agreement with the present results. 

High resolution triple-axis results at non-zero pressures are scarce. Those by Talbot \textit{et al.} \cite{Talbot-Glyde}, only available at 20.0\,bar, with large error bars, agree well with our data. Stirling's data \cite{stirling-83,GlydeBook,StirlingExeter} are only available at 15.2 and 24\,bar (T=0.9\,K), but their uncertainty in the pressure of 1\,bar, unfortunately, translates into an energy uncertainty of about 6\,$\mu$eV. As seen in Fig. \ref{fig:RotonVsPressure}, the data point at about 15\,bar agrees with ours, but this is not the case for the high pressure one. The latter is definitely very low in energy, a discrepancy that cannot be explained by errors on the pressure measurement, since solidification takes place at 25.32\,bar. It is closer to the much higher temperature result of Dietrich \textit{et al.} \cite{Dietrich72}, than to our high pressure data. We should point out here that the polynomial fit to our data does not change significantly if our point at 24\,bar is omitted.   

The only recent source of good resolution data at non-zero pressures is Gibbs \textit{et al.} \cite{GibbsThesis,AndersenRoton}. Fig. \ref{fig:RotonVsPressure} shows that there is a good agreement between these results and ours in the pressure dependence of the roton gap. The deviations are probably explained by the larger uncertainties of the data by Gibbs \textit{et al.}, including a likely error in their pressures, measured with a Bourdon gauge (uncertainties not quoted).

\begin{figure}[h]
	\centering
		\includegraphics[width=1.0\columnwidth]{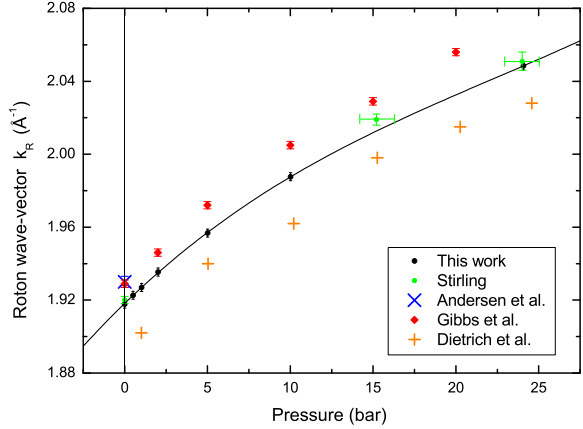}
	\caption{Pressure dependence of the roton wave-vector. Small black dots: this work; the solid line is a guide to the eye. The energy error bars correspond to the systematic uncertainties, statistical ones are not visible in this plot (see text). Green dots: Stirling \cite{stirling-83,StirlingExeter}. Red diamonds: Gibbs \textit{et al.} \cite{GibbsThesis,AndersenRoton}. Blue $\times$: Andersen \textit{et al.} \cite{Andersen94a}. Orange +: Dietrich \textit{et al.} \cite{Dietrich72} (at T=1.3\,K).}
	\label{fig:RotonWavevector}
\end{figure}

The pressure dependence of the roton wave-vector $k_R$ is shown in Fig. \ref{fig:RotonWavevector}. The present data display a smoother behavior, with small error bars, compared to former results. The statistical uncertainties from the fits at constant scattering angle are one order of magnitude smaller than the systematic errors (see the discussion at the end of Section \ref{sec:pixelanalysis}). The latter, on the order of 0.002\,\AA$^{-1}$, are due to the uncertainty in the detector angles, and to the conversion from scattering angle to wave-vector, which involves the systematic uncertainty of the energies. 
We observe a good agreement with Stirling's triple-axis data \cite{stirling-83,StirlingExeter}. TOF data by Dietrich \textit{et al.} \cite{Dietrich72}, Andersen \textit{et al.} \cite{AndersenThesis,Fak-Andersen-91,Andersen92,Andersen94a,Andersen94b} and Gibbs \textit{et al.} \cite{GibbsThesis,AndersenRoton} are systematically shifted, but on both sides of our curve. 

\begin{figure}[t]
	\centering
		\includegraphics[width=1.0\columnwidth]{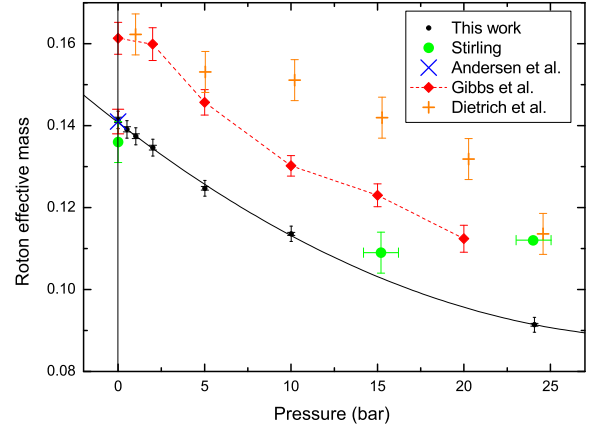}
	\caption{Pressure dependence of the roton effective mass. Small black dots: this work. The large (small) energy error bars correspond to the systematic (statistical) uncertainties, respectively (see text). Green dots: Stirling \cite{stirling-83,StirlingExeter}. Red diamonds: Gibbs \textit{et al.} \cite{GibbsThesis,AndersenRoton}. Blue $\times$: Andersen \textit{et al.} \cite{Andersen94a}. The lines are guides to the eye.}
	\label{fig:RotonMass}
\end{figure}

The pressure dependence of the roton effective mass $\mu_R$ is shown in Fig. \ref{fig:RotonMass}. Stirling's data points \cite{stirling-83,StirlingExeter} at SVP and 15\,bar agree with ours, but this is not the case for the point at 24 bar. The data by Dietrich \textit{et al.} \cite{Dietrich72} and those by Gibbs \textit{et al.} \cite{GibbsThesis,AndersenRoton} are shifted with respect to ours, but follow the same trend. 
We also note that the roton effective mass is strongly non-linear as a function of pressure, but it is an almost linear function of the density. This will be discussed in more detail below, in the comparison of our data with DMBT calculations.  

\subsection{Maxons}
\label{sec:maxon}

\begin{figure}[h]
	\centering
		\includegraphics[width=1.0\columnwidth]{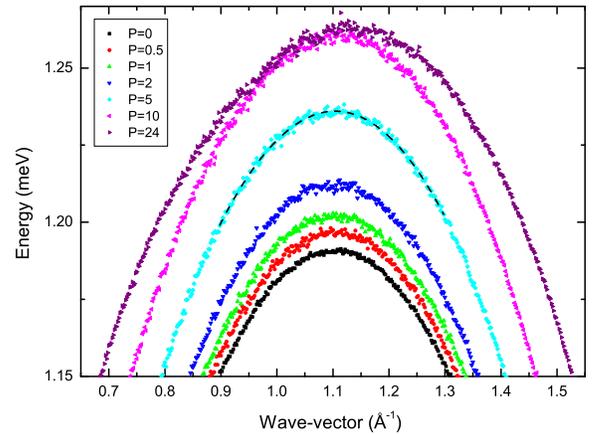}
	\caption{The dispersion relation in the vicinity of the maxon at different pressures (accurate values of P are given in Table \ref{table:Pressures}). The dashed line on the P=5 bar curve shows a typical cubic fit (see text). At 24 bar, the maxon is strongly damped.}
	\label{fig:maxonfit}
\end{figure}

The properties of the dispersion curve around the maxon (Fig. \ref{fig:maxonfit}) can be studied in a similar way.  
Fits have been made using the cubic polynomial expression:
\begin{equation}
	\epsilon_M(k)=\Delta_M+\frac{\hbar^2}{2m_4\mu_M}(k-k_M)^2+B_M(k-k_M)^3
	\label{Eq:fitmaxon}
\end{equation}

where the parameters are the maxon energy $\Delta_M$, the maxon wave-vector $k_M$, and the (negative) maxon effective mass $\mu_M$. 
Different fitting ranges were tested in order to evaluate the influence of systematic errors.
The maxon curves display a much smaller asymmetry than the roton ones. In the fits, it is possible to limit the polynomial expression to the cubic term in the wave-vector range spanning 0.2\AA$^{-1}$ around $k_M$. 
A typical fit is shown in Fig. \ref{fig:maxonfit} on the 5\,bar curve, results for all pressures are given in Table \ref{tab:maxon}.

\begin{table}[h]
\begin{ruledtabular}
\begin{tabular}{lllllll}
$P$ (bar) & & $\Delta_M$ (meV)& &  $k_M$ (\AA$^{-1}$) & & $\mu_M$   \\ \hline
0       & & 1.191(1) & & 1.103(2) & & -0.545(2) \\ 
0.51    & & 1.197(1) & & 1.104(2) & & -0.547(2) \\ 
1.02    & & 1.202(1) & & 1.102(2) & & -0.552(2) \\ 
2.01    & & 1.212(1) & & 1.102(2) & & -0.561(2) \\ 
5.01    & & 1.236(1) & & 1.104(2) & & -0.591(2) \\ 
10.01   & & 1.260(1) & & 1.110(2) & & -0.666(2) \\ 
24.08   & & 1.263(1) & & 1.134(2) & & -0.874(2) \\ 
\end{tabular}
\end{ruledtabular}
\caption{Maxon parameters deduced from fits using Eq. \ref{Eq:fitmaxon}. }
\label{tab:maxon}
\end{table}

\begin{figure}[h]
	\centering
		\includegraphics[width=1.0\columnwidth]{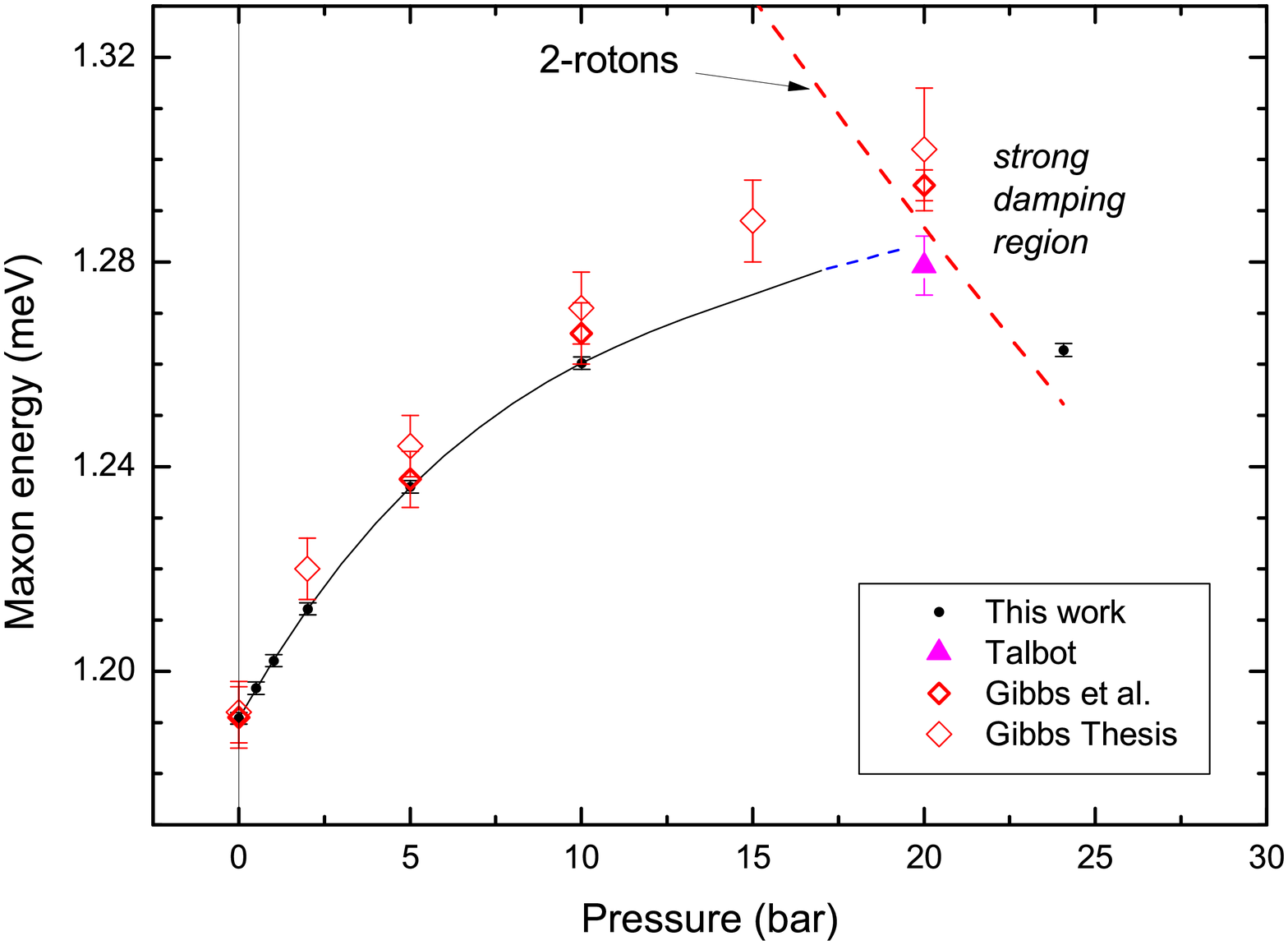}
	\caption{Pressure dependence of the maxon energy. Small black dots: this work. The energy error bars correspond to the systematic uncertainties (see text). The solid line is a polynomial fit of order 3 through the low pressure data. The point at 24 bar is in a different regime (see text). Triangle: Talbot \textit{et al.} \cite{Talbot-Glyde}. Red diamonds: Gibbs \textit{et al.} \cite{AndersenRoton} and Gibbs (Thesis) \cite{GibbsThesis}. The red dashed line indicates twice the measured roton energy. }
	\label{fig:MaxonVsPressure}
\end{figure}

There is an excellent agreement between our results for the maxon energy at saturated vapor pressure and those from Gibbs \textit{et al.} \cite{AndersenRoton} and Gibbs (Thesis) \cite{GibbsThesis}. At finite pressures, however, there is a clear discrepancy between these data and ours. The published version \cite{AndersenRoton} is closer to our result than the earlier (but more detailed) version of the same work, found in Gibbs's thesis manuscript \cite{GibbsThesis}. 

The behavior at high pressures is interesting, because a different regime is reached when $\Delta_M$$\sim$$2\Delta_R$, the maxon being spectacularly damped by three-particle processes \cite{Beauvois2018}. This happens according to Fig. \ref{fig:MaxonVsPressure} at P$\sim$20\,bar. Results at the same pressure by Talbot \textit{et al.} \cite{Talbot-Glyde} and Gibbs \textit{et al.} \cite{GibbsThesis,AndersenRoton} are considerably shifted, on both sides, with respect to the present data. 
The large discrepancy between former data may be due to the effect of damping on the maxon energy, as seen in Fig. \ref{fig:maxonfit}. 
A third order polynomial fit can describe the pressure dependence of the maxon energy in our data, within their very small statistical uncertainty. The curve is shown in Fig. \ref{fig:MaxonVsPressure}. Extrapolation to high pressures is delicate, and a better description, in terms of densities, will be presented below.

The pressure dependence of the maxon wave-vector $k_M$ is given in Table \ref{tab:maxon}. This parameter is, surprisingly, rather constant at low pressures.  A substantial increase is observed at 24\,bar, related to the damping, as seen in Fig. \ref{fig:maxonfit}. A small increase of $k_M$ is already observed at 10\,bar.
The maxon effective mass $\mu_M$, on the other hand, has a smooth variation with pressure, that we can fit by a simple second order polynomial expression. Its values can be found in Table \ref{tab:maxon}. A discussion of these results is given below.

\subsection{Theory: Dispersion relation at SVP}
\begin{figure}[h]
	\centering
		\includegraphics[width=0.95\columnwidth]{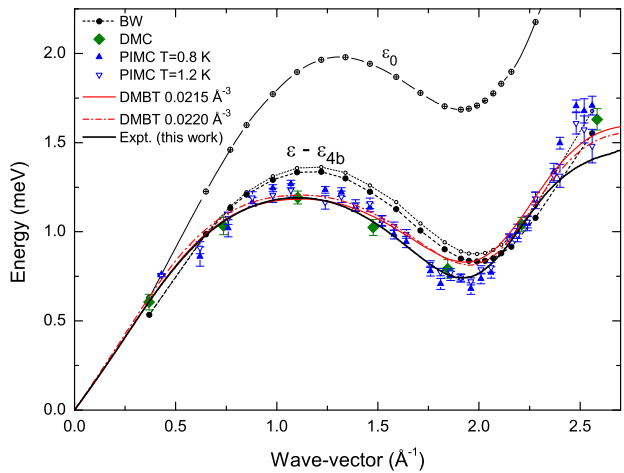}
	\caption{Dispersion relation $\epsilon(k)$ at saturated vapor pressure. Upper curves: Bijl-Feynman spectrum  \cite{Feynman1954} ($\epsilon_0$); BW calculation \cite{LeeLee} (black dots); BW calculation, excluding the $\epsilon_{4b}$ term \cite{LeeLee} (circles). Red curves: DMBT \cite{eomIII} at two densities around SVP (see legend).  Green lozenges: DMC  \cite{BoronatRoton,BoronatCasulleras}; triangles: PIMC \cite{2016FerreBoronat} at T=0.8 and 1.2\,K. Thick black line: experiment (this work).}
	\label{fig:DispersionTheory}
\end{figure}

Among the numerous theoretical calculations of the dispersion relation of superfluid $^4$He to be found in the literature, we have chosen four examples, especially appropriate for this manuscript (see Fig. \ref{fig:DispersionTheory}):    
the Brillouin-Wigner (BW) perturbative calculation by Lee and Lee \cite{LeeLee}, two different types of Monte-Carlo (MC) calculations \cite{JordiQFSBook}, and the variational dynamical many-body theory \cite{eomIII} (DMBT). 
Starting from the Bijl-Feynman spectrum \cite{Feynman1954}, which is clearly very far from the experimental result, the perturbative calculation by Lee and Lee has, first of all, the merit to bring theory closer to the experiment. In addition, it provides a quantitative estimate of the corrections due to the different Feynman diagrams involved in microscopic calculations. The effect of the $\epsilon_{4b}$ term \cite{LeeLee} is shown, as an example, in Fig. \ref{fig:DispersionTheory}. In spite of the large number of diagrams included, the BW approach is not satisfactory: strong departures from the experimental results are seen in the whole wave-vector range. 

DMBT provides accurate results from low wave-vectors to somewhat beyond the maxon. The discrepancy observed at high  wave-vectors is, according to the BW calculation described above, consistent with fact that the $\epsilon_{4b}$ diagram is not included in the DMBT calculation (see the discussion in Ref.  \onlinecite{eomIII}). Improving the accuracy would imply an additional computational effort which is not necessary, in particular, to investigate the density dependence of the dispersion.    
The dispersion relation calculated by DMBT at the present level is already in good agreement with the experiment in the whole wave-vector range, including the  plateau region, the calculation of which constitutes a severe theoretical challenge. 

Monte Carlo calculations \cite{CeperleyRMP,CeperleyExcitationPIMC,JordiQFSBook} constitute a very different approach to the microscopic description of quantum fluids. We show in Fig. \ref{fig:DispersionTheory} the results of Diffusion Monte Carlo (DMC) calculations by Boronat and coworkers \cite{BoronatRoton,BoronatCasulleras}, that clearly provide values of the dispersion relation at zero temperature in excellent agreement with the experiment. 

Path Integral Monte Carlo (PIMC) calculations yield results on the dispersion relation at finite temperatures. The data \cite{2016FerreBoronat} at T=0.8 and 1.2\,K (Fig. \ref{fig:DispersionTheory}) (identical within uncertainties, since the dispersion relation is not strongly temperature dependent below 1.25\,K) are in good agreement with the experimental values. Both MC methods, however, experience difficulties in observing  the plateau of the dispersion relation, essentially because of its very low weight; calculations in this region capture instead a multiexcitation `branch' also seen in the experiments (see Ref. \onlinecite{Beauvois2018} and Section \ref{sec:veryhighQ}).

Since constant progress is made in numerical methods and techniques, both variational and Monte Carlo methods are expected to yield further important developments in this field.  

\subsection{DMBT: density dependence}
\label{sec:DMBT}
The experimental properties of the roton and the maxon described above can be compared to DMBT predictions.
Since a small shift in density \cite{eomIII} is often needed in order to compare quantitatively the DMBT calculations and the experiments, we shall use in the following, instead of pressures, atomic densities. We thus avoid introducing in the comparison the theoretical equation of state.  
The pressure-density relations used here to convert the experimental pressures in atomic densities are known with excellent accuracy, they are found in Abraham \textit{et al.} \cite{Abraham} and  Greywall \cite{Greywall78} (see Section \ref{sec:sound}). 

\begin{figure}[h]
	\centering
		\includegraphics[width=0.9\columnwidth]{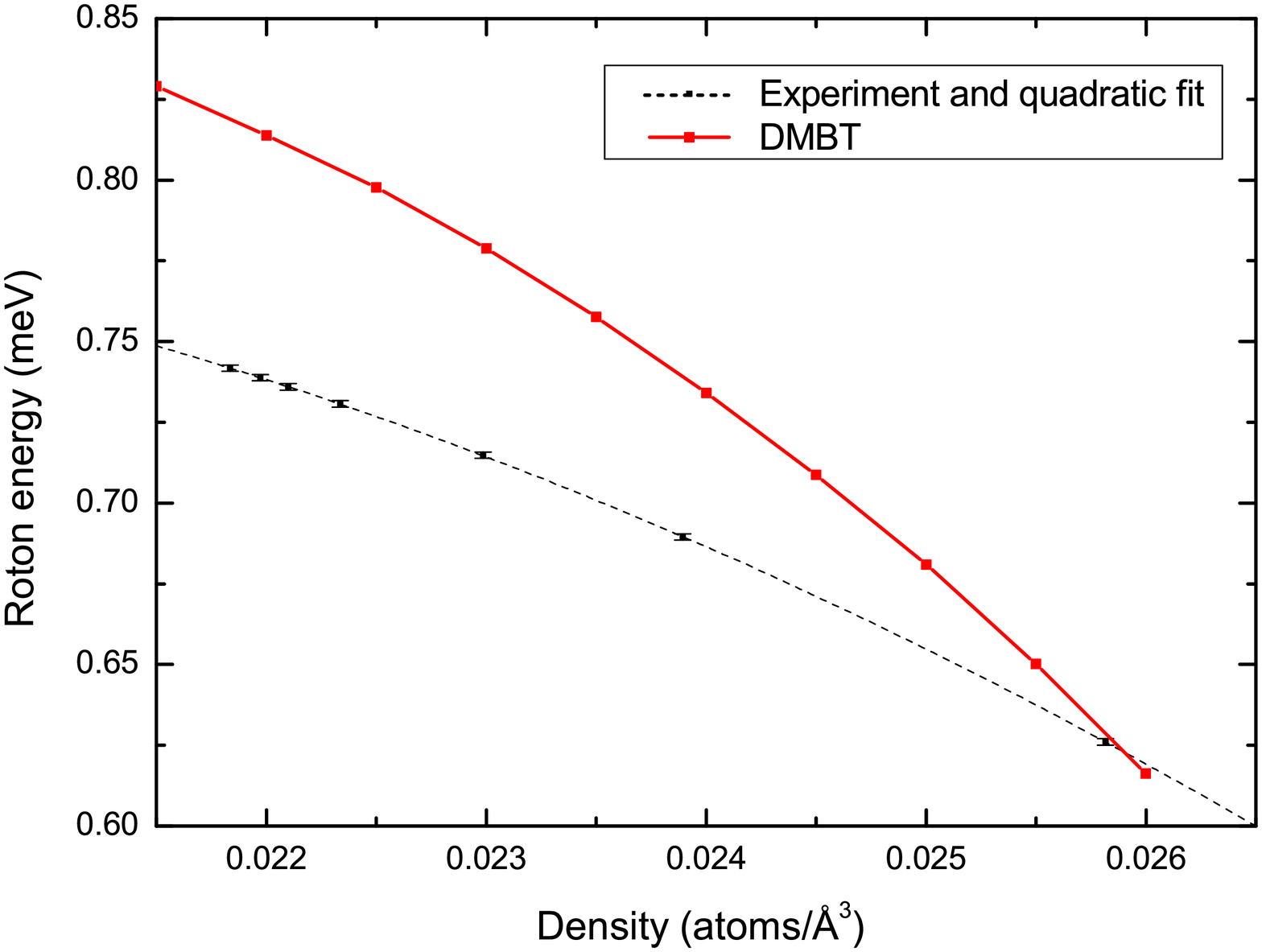}
	\caption{Density dependence of the roton energy. Present results (black dots with total uncertainty error bars, see text) are compared to the DMBT calculation \cite{eomIII} (red squares). }
	\label{fig:RotonEvsDensity}
\end{figure}

The dependence of the roton energy on density is shown in Fig. \ref{fig:RotonEvsDensity}. A good agreement is found within the expected accuracy of the theoretical calculation, estimated to be on the order of 10\%. This is not due to a shortcoming of the theory, but to the choice of the  diagrams included in the calculation, limited to the most significant ones, as far as the physics is concerned. As seen in the previous section, an estimate of the energy correction \cite{eomIII} can be made using the Brillouin-Wigner perturbation calculation by Lee and Lee \cite{LeeLee}. The first omitted diagram would decrease the roton energy by 0.05\,meV. This brings the (corrected) theory close to the experimental result at low pressures and, as expected, the deviation grows in fact at high pressures, where correlations are strongest. 

\begin{figure}[h]
	\centering
		\includegraphics[width=1.0\columnwidth]{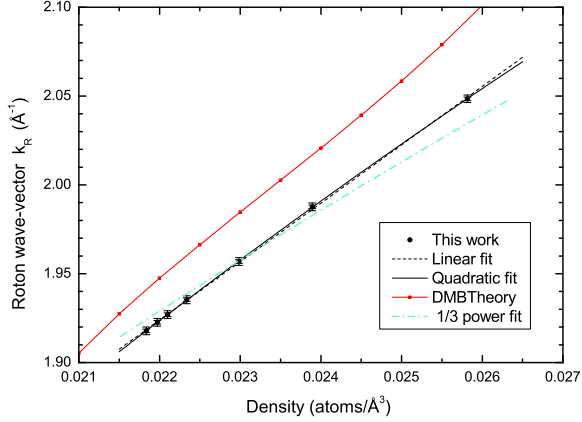}
	\caption{Density dependence of the roton wave-vector. Black dots: this work (small error bars: statistical uncertainty, larger bars: systematic uncertainty, see text). The deviations from a linear dependence are small: the short-dashed and the solid red lines are, respectively, a linear and a quadratic polynomial fit to the data. The expression \cite{Dietrich72} $k_R=a\rho{^{1/3}}$ (green dash-dotted line) clearly does not fit the data. Red squares: DMBT calculation  \cite{eomIII}.  }
	\label{fig:RotonQvsDensity}
\end{figure}

The calculated roton wave-vector $k_R$ and its density dependence (Fig. \ref{fig:RotonQvsDensity}) are quantitatively very close to the experimental result. 
It is clear that the diagrams included in the DMBT calculation capture all the essential features. The omitted diagrams 
have a smaller effect on this parameter, than on the roton energy. 

It has been suggested by Dietrich \textit{et al.} \cite{Dietrich72} that the density dependence of the roton wave-vector obeyed a simple law, 
$k_R=a\rho{^{1/3}}$, expected if the system is homothetically transformed with pressure. 
Clearly, as seen in Fig. \ref{fig:RotonQvsDensity}, neither theory nor experiment follow this law. Indeed, the density dependence of $k_R$ is almost linear, even within the very small statistical error bars of the fits, and \textit{a fortiori} within the somewhat larger total error bars including systematic uncertainties. 

\begin{figure}[t]
	\centering
		\includegraphics[width=0.9\columnwidth]{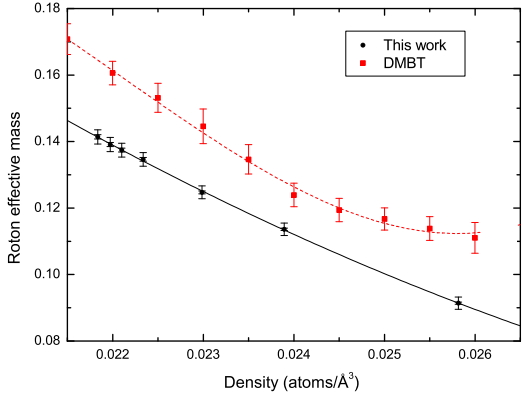}
	\caption{Density dependence of the roton effective mass. Black dots: this work (small error bars: statistical uncertainty, larger bars: systematic uncertainty, see text). The solid line is a quadratic fit to the data. Red squares: DMBT calculation. The dashed red line is a guide to the eye. }
	\label{fig:RotonMvsDensity}
\end{figure}

The density dependence of the roton effective mass $\mu_R$ is shown in Fig. \ref{fig:RotonMvsDensity}, where the experimental data are compared to the DMBT calculations. 
The curves were found to be very similar, and the analysis could be carried out in the same way: the same function and wave-vector range already applied (see above) to the experimental data were used to fit the DMBT results. As seen in the figure, the predicted magnitude as well as the density dependence are confirmed by the experiment. The slightly higher values of the theory are expected, since the theoretical roton minimum, calculated with a limited number of diagrams, is not as deep as the experimental one. 

\begin{figure}[h]
	\centering
		\includegraphics[width=0.9\columnwidth]{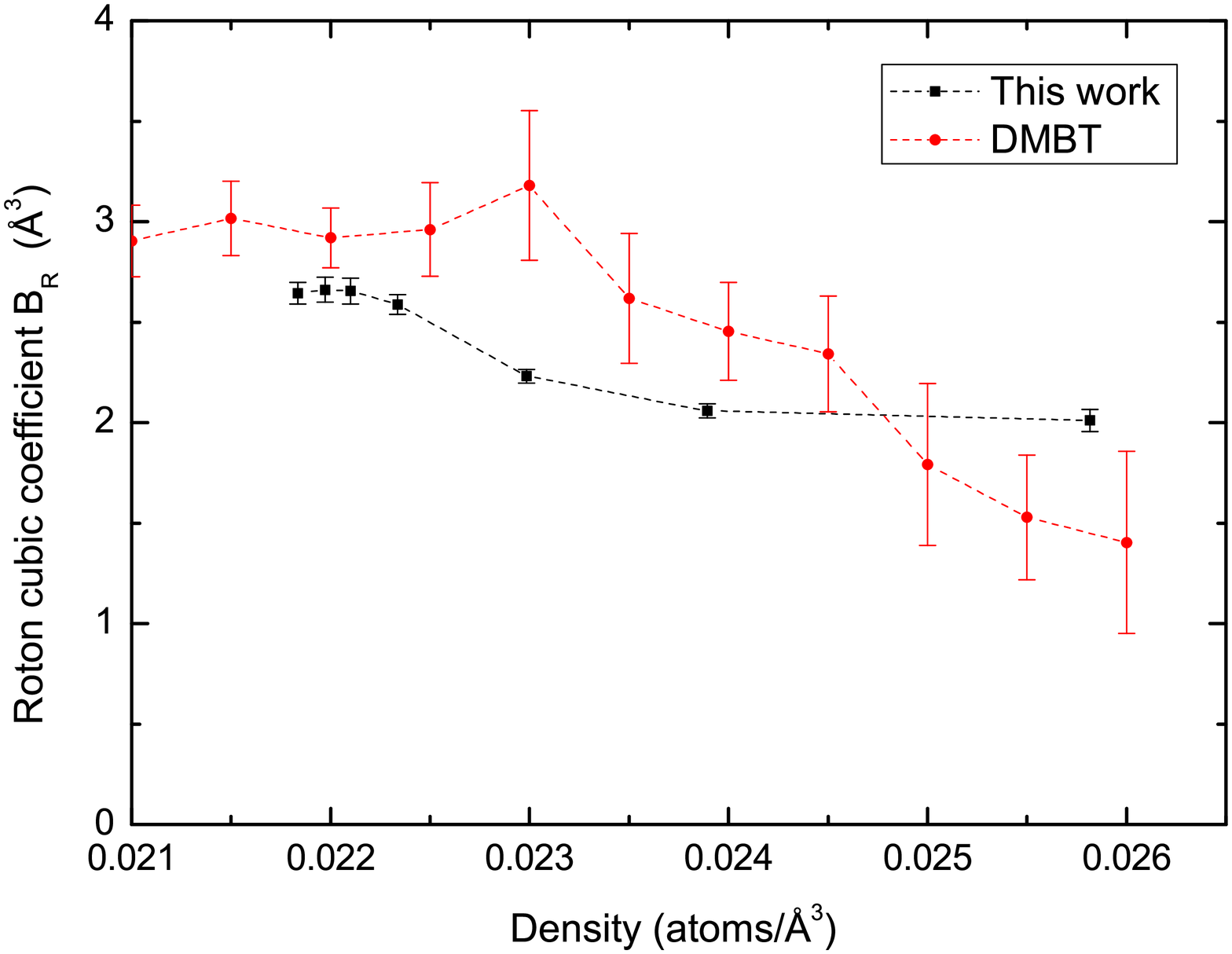}
	\caption{Density dependence of the roton cubic term B$_R$ (asymmetry coefficient). Black dots: this work. Red squares: DMBT \cite{eomIII}calculation \cite{eomIII}. Lines are guides to the eye. }
	\label{fig:RotonBetaRcoeff}
\end{figure}

\begin{figure}[t]
	\centering
		\includegraphics[width=0.9\columnwidth]{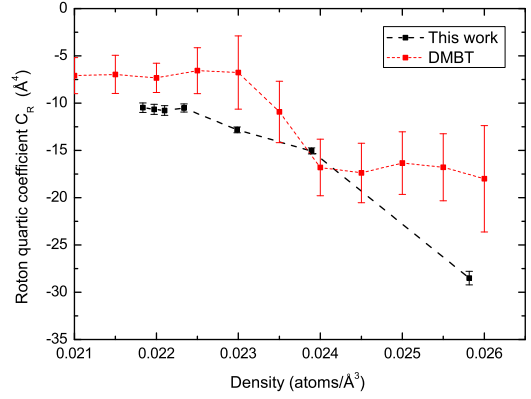}
	\caption{Density dependence of the roton quartic coefficient C$_R$. Black dots: this work. Red squares: DMBT \cite{eomIII} calculation. Lines are guides to the eye. }
	\label{fig:RotonGammaRcoeff}
\end{figure}

The shape of the dispersion curve around the roton minimum deviates rapidly from a simple parabola, and it also changes substantially with density. Higher order terms in the polynomial expansion (see Eq. \ref{Eq:fitroton}) are not at all negligible, unless fits are limited to a very small range around the minimum, reducing the accuracy of the fits. With a large number of independent data points, we have access to higher order coefficients: the cubic term (B$_R$) and the quartic term  (C$_R$) defined by Eq. \ref{Eq:fitroton}. The results are shown in Figs. \ref{fig:RotonBetaRcoeff} and \ref{fig:RotonGammaRcoeff}.

\begin{figure}
	\centering
	\includegraphics[width=0.9\columnwidth]{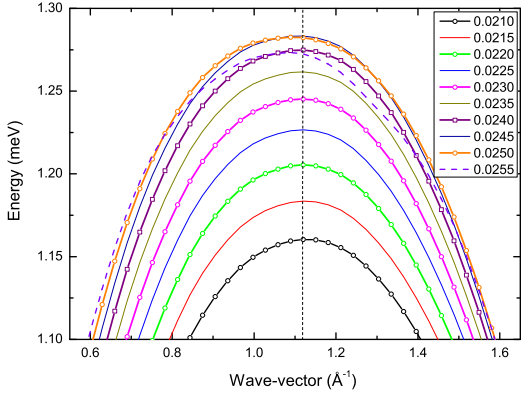}
	\caption{The dispersion relation calculated (DMBT \cite{eomIII}) in the maxon region, for different densities indicated in the label in atoms/\AA$^{3}$. See also experimental the data in Fig. \ref{fig:maxonfit}.}
	\label{fig:MaxonAllDensities}
\end{figure}

We consider now the maxon properties, comparing our data to the predictions of the DMBT.
The calculated dispersion relation in the vicinity of the maxon is shown in Fig.  \ref{fig:MaxonAllDensities} for several densities, directly comparable to our experimental result shown in Fig. \ref{fig:maxonfit}. 

\begin{figure}[t]
	\centering
		\includegraphics[width=1.0\columnwidth]{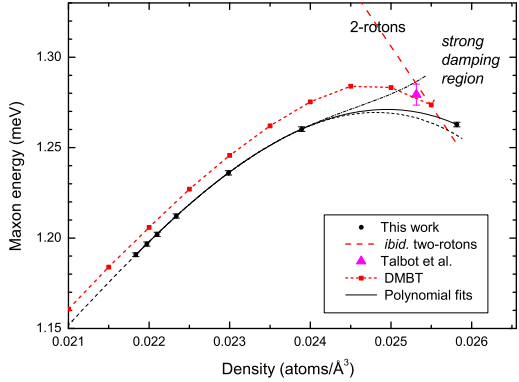}
	\caption{ Density dependence of the maxon energy. Black dots: this work. Error bars: systematic uncertainty; statistical error bars are not visible at this scale (see text). The solid line is a quadratic fit to the data. A quadratic and a cubic fit excluding the high pressure point are also shown. The long-dash red line is the 2-roton energy. Triangle: Talbot \textit{et al.} data \cite{Talbot-Glyde}. Red squares: DMBT \cite{eomIII} calculation. The short-dash red line is a guide to the eye.}
	\label{fig:MaxonEvsDensity}
\end{figure}

The maxon energy, represented in Fig. \ref{fig:MaxonEvsDensity} as a function of density, displays a much weaker variation than that observed for the roton. As described above, the high density data point is beyond the 2-roton limit, the corresponding maxon is damped, and this point is in a different regime compared to the lower pressure ones. Polynomial fits of order 2 and 3 excluding the high density data point encompass a relatively small portion of the 2-roton line, 
indicating that the maxon damping begins at a density n$_c$=0.0255$\pm$0.0002\,\AA$^{-3}$ (P=21.5$\pm$1.6\,bar).  

\begin{figure}[htbp]
	\centering
		\includegraphics[width=1.0\columnwidth]{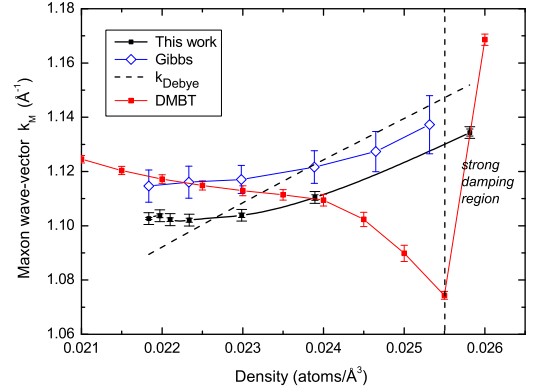}
	\caption{The maxon wave-vector as a function of density. Black squares: this work. Open lozenges: our fits to Gibbs's data \cite{GibbsThesis}. Theory: DMBT \cite{eomIII}. Lines are guides to the eye.  The Debye wave-vector calculated for the longitudinal phonon of a periodic system (see text) is shown for comparison.}
	\label{fig:MaxonQvsDensity}
\end{figure}

The maxon wave-vector k$_M$ is essentially constant at low densities, as seen in Fig. \ref{fig:MaxonQvsDensity}. 
We have fitted Gibbs's data \cite{GibbsThesis}, and we find a systematic difference, somewhat outside their relatively large error bars.

Also shown is the dependence of the Debye wave-vector on the  number density n. 
k$_D$=(6$\pi^2$n)$^{1/3}$ is defined by assigning one degree of freedom per atom for the longitudinal mode, integrated with an upper limit k$_{D}$: 
$N_a=(V/(2\pi^2))\int^{k_{D}}_{0}k^2dk$, 
where $N_a$ is Avogadro's number and V the molar volume. 
The Debye model \cite{FetterWalecka} usually describes a solid, where the wave-vector is limited by the inverse of the lattice spacing. In the liquid, such limitation does not exist. The hard core of the helium atoms, however, introduces a characteristic length and solid-like properties, like the very existence of roton excitations \cite{SmithSolid,NozSolid}. The Debye wave-vector is quantitatively similar to the maxon wave-vector, the maxon being analogous to a zone-boundary phonon. At high pressures, one can expect this analogy to work even better, and k$_M$ closely follows, indeed, the density dependence of k$_D$. In the liquid, the states with higher wave-vectors  progressively loose the relation with their `first Brillouin zone' analogues, and at k$\approx$2k$_D$ one observes a roton minimum, instead of a zero.  

The values of k$_{M}$ calculated by the DMBT agree well in magnitude with the experimental ones. A deviation is seen at relatively high densities, just before entering the strong damping region of the continuum. The experimental data discussed above display a small increase at high densities, and the opposite is found in the theoretical calculations. It must be pointed out that the maxon is very flat, and hence systematic errors can easily shift its position:  this very small effect (about 2\%) may come from the approximations in the theory, or from experimental resolution problems. In the theory, the decrease of the maxon wave-vector is clearly correlated with the damping at the maxon (see Figs. \ref{fig:MaxonAllDensities} and \ref{fig:MaxonEvsDensity}).

In a previous publication \cite{Beauvois2018} we provided tables of roton and maxon parameters obtained with the standard data analysis (see Section \ref{sec:data-acquisition}). A uniform shift of 9.2\,$\mu$eV applied to the energies    
yielded the correct sound velocities and the expected roton gap. This was accurate enough for the study of multi-excitations, but insufficient to establish the dispersion curve of single-excitations. 
Here, the energy scale has been \emph{calibrated} at the roton energy. The energies as a function of pressure obtained from both  analysis differ by a small amount. Also, the wave-vectors (in particular at the roton and the maxon) are slightly reduced by the energy recalibration. The values of the effective masses agree reasonably well after correcting an error (a missing factor 1.0546$^2$ from $\hbar^2$) in the previous publication. The new roton and maxon data tables are more accurate, they have been obtained by a consistent procedure entirely based on neutron data, and error bars including detailed systematic uncertainties are given. 
     
\subsection{Beyond the roton}
\label{sec:veryhighQ}
The very high wave-vector region, usually referred to as `beyond the roton' in the literature \cite{GlydeBook,Glyde2017,Fak98b,faak1998excitations,Pistolesi,PistolesiJLTP}, is also of interest. 
The dispersion relation at high wave-vectors increases up to an energy $\epsilon_{max}$$\approx$2$\Delta_R$, the Pitaevskii plateau. Since high-k excitations can decay into roton pairs, they cannot remain sharp above the plateau. 
The experimental investigation of this region \cite{GlydeBook,Glyde2017,Fak98b,faak1998excitations} is difficult, due to the limited  energy resolution of the spectrometers at the relevant energies. Early experiments placed the curve above the plateau, even at SVP, but this has been shown by Pistolesi to be due to experimental resolution effects \cite{GlydeBook,Glyde2017,Pistolesi,PistolesiJLTP}. 

According to high resolution work \cite{Glyde-Gibbs-98} at SVP,  
single excitations reach twice the roton gap at Q = 2.8\,\AA$^{-1}$,  
 the energy remains constant in the vicinity of 2$\Delta_R$ between Q=3.0\,\AA$^{-1}$,
and the end point of the dispersion curve is at Q =3.6\,\AA$^{-1}$. 
At high pressures (20 bar), the dispersion curve has been shown to lie essentially just below the plateau \cite{PearceAzuah}, and to also end at about k=3.6\,\AA$^{-1}$.

From the theoretical point of view, according to the calculations of Pitaevskii and their extension by Zawadowski, Ruvalds, Solana (ZRS theory) and Pistolesi \cite{GlydeBook,Glyde2017,Pistolesi,PistolesiJLTP}, the dispersion curve is believed to be slightly below Pitaevskii's plateau. 

\begin{figure}[htbp]
	\centering
		\includegraphics[width=1.0\columnwidth]{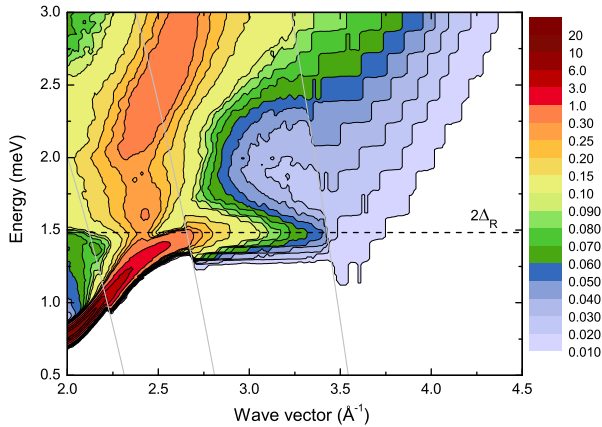}
	\caption{The dynamic structure factor S($Q$,$\omega$) in Pitaevskii's plateau region, measured using different incident neutron energies (E$_i$ = 3.520, 5.071, 7.990, and 20.45 meV), with energy resolutions (FWHM) at elastic energy transfer 0.07, 0.12, 0.23 and 0.92\,meV, respectively  (thin gray lines indicate limits of kinetic ranges). The color-coded intensity scale is in units of meV$^{-1}$. Dashed line: 2$\Delta_R$. }
	\label{fig:PitaPlateauContour}
\end{figure}

The results of the present measurements, obtained with different incident neutron energies in order to optimize the  energy resolution, are shown in Fig. \ref{fig:PitaPlateauContour}.  

\begin{figure}[htbp]
	\centering
		\includegraphics[width=1.0\columnwidth]{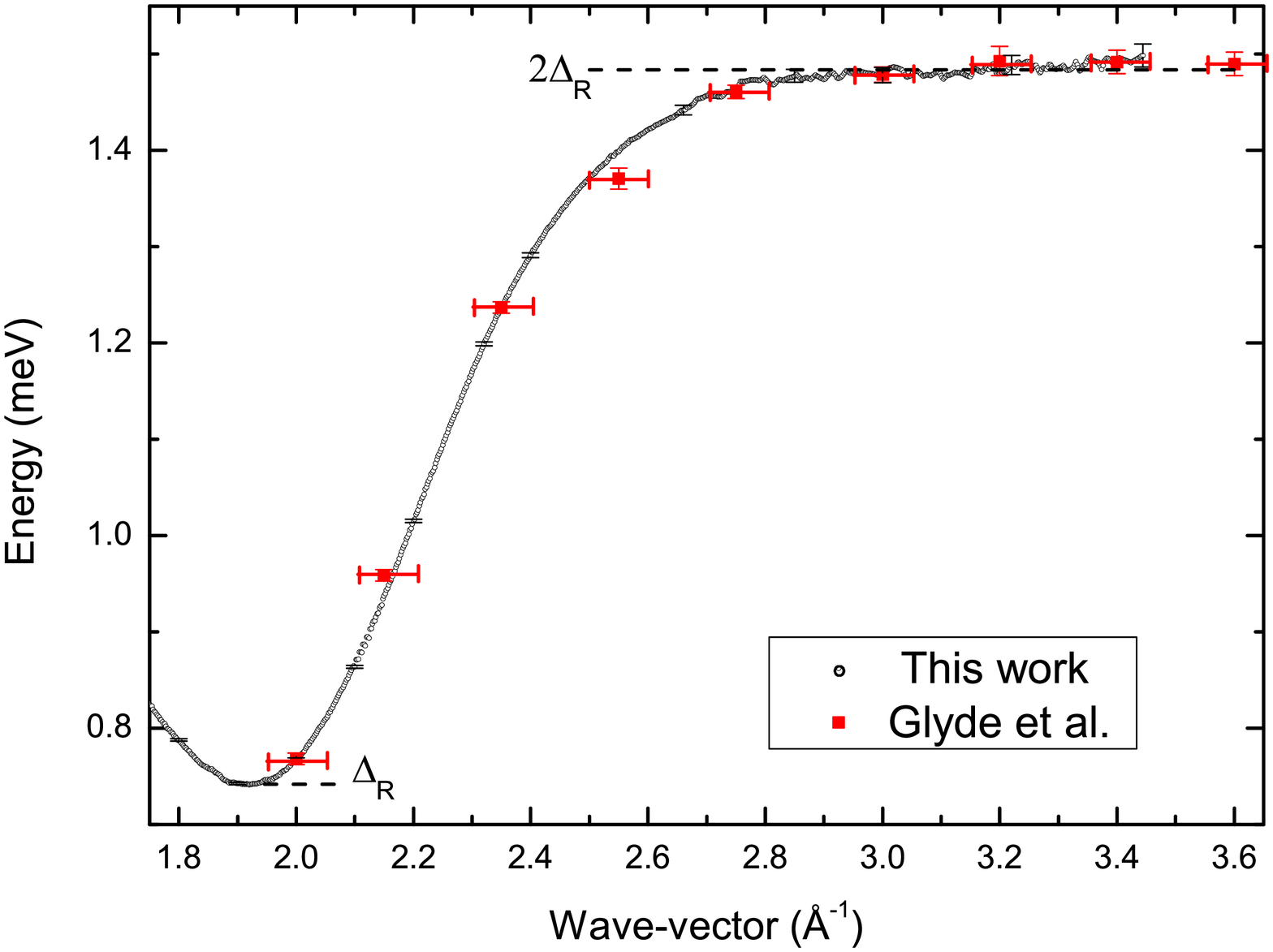}
	\caption{The dispersion relation measured beyond the roton. Open circles: present work at T$<$0.1\,K (error bars, shown for some points, are better seen on-line). Red squares: Glyde \textit{et al.} \cite{Glyde-Gibbs-98} (T=1.35\,K)}
	\label{fig:PitaPlateau}
\end{figure}

We show in Fig. \ref{fig:PitaPlateau} our data for the dispersion relation at T$<$0.1\,K. We also show the results obtained by Glyde \textit{et al.} at a higher temperature (1.35\,K) using the IRIS spectrometer; error bars are similar to ours in energy, but 25 times larger in wave-vector. There is an excellent agreement between these data where they can be compared.  

\begin{figure}[htbp]
	\centering
		\includegraphics[width=1\columnwidth]{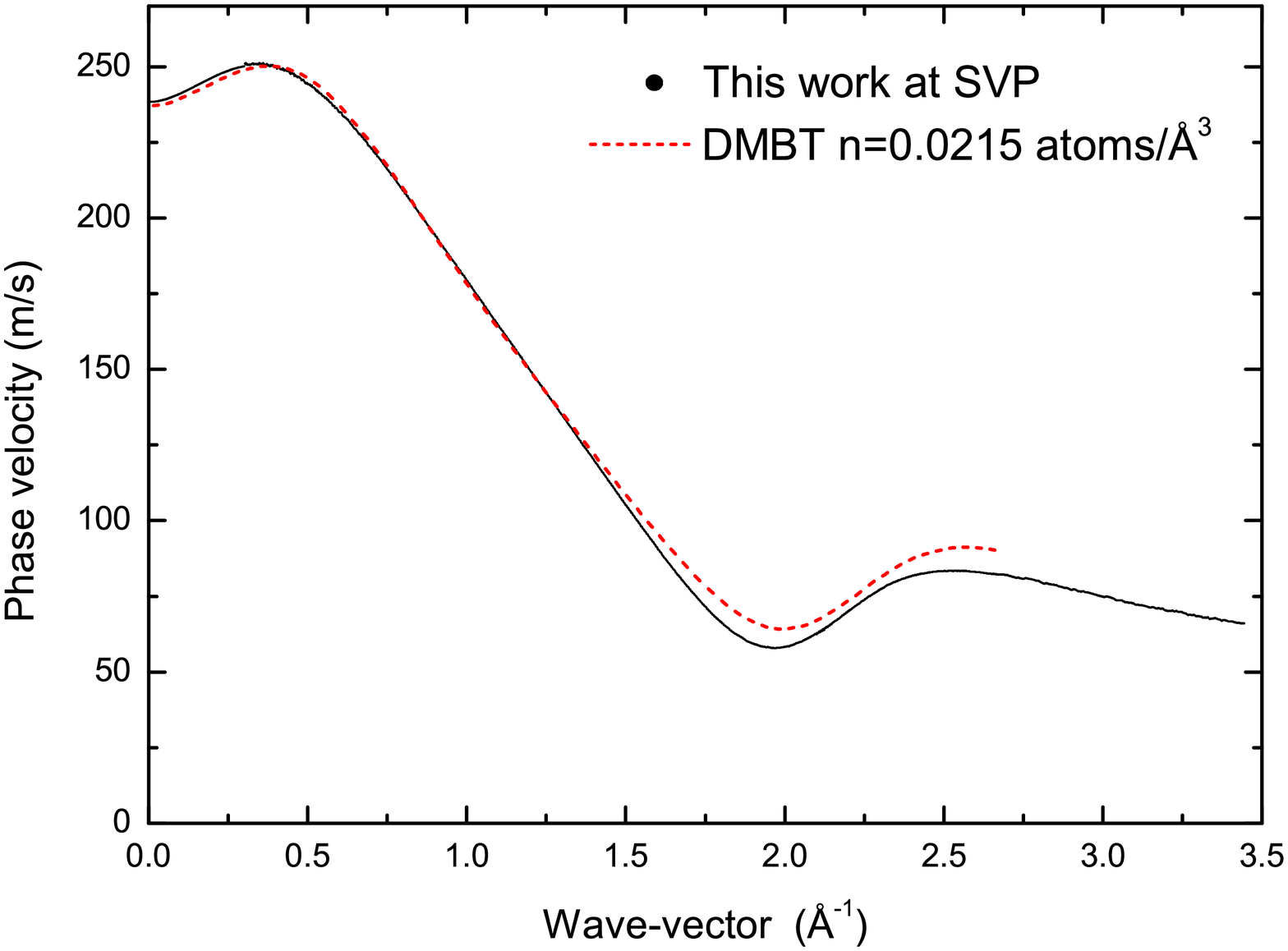}
	\caption{The single-excitations phase velocity at saturated vapor pressure (black dots, almost continuous line), is compared to the calculated values (dashed line, DMBT \cite{eomIII}) at a neighboring density (see Table \ref{table:soundveloc}).}
	\label{fig:PhaseVelocAllRange}
\end{figure}

\begin{figure}[htbp]
	\centering
		\includegraphics[width=1\columnwidth]{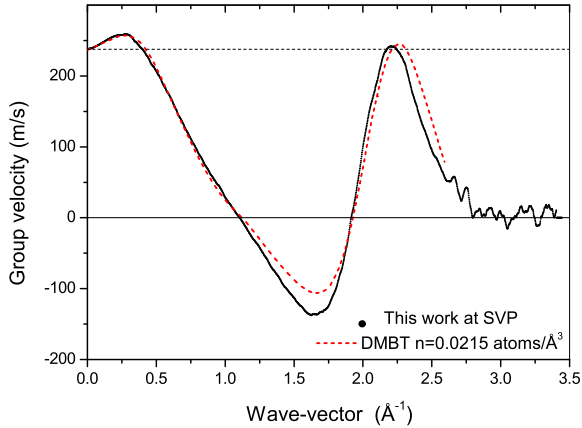}
	\caption{ The single-excitations group velocity at saturated vapor pressure  (black dots, almost continuous line), is compared to the calculated values (dashed line, DMBT \cite{eomIII}) at a neighboring density (see Table \ref{table:soundveloc}).}
	\label{fig:GroupVeloc-all-range}
\end{figure}

The single excitations phase velocity (Fig. \ref{fig:PhaseVelocAllRange}), does not display particular features in this range. The group velocity (Fig. \ref{fig:GroupVeloc-all-range}), on the other hand, vanishes at Q = 2.8\,\AA$^{-1}$ and, as seen in Fig. \ref{fig:PitaPlateau}, the dispersion curve becomes flat. 
The intensity of the single excitation decreases approximately exponentially (Fig. \ref{fig:PitaPlateauContour}) for k$>$3\,\AA$^{-1}$. It is difficult to define an `end wave-vector' of the single excitation. We observe however a change in the exponential-like intensity decrease at k$\approx$3.7\,\AA$^{-1}$, close to the value Q =3.6\,\AA$^{-1}$ quoted in the only other high resolution data at SVP in this range \cite{Glyde-Gibbs-98}.

\begin{figure}[htbp]
	\centering
	\begin{turn}{-90}
		\includegraphics[width=0.85\columnwidth]{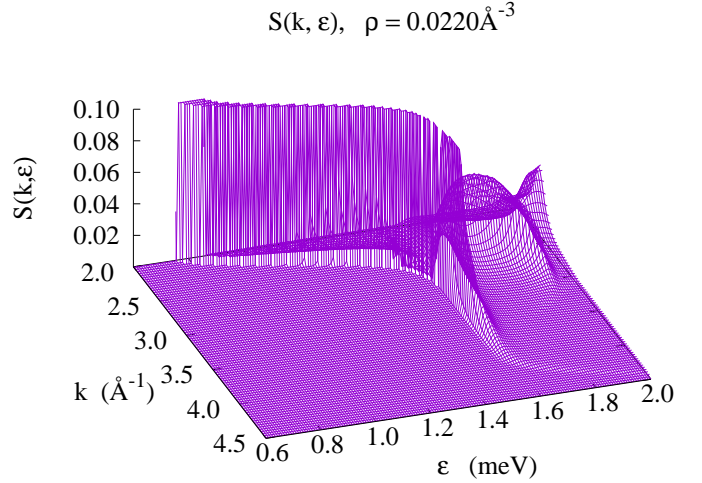}
		\end{turn}
	\caption{The dynamic structure factor calculated (DMBT \cite{eomIII}) for the density n=0.0220\,\AA$^{-3}$ (close to SVP), in the plateau region. The solid lines are cuts of S($k$,$\epsilon$) for fixed wave-vectors.}
	\label{fig:PitaPlateauTheory}
\end{figure}

The calculations in the framework of DMBT \cite{eomIII} predict that the dispersion curve remains almost constant below the plateau at high densities, but the behavior is different at low densities: at  saturated vapor pressure, after reaching the continuum at k$\approx$2.8\,\AA$^{-1}$, the curve remains at the edge of the continuum with some undulations. An asymmetric peak is the formed, and its shape varies with k, as seen in Fig. \ref{fig:PitaPlateauTheory} (the figure can be expanded in the on-line version).
At all densities, the single excitation intensity finally vanishes at a wave-vector on the order of 3.6\,\AA$^{-1}$, slightly increasing with density. 

Disentangling the single excitation from the multi-excitations contribution cannot be done unambiguously at the largest wave-vectors. The good general agreement between experiment and theory supports the hypothesis of an end of the dispersion curve in this range.

\subsection{Comparison to previous works}
\label{DispersionFinal}

The measured dispersion relation at saturated vapor pressure has been represented in Fig. \ref{fig:dispersion}. A table summarizing the results is given below (Table \ref{tab:dispersionP0}). Raw data and tables with a finer wave-vector grid (0.002\AA$^{-1}$) are provided, see Supplemental Material at [URL will be inserted by publisher].  

\begingroup
\begin{table}[h]
\begin{ruledtabular}
\begin{tabular}{llll|llll|lll}
$k$    &$\epsilon(k)$ & err $\epsilon$   &  & $k$    &$\epsilon(k)$ & err $\epsilon$    &  & $k$    &$\epsilon(k)$ & err $\epsilon$    \\  
\AA$^{-1}$ & meV & meV &  & \AA$^{-1}$ & meV & meV &  & \AA$^{-1}$ & meV & meV \\
\hline 
0    & 0.0000 &         &  & 1.25 & 1.1698 & 0.0020 &  & 2.5  & 1.3718 & 0.0035 \\
0.05 & 0.0787 &         &  & 1.3  & 1.1542 & 0.0020 &  & 2.55 & 1.3985 & 0.0039 \\
0.1  & 0.1587 &         &  & 1.35 & 1.1310 & 0.0019 &  & 2.6  & 1.4214 & 0.0044 \\
0.15 & 0.2407 & 0.0014  &  & 1.4  & 1.1055 & 0.0019 &  & 2.65 & 1.4375 & 0.0049 \\
0.2  & 0.3244 & 0.0012  &  & 1.45 & 1.0743 & 0.0018 &  & 2.7  & 1.4571 & 0.0053 \\
0.25 & 0.4092 & 0.0010  &  & 1.5  & 1.0381 & 0.0018 &  & 2.75 & 1.4617 & 0.0058 \\
0.3  & 0.4938 & 0.0010  &  & 1.55 & 0.9999 & 0.0017 &  & 2.8  & 1.4746 & 0.0062 \\
0.35 & 0.5782 & 0.0011  &  & 1.6  & 0.9563 & 0.0016 &  & 2.85 & 1.4776 & 0.0066 \\
0.4  & 0.6581 & 0.0012  &  & 1.65 & 0.9113 & 0.0015 &  & 2.9  & 1.4737 & 0.0071 \\
0.45 & 0.7338 & 0.0013  &  & 1.7  & 0.8672 & 0.0015 &  & 2.95 & 1.4818 & 0.0075 \\
0.5  & 0.8040 & 0.0014  &  & 1.75 & 0.8242 & 0.0014 &  & 3    & 1.4784 & 0.0080 \\
0.55 & 0.8701 & 0.0015  &  & 1.8  & 0.7877 & 0.0013 &  & 3.05 & 1.4793 & 0.0085 \\
0.6  & 0.9279 & 0.0016  &  & 1.85 & 0.7584 & 0.0013 &  & 3.1  & 1.4781 & 0.0089 \\
0.65 & 0.9812 & 0.0017  &  & 1.9  & 0.7427 & 0.0013 &  & 3.15 & 1.4767 & 0.0094 \\
0.7  & 1.0267 & 0.0018  &  & 1.95 & 0.7459 & 0.0013 &  & 3.2  & 1.4805 & 0.0098 \\
0.75 & 1.0667 & 0.0018  &  & 2    & 0.7681 & 0.0013 &  & 3.25 & 1.4920 & 0.010  \\
0.8  & 1.1002 & 0.0019  &  & 2.05 & 0.8094 & 0.0014 &  & 3.3  & 1.4855 & 0.011  \\
0.85 & 1.1280 & 0.0019  &  & 2.1  & 0.8637 & 0.0015 &  & 3.35 & 1.4954 & 0.011  \\
0.9  & 1.1518 & 0.0020  &  & 2.15 & 0.9374 & 0.0016 &  & 3.4  & 1.4940 & 0.012  \\
0.95 & 1.1678 & 0.0020  &  & 2.2  & 1.0154 & 0.0017 &  & 3.45 & 1.5012 & 0.012  \\
1    & 1.1809 & 0.0020  &  & 2.25 & 1.0944 & 0.0019 &  & 3.5  & 1.5199 & 0.012  \\
1.05 & 1.1882 & 0.0020  &  & 2.3  & 1.1701 & 0.0020 &  & 3.55 & 1.5377 & 0.013  \\
1.1  & 1.1907 & 0.0020  &  & 2.35 & 1.2370 & 0.0023 &  & 3.6  & 1.5538 & 0.013  \\
1.15 & 1.1884 & 0.0020  &  & 2.4  & 1.2914 & 0.0025 &  &      &        &        \\
1.2  & 1.1814 & 0.0020  &  & 2.45 & 1.3363 & 0.0030 &  &      &        &        \\ 
\end{tabular}
\end{ruledtabular}
\caption{The dispersion relation of $^4$He at saturated vapor pressure and very low temperatures (T$<$100\,mK). Below k=0.15\,\AA$^{-1}$: ultrasound data \cite{Rugar1984} (see text). From 0.15 to 0.3\,\AA$^{-1}$: combined ultrasound and present neutron data; above 0.3\,\AA$^{-1}$, present neutron data. See Supplemental Material at [URL will be inserted by publisher] for more detailed tables. }
\label{tab:dispersionP0}
\end{table}
\endgroup

\begin{figure}[htbp]
	\centering
		\includegraphics[width=1.0\columnwidth]{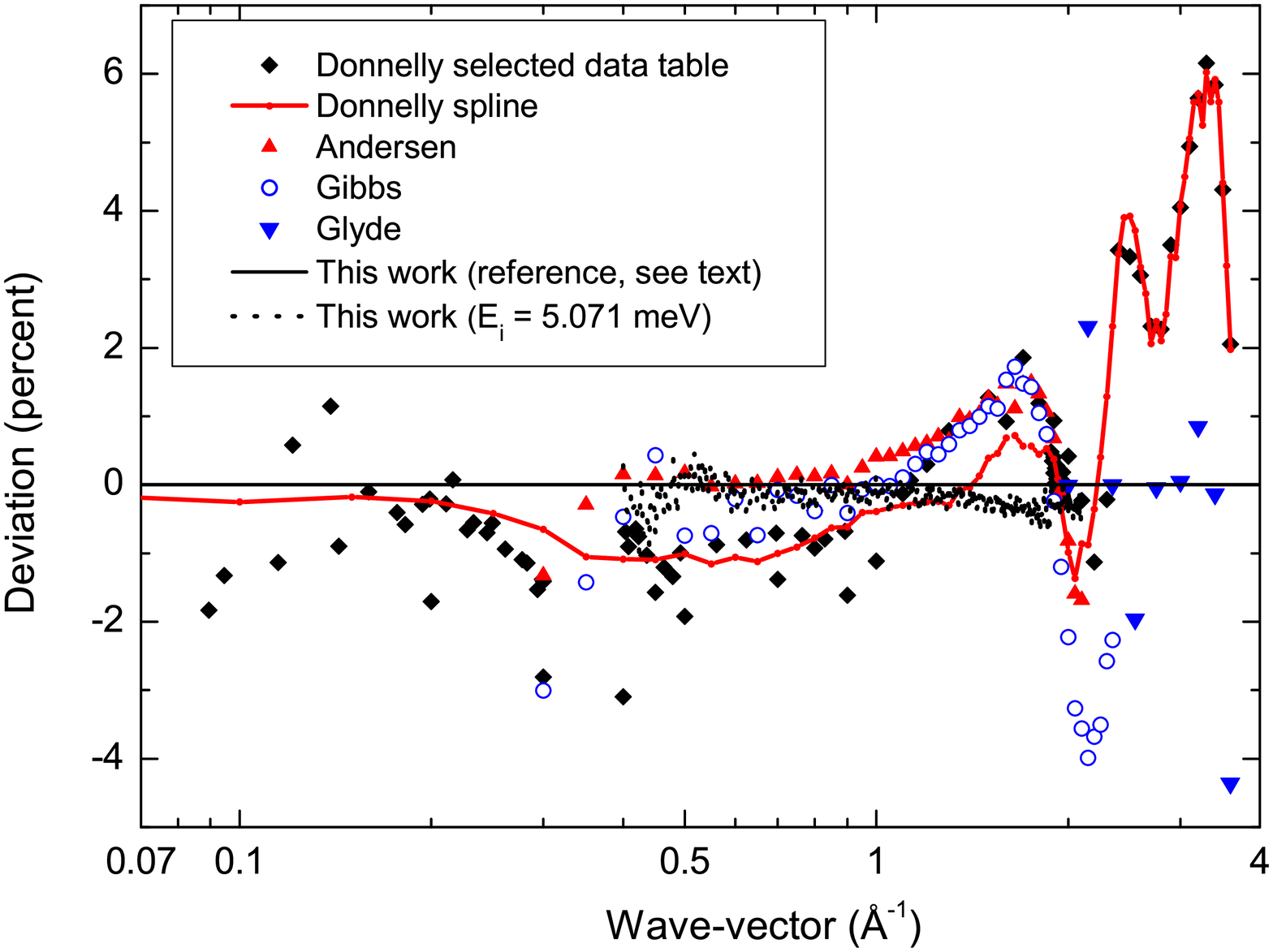}
	\caption{Percent deviation plot comparing different data to our measured dispersion curve. Black diamonds: data-base selected by Donnelly \textit{et al.}  \cite{BrooksDonnelly,DonnellyBarenghi}, and their spline fit (red curve);  triangles: Andersen \textit{et al.} \cite{AndersenThesis,Andersen94a}; open circles: Gibbs \textit{et al.} \cite{GibbsThesis,AndersenRoton}; inverted triangles: Glyde \textit{et al.} \cite{Glyde-Gibbs-98}. Small black dots: this work, additional data measured at a different energy, E$_i$=5.071\,meV. }
	\label{fig:Dispersion-deviation}
\end{figure}

The present results are compared to previous experimental data in Fig. \ref{fig:Dispersion-deviation}, a percent deviation plot where our data are taken as the reference. We examine first the data-base carefully selected by Brooks, Donnelly and Barenghi \cite{BrooksDonnelly,DonnellyBarenghi}, where several results of lower accuracy or considered less trustworthy, have already been  discarded by these authors. Large deviations with respect to our data are seen, in particular around 0.4\,\AA$^{-1}$ and at high wave-vectors. The thick red line in Fig. \ref{fig:Dispersion-deviation} shows Donnelly's spline fit of this set of data (see Ref. \onlinecite{DonnellyBarenghi} and references therein). Also shown are results by Andersen \textit{et al.} \cite{AndersenThesis,Andersen94a}, Gibbs \textit{et al.} \cite{GibbsThesis,AndersenRoton}, and Glyde \textit{et al.} \cite{Glyde-Gibbs-98}, which were not included in the data-base mentioned above. 

One can distinguish several regions. For k$<$0.2\,\AA$^{-1}$, the dispersion is essentially obtained from ultrasound data, and the very small difference between our data and Donnelly's spline fit is within error bars. For 0.2$<$k$<$1\,\AA$^{-1}$ our data agree well with those of Andersen \textit{et al.}. Furthermore, our data measured at a different incident neutron energy (5.071\,meV), are  in excellent agreement with those measured at 3.52\,meV, our reference in this wave-vector range. The data of Gibbs \textit{et al.}, with a larger statistical uncertainty, lie in-between our curve and Donnelly's. 

Between 1 and 2\,\AA$^{-1}$, our data at the two different incident energies are again found to be consistent. It is important to realize that these two sets of data are independent, in particular the angles and distances corresponding to a given wave-vector are different. The deviation between these data sets is very small, it does not display accidents or inconsistencies, and this is observed in a very large wave-vector range. This result constitutes an important verification of the consistency of the present data.
The results of Andersen \textit{et al.} agree with those of Gibbs \textit{et al.}. They are consistent with Donnelly's data-base, but all these points are ignored by Donnelly's `fit' to a great extent, the curve is in fact closer to our result.  

At the roton, the agreement is good between the various data sets. This point, of course, has been used in order to calibrate our energy scale. At wave-vectors beyond the roton, previous results display strong statistical scatter and systematic inconsistencies; our data are just in the middle, and very close, as noted above, to the data of Glyde \textit{et al.}.

Our data for the dispersion curve constitute therefore a new, comprehensive and coherent data base. In the following, we use this result to calculate the thermodynamical properties, a stringent test for the measured dispersion curve.

\section{Specific heat and other thermodynamical properties}
\label{sec:Cv}
Considerable effort has been devoted to the calculation of the thermodynamical properties of superfluid $^4$He starting from the dispersion curve. 
Different kinds of experiments were performed, and subsequently analyzed using Landau's model 
\cite{Landauroton,Landauroton2}. 
In the simplest approximation, applicable at low temperatures, the thermal population of the states
described by Landau's spectrum is significant in the low-k phonon region, 
and around the roton minimum. The model has 4 parameters: the sound velocity, the roton gap, the roton wave-vector and 
the roton effective mass. 
The heat capacity calculated by Landau \cite{Landauroton,Landauroton2} was in good
agreement with early specific heat measurements. 
When high accuracy data became available, substantial quantitative deviations from Landau's model were observed, and several attempts were made to improve this formalism. 
Phillips \textit{et al.} \cite{Phillips1970} and Greywall \cite{Greywall78,Greywall79ERRATUM} calculated low temperature series expansions for the specific heat based on the series expansion of the dispersion curve given by Eq. \ref{eq:dispfit}. 
Since published formulas contain errors \cite{Phillips1970} or misprints \cite{Greywall78}, we provide below the correct results. We also expand the series to higher order, which is found necessary to obtain thermodynamically consistent series expansions.

In the following section, we first provide a \emph{numerical} calculation of the thermodynamic properties using the dispersion relation determined in the present work. We then compare these results with those obtained with our \emph{analytical} formulas. 

\subsection{Specific heat: complete numerical calculation.}
The internal energy E and other thermodynamical properties, in particular the specific heat $C_V = \left(\frac{\partial E}{\partial T}\right)_{V}$, can be calculated as a function of temperature by elementary statistical physics \cite{FetterWalecka}, using the dispersion relation $\epsilon(k)$, the Bose distribution function $n(\epsilon(k))$, and the density of states of an isotropic 3-dimensional system, 
$ D(\omega) d\omega=(V/2\pi^2)k^2dk$: 

\begin{equation}
	\label{equ:energy}
E=(k_B V/2\pi^2)\int_{0}^{k_{end}}\frac{\epsilon(k)}{e^{\frac{\epsilon(k)}{k_B T}}-1}k^2dk
\end{equation}

Integration is performed over all the states; in Landau's model, the upper limit is taken as infinity, the exponential factor making  this choice possible in the very low temperature limit; for our higher temperature approximations we introduce a finite upper limit of the spectrum, k$_{end}$. 

\begin{figure}[htbp]
	\centering
		\includegraphics[width=1.0\columnwidth]{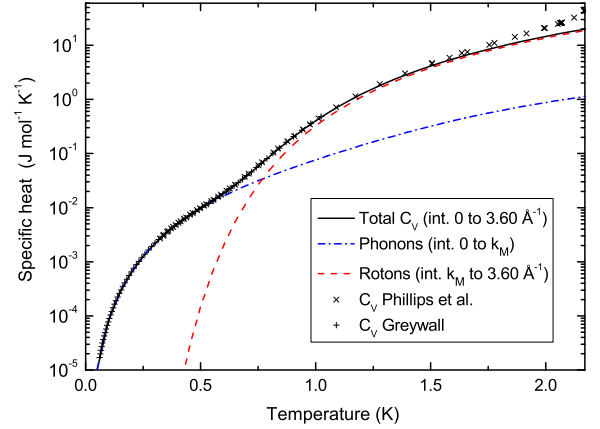}
	\caption{Total specific heat, and phonon and roton contributions, calculated numerically with the measured dispersion relation. Specific heat data by Phillips \textit{et al.} \cite{Phillips1970} and Greywall \cite{Greywall78,Greywall79ERRATUM} are shown for comparison.} 
	\label{fig:Cv-fullRange}
\end{figure}

Donnelly \textit{et al.} \cite{DonnellyDonnellyHills} concluded that the early neutron data were `consistent' with high resolution specific heat results (see previous section). 
Given the large dispersion of the neutron data around their proposed curve, they emphasized the real need for further neutron scattering studies. 
We show in Fig. \ref{fig:Cv-fullRange} and in Table \ref{tab:CvTable} the specific heat calculated numerically in all the temperature range below T$_\lambda$ using the present neutron scattering data.

\begingroup
\begin{table}[h]
\begin{ruledtabular}
\begin{tabular}{clllll}
$T$   & $C_v$ (total)  & 	$C_v$ (phonons)	  & $C_v$ (rotons)	 &   & $S$ (total)  \\ 
$K$   & $J/(K.mol)$    &  $J/(K.mol)$       &  $J/(K.mol)$     &   &  $J/(K.mol)$  \\ \hline
0.05	&	1.037E-05	&	1.037E-05	&	2.270E-70	&	&	3.459E-06	\\
0.10	&	8.260E-05	&	8.260E-05	&	1.920E-33	&	&	2.760E-05	\\
0.15	&	2.770E-04	&	2.770E-04	&	3.030E-21	&	&	9.276E-05	\\
0.20	&	6.513E-04	&	6.513E-04	&	3.360E-15	&	&	2.187E-04	\\
0.25	&	0.001261	&	0.001261	&	1.330E-11	&	&	4.247E-04	\\
0.30	&	0.002157	&	0.002157	&	3.170E-09	&	&	7.290E-04	\\
0.35	&	0.003392	&	0.003392	&	1.531E-07	&	&	0.001149	\\
0.40	&	0.005016	&	0.005013	&	2.738E-06	&	&	0.001704	\\
0.45	&	0.007094	&	0.007069	&	2.532E-05	&	&	0.002409	\\
0.50	&	0.009756	&	0.009607	&	1.480E-04	&	&	0.003288	\\
0.55	&	0.01330	&	0.01268	&	6.203E-04	&	&	0.004376	\\
0.60	&	0.01837	&	0.01634	&	0.002028	&	&	0.005736	\\
0.65	&	0.02613	&	0.02064	&	0.005487	&	&	0.007489	\\
0.70	&	0.03846	&	0.02566	&	0.01279	&	&	0.009841	\\
0.75	&	0.05784	&	0.03147	&	0.02651	&	&	0.01310	\\
0.80	&	0.08812	&	0.03816	&	0.04994	&	&	0.01774	\\
0.85	&	0.1329	&	0.04586	&	0.08702	&	&	0.02434	\\
0.90	&	0.1968	&	0.05469	&	0.1421	&	&	0.03364	\\
0.95	&	0.2848	&	0.06481	&	0.2200	&	&	0.04653	\\
1.00	&	0.4018	&	0.07641	&	0.3253	&	&	0.06397	\\
1.05	&	0.5527	&	0.08969	&	0.4630	&	&	0.08708	\\
1.10	&	0.7424	&	0.1049	&	0.6375	&	&	0.1170	\\
1.15	&	0.9755	&	0.1221	&	0.8533	&	&	0.1549	\\
1.20	&	1.256	&	0.1418	&	1.115	&	&	0.2022	\\
1.25	&	1.589	&	0.1639	&	1.425	&	&	0.2600	\\
1.30	&	1.977	&	0.1889	&	1.788	&	&	0.3297	\\
\end{tabular}
\end{ruledtabular}
\caption{Specific heat (total, phonon and roton contributions), and total molar entropy, at the saturated vapor pressure, as a function of temperature, calculated numerically using the measured dispersion relation (this work). The uncertainties on the specific heat are $\leq$1\%, see Supplemental Material at [URL will be inserted by publisher] for more detailed tables. }
\label{tab:CvTable}
\end{table}
\endgroup

The figure also shows the phonon and the roton contributions, defined here as the specific heat obtained by integration over wave-vector ranges separated by the maxon wave-vector k$_M$ (see Section \ref{sec:results}). The `phonons' thus correspond to the range (0-k$_M$) and the `rotons' to (k$_M$-k$_{end}$). 

A clear deviation between the calculated and measured specific heats is seen 
above 1.3\,K. This is not surprising: at high temperatures the dispersion curve itself changes rapidly as the temperature increases, and the excitations broaden substantially \cite{GlydeBook,Glyde2017,DonnellyRoberts}. This regime is outside the scope of this study, we concentrate here on the low temperature properties, where the dispersion curve is unique. 

\begin{figure}[htbp]
	\centering
		\includegraphics[width=1.0\columnwidth]{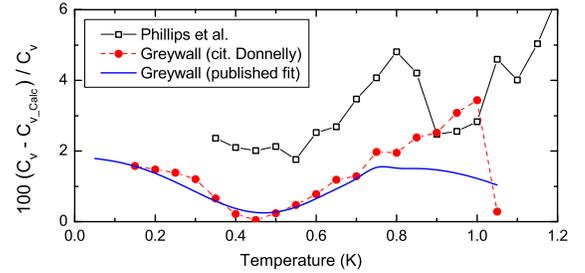}
	\caption{Percent difference between specific heat measurements, and the values inferred from the dispersion relation measurement (present work). Open squares: Phillips \textit{et al.} \cite{Phillips1970}. Red dots: Greywall data, listed in Refs. \onlinecite{BrooksDonnelly,DonnellyBarenghi}. Blue solid line: Greywall's fit of his data \cite{Greywall78,Greywall79ERRATUM}.  }
	\label{fig:Difference-Cv-neutrons}
\end{figure}

\begin{figure}[htbp]
	\centering
		\includegraphics[width=1.0\columnwidth]{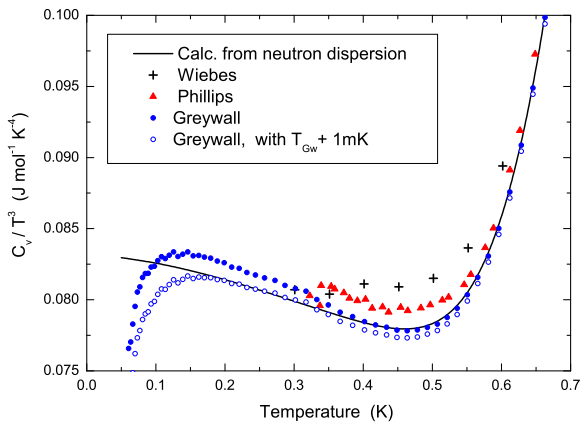}
	\caption{C$_V$/T$^3$ in the phonon range. The values  calculated from  the dispersion curve (solid line, this work), compared to heat capacity data. Crosses: Wiebes \cite{WiebesThesis,DonnellyBarenghi}. Triangles: Phillips \textit{et al.} \cite{Phillips1970}. Dots: Greywall \cite{Greywall78,Greywall79ERRATUM}. Circles: Greywall's data, temperatures corrected by adding 1\,mK; this correction should apply below 0.3\,K (see text). }
	\label{fig:Cv-PhononRange}
\end{figure}

Deviations are also observed at low temperatures, they are rather small (Fig. \ref{fig:Difference-Cv-neutrons}). Their probable origin can be understood by inspection of  Fig. \ref{fig:Cv-PhononRange}, where the phonon regime is emphasized. The expanded scale reveals a clear accident in Greywall's data around 0.3\,K. The curve calculated from the neutron data, however, does not display any accident in this range. An excellent agreement is obtained with Greywall's data if we correct the latter in an obvious way: above 0.33\,K we keep his temperature scale, based on the $^3$He vapor pressure thermometer, and below this temperature we correct the Curie temperature of his CMN thermometer by simply adding 1\,mK. Such a correction is within the possibilities considered by Greywall in his error handling discussion. It would be desirable to repeat the heat capacity experiment using modern  thermometric techniques, but this kind of experiment remains difficult. Finally, let us mention that the drop of C$_V$/T$^3$ observed in Greywall's data below 0.15\,K is clearly an artifact due to thermal decoupling.     
 
\begin{figure}[htbp]
	\centering
		\includegraphics[width=1.0\columnwidth]{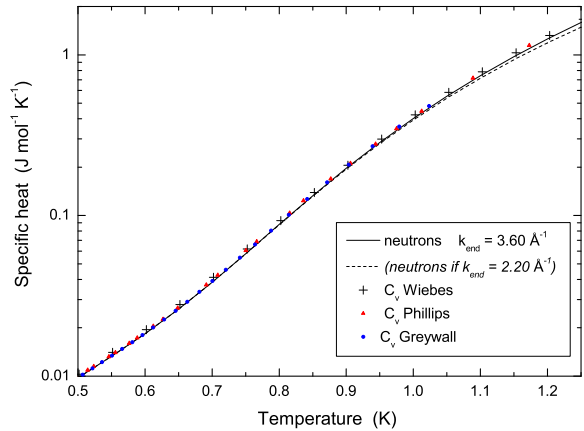}
	\caption{Specific heat calculated numerically with the measured dispersion relation, for an  integration limit k$_{end}$=3.6 \AA$^{-1}$. The result obtained using k$_{end}$=2.2 \AA$^{-1}$ is shown for comparison. Specific heat measurements by Wiebes \cite{WiebesThesis,DonnellyBarenghi}, Phillips \textit{et al.} \cite{Phillips1970} and Greywall \cite{Greywall78,Greywall79ERRATUM} are also shown. }
	\label{fig:Cv-RotonRange}
\end{figure}

At higher temperatures, the behavior of the heat capacity is dominated by the exponential growth associated to the roton gap. Fig. \ref{fig:Cv-RotonRange} shows that the influence of k$_{end}$, the upper integration limit of the dispersion relation, is small below 1.25\,K, the maximum temperature where the low temperature dispersion relation can be safely used. We find that values of k$_{end}$$\approx$3.6$\pm$1\,\AA$^{-1}$ provide good fits. A curve calculated with 2.20 \AA$^{-1}$, too small a value, is shown for comparison purposes only. A  termination of the spectrum in this wave-vector region is therefore in rough agreement with specific heat data. 

To conclude this section, we should say that the systematic difference on the order of 1 to 2\% observed between the specific heat calculated from the dispersion curve in the present work, and the results of heat capacity measurements, is larger than the estimated uncertainty of the former ($\leq$1\%, see Supplemental Material at [URL will be inserted by publisher] for uncertainty data tables and files), but consistent with the quoted uncertainty of the latter (2\%). Neutron data (combined with ultrasound in the phonon region) provide therefore a better determination of this thermodynamic parameter than the direct measurement.  
 
\subsection{Specific heat: analytical calculation} 
It is often convenient to have analytical expressions for the thermodynamical properties. We calculate in this  section the low temperature polynomial series expansions describing phonon and roton contributions to the  specific heat, and to other thermodynamical properties. 

\subsubsection{Phonon series expansions} 
The properties related to the phonons are calculated with Eq. \ref{equ:energy} and the following series expansion of the phonon excitation energies: 

\begin{equation}
\epsilon(k)=ck(1+ \alpha_1 k +\alpha_2 k^2 + \alpha_3 k^3 + \alpha_4 k^4+ \alpha_5 k^5+ \alpha_6 k^6) 
\end{equation}

The integration upper limit is taken as infinity, since at temperatures T$<<$$\Delta_R$$/$k$_B$ the exponential thermal population factor suppresses the contribution of the high energy parts of the spectrum. In practice, the analytic calculation is done by integration over the energy, rather than over wave-vector, and we introduce the inverse series describing the single-valued (phonon) branch of the spectrum:

\smallskip
\begin{align}
k=&\frac{\omega }{c}-\alpha_1\left(\frac{\omega}{c}\right)^2+\left(2\alpha_1^2-\alpha_2\right)\left(\frac{\omega}{c}\right)^3 \nonumber\\
&+\left(-5\alpha_1^3+5\alpha_1\alpha_2-\alpha_3\right)\left(\frac{\omega}{c}\right)^4 \nonumber\\
&+\left(14 \alpha_1^4-21 \alpha_1^2 \alpha_2+6 \alpha_1 \alpha_3+3 \alpha_2^2-\alpha_4\right) \left(\frac{\omega}{c}\right)^5 \nonumber\\
&+\bigl(-42 \alpha_1^5+84 \alpha_1^3 \alpha_2-28 \alpha_1^2 \alpha_3-28 \alpha_1 \alpha_2^2\nonumber\\
&\qquad+7 \alpha_1 \alpha_4+7 \alpha_2 \alpha_3-\alpha_5\bigr)\left(\frac{\omega}{c}\right)^6\nonumber\\
&+ \eta\left(\frac{\omega}{c}\right)^7 +{\cal O}\left(\omega^8\right)\nonumber
\end{align}
where
\begin{align}
  \eta&=132 \alpha_1^6-330 \alpha_1^4 \alpha_2+120 \alpha_1^3 \alpha_3+
  180 \alpha_1^2 \alpha_2^2\nonumber\\
  &-36 \alpha_1^2 \alpha_4-72 \alpha_1 \alpha_2 \alpha_3+8 \alpha_1 \alpha_5
  \nonumber\\
  &-12 \alpha_2^3+8 \alpha_2 \alpha_4+4 \alpha_3^2-\alpha_6\,.
  \nonumber
\end{align}

The complete analytical formulas for the specific heat are cumbersome, and we only give below those applicable when $\alpha_1$=0, which is the case in practice (see Section \ref{sec:sound}). 
The corresponding expression  is 
\begin{equation}
C_{V}^{phonon}=\textbf{A} T^3+ \textbf{C} T^5+ \textbf{D} T^6+ \textbf{E} T^7+ \textbf{K} T^8+ \textbf{L} T^9
\end{equation}

where the term \textbf{B}T$^4$ is absent because $\alpha_1$=0, and the relevant coefficients are given by: 

\noindent $\textbf{A}=\frac{2 \pi^2 k_B^4 V}{15 c^3 \hbar ^3}$ \\
$\textbf{C}=-\frac{40 \left(\pi^4 \alpha_2 k_B^6 V\right)}{21 \left(c^5 \hbar^5\right)}$\\
$\textbf{D}=-\frac{15120 \left(\alpha_3 k_B^7 V \zeta (7)\right)}{\pi^2 c^6 \hbar^6}$ \\
$\textbf{E}=\frac{224 \pi^6 k_B^8 V \left(4 \alpha_2^2-\alpha_4\right)}{15 c^7 \hbar^7}$ \\
$\textbf{K}=\frac{1451520 k_B^9 V \zeta (9) (9 \alpha_2 \alpha_3-\alpha_5)}{\pi ^2 c^8 \hbar^8}$ \\ 
$\textbf{L}=-\frac{640 \left(\pi^8 k_B^{10} V \left(55 \alpha_2^3-30 \alpha_2 \alpha_4-15 \alpha_3^2+3 \alpha_6\right)\right)}{11 \left(c^9 \hbar^9\right)}$\\
where $\zeta$ is the Riemann zeta function. 

It is clear from this expansion that, contrarily to what is generally believed, the coefficients of the specific heat series expansion are not associated to a single coefficient of the dispersion relation series; \textbf{E}, for instance, contains both $\alpha_2$ and $\alpha_4$.   

The $\alpha_i$ coefficients have been determined in Section \ref{sec:lowQ} by fits at low wave-vectors. 
For the dispersion curve at saturated vapor pressure, the coefficients    
 c=238.3 m/s, $\alpha_1$=0, $\alpha_2$=1.55\AA$^{-2}$, $\alpha_3$=-4.04\AA$^{-3}$, $\alpha_4$=2.30\AA$^{-4}$, $\alpha_5$=0, $\alpha_6$=0 provide a good fit of the dispersion curve for k$<$0.5\,\AA$^{-1}$.  
The corresponding coefficients of the specific heat series are therefore:   
\textbf{A}=0.0831, \textbf{C}=-0.0548, \textbf{D}=0.0653, \textbf{E}=0.0603, \textbf{K}=-0.262, \textbf{L}=0.141, 
when temperatures are expressed in kelvin and C$_V$ in J/(mol.K).

\begin{figure}[htbp]
	\centering
		\includegraphics[width=1.0\columnwidth]{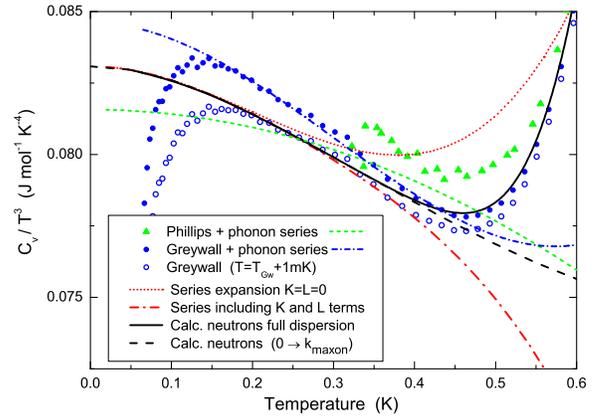}
	\caption{C$_V$/T$^3$ showing the phonon contribution to the specific heat and the onset of the roton contribution. Dots: C$_V$ measurements \cite{Phillips1970, Greywall78,Greywall79ERRATUM} with their fitted phonon contributions (see legend). Present work: a) red dotted line: analytical result with K=L=0; b) red dash-dotted line: analytical result with K$\neq$0 and L$\neq$0; c) black solid line: numerical integration of full dispersion curve; d) dashed line: numerical integration over the phonon region (0$<$k$<$k$_M$). See Fig. \ref{fig:Cv-PhononRange} for additional details.  }
	\label{fig:PhononContribution}
\end{figure}

Strong oscillations in the amplitude of successive terms of the series expansions are present for temperatures $\approx$1\,K. They originate from an attempt to describe the dispersion relation by a power series expansion around  k=0 while (as seen  in Fig. \ref{fig:phasevelocLargeRange}) the phase velocity has a very linear portion, quite extended, far from k=0. 
This quite unusual behavior has several consequences. Clearly, for k$>$0.5\AA$^{-1}$, the higher order coefficients will vary with the order of the polynomial, and once a polynomial providing a good description of the dispersion relation is chosen, it must be consistently used in the thermodynamic calculations. The previous discussion may appear as academic, but it is at the root of the problems encountered in the analysis of specific heat measurements \cite{Phillips1970,Greywall78,Greywall79ERRATUM}, as is illustrated in Fig. \ref{fig:PhononContribution}. 
The partial contribution due to the phonons estimated by these authors (green dashed line for Phillips \textit{et al.}, blue dash-dotted line for Greywall), has the right order of magnitude, but it deviates from the present neutron scattering result (black dashed line) above 0.5\,K; in particular, the temperature dependence is clearly different. 

The \emph{analytical} calculations of the specific heat using series expansions of the dispersion curve display strong deviations with respect to the full numeric result, beginning at temperatures as low as 0.4\,K. Furthermore, the series expansions used in the previous analysis of specific heat data \cite{Phillips1970, Greywall78,Greywall79ERRATUM}, being truncated, cannot be directly used to extract the dispersion relation coefficients. As shown in the figure, removing the terms K and L strongly affects the fits. 
Handled with care, the analytical expansions are still very useful for the calculation of phonon thermal properties in different contexts. 
 
\subsubsection{Roton region} 

The analytical calculation of the roton contribution to the thermodynamical properties was already performed by Landau in his seminal papers on roton excitations \cite{Landauroton,Landauroton2}. 
In the vicinity of the roton minimum, the excitation energies are described by Eq. \ref{Eq:fitroton}. 
In the limiting case where B$_R$=C$_R$=0, Landau obtains the expression 

\begin{equation}
C_V^R=C_{\textit{L}} e^{-\frac{\Delta_R}{k_B T}} \left(1 +\frac{k_B T}{\Delta_R} +\frac{3}{4} \left(\frac{k_B T}{\Delta_R}\right)^2\right)
\label{eq:CvRotonLandau}
\end{equation}

with  $C_{\textit{L}}=\frac{\Delta_R^2 k_R^2 V \sqrt{\mu_R m_4}}{\sqrt{2} \pi^{3/2} \sqrt{k_B} T^{3/2} \hbar}$\\

The analysis of the measured specific heats with Landau's analytical expression is therefore sensitive to a combination of several roton parameters (energy, wave-vector and effective mass). 

Landau also assumes that $\left|k-k_R\right|<<k_R$. 
The analytical calculation, however, can be done without this last assumption. 
We calculate here the average energy of the rotons E$_R$ and their specific heat C$_V^R$ with Eq. \ref{equ:energy} using Eq. \ref{Eq:fitroton} with B$_R$=C$_R$=0, including in the integrand all terms of the expansion of $(k-k_R)^2$. The calculation is performed by integration over wave-vectors. We find an additional term 

\begin{equation}
C_V^{addit}=C_{\textit{L}} e^{-\frac{\Delta_R}{k_B T}}\left(\frac{T }{2\tau_R}\right)\left(1+\frac{3 k_B T}{\Delta_R}+\frac{15}{4}\left(\frac{k_B T}{\Delta_R}\right)^2\right)
\label{eq:CvRotonHG}
\end{equation}

where we introduced the parameter $\tau_R$ which has the dimensions of temperature:
$\tau_R$ $=\frac{k_R^2 \hbar^2}{2 k_B \mu_R m_4}$. 
The additional term is similar to Landau's expression, with an additional coefficient T/(2$\tau_R$). Since $\tau_R$ $\approx$ 158\,K, this additional term is of the same magnitude as the third term in Eq. \ref{eq:CvRotonLandau} for temperatures above 1\,K, and they are both very small.

\begin{figure}[htbp]
	\centering
		\includegraphics[width=1.0\columnwidth]{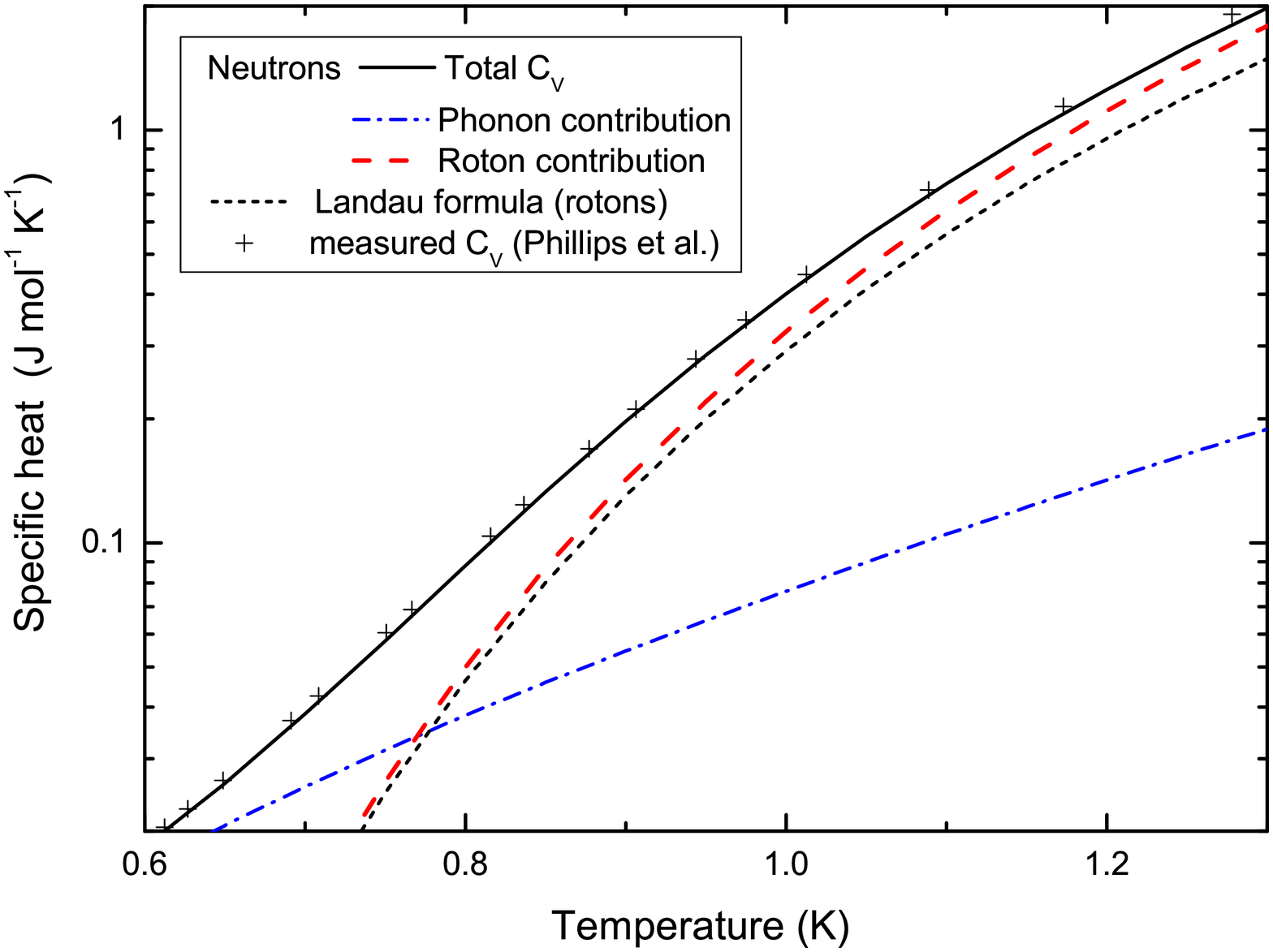}
	\caption{The total heat capacity (black solid line), and phonon (blue dash-dotted line), roton (red dashed line) contributions to the specific heat calculated numerically with the measured dispersion relation (present work). Short-dash black line: Landau formula for the rotons contribution. Crosses: specific heat measurements \cite{Phillips1970}. }
	\label{fig:rotonCompLandau}
\end{figure}

In order to improve the accuracy of Landau's analytic description, it is necessary to take into account the asymmetry (cubic term) and  the deformation of the minimum (quartic term). Unfortunately, the corresponding expressions are extremely complex and cumbersome, they are not of practical interest. At the analytical level, we are therefore left with equations providing a rough approximation. As seen in Fig. \ref{fig:rotonCompLandau}, the difference between Landau's approximation and the roton contribution calculated numerically with the dispersion curve grows substantially with the temperature. The figure also illustrates the fact that phonons, defined as excitations below the maxon wave-vector, have a non-negligible contribution to the thermodynamic properties at higher temperatures than commonly believed: the phonon and roton contributions become equal at 0.77\,K for the heat capacity, for instance.

\subsection{Normal fluid density} 

The normal fluid density is given \cite{Landauroton,Landauroton2,FetterWalecka} by the expression
\begin{equation}
\rho_n=\left(\frac{\hbar^2}{6\pi^2 k_B T}\right) \int_{0}^{k_{end}} \frac{e^{\frac{\epsilon(k)}{k_B T}}}{e^{\frac{\epsilon(k)}{k_B T}}-1}k^4dk 
\label{eq:rho_n}
\end{equation}

\begin{figure}[htbp]
	\centering
		\includegraphics[width=1.0\columnwidth]{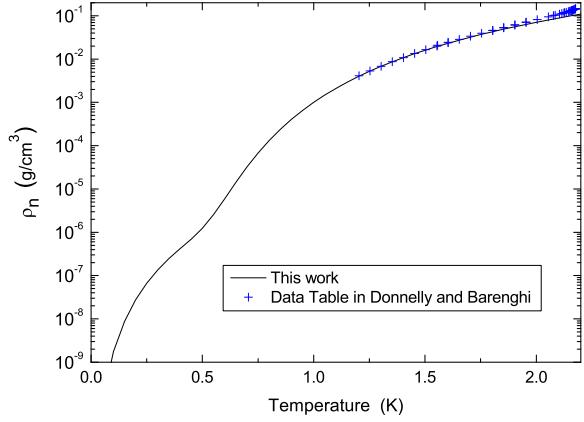}
	\caption{Normal fluid density calculated using the measured dispersion relation. Experimental points by Maynard, Tam and Ahlers, and Singass and Ahlers (data-base in Ref. \onlinecite{DonnellyBarenghi})}.
	\label{fig:Rho_n}
\end{figure}

One can obviously integrate analytically the series expansions within the  limitations discussed previously. In order to take full advantage of the  measured dispersion curve $\epsilon(k)$, it is convenient to perform a numerical integration, as done above for the specific heat.
The normal fluid densities calculated using the dispersion curve are given in Table \ref{tab:rho_n} in the temperature range up to 1.3\,K. 
The result is compared in Fig. \ref{fig:Rho_n} to selected experimental data \cite{DonnellyBarenghi}. The latter cover the  temperature range from 1.2\,K to the lambda point. 
   
\begin{figure}[htbp]
	\centering
		\includegraphics[width=1.0\columnwidth]{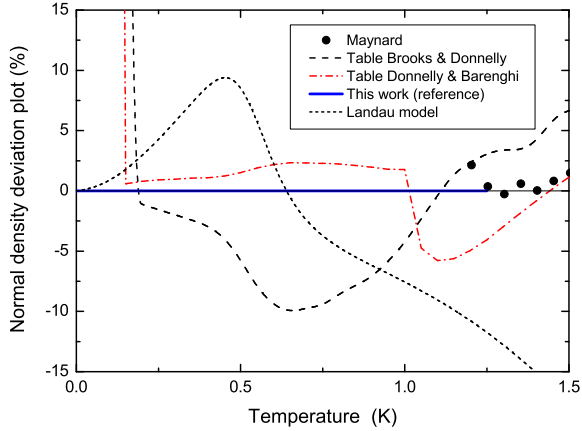}
	\caption{Deviation plot for the normal fluid density. We show the percent difference between Maynard's experimental data, or tabulated (essentially, calculated) values \cite{BrooksDonnelly,DonnellyBarenghi}, and the values calculated using the  dispersion relation, taken as the reference (this work). The `simple Landau model' (see text) deviates substantially from our data. }
	\label{fig:deviationRho_n}
\end{figure}

The percent difference between previous data \cite{BrooksDonnelly,DonnellyBarenghi} and our calculated curve is shown in Fig. \ref{fig:deviationRho_n}, on the one hand for the experimental data-base, and on the other hand for the  tabulated values. The latter were calculated  below 1\,K using spline approximations to different neutron data sets. We see that for the purposes of calculating $\rho_n$, our result matches accurately the direct measurements at about 1.25\,K. Above this temperature, the dispersion curve itself becomes temperature dependent, a problem that is beyond the scope of the present article. It is therefore extremely satisfactory to see that these two sets of data agree particularly well. It is also clear that neutron techniques extend considerably towards lower temperatures the range where this parameter has been accurately determined. The temperature range below 1\,K is a particularly fertile playground for objects all possible sizes immersed in helium and sensitive to its excitations, from nano-oscillators to huge particle detectors \cite{Guenault2019,2020-PRB-nano-osc,Guenault2019,2020-PRB-nano-osc,Guo2013,morishita1989mean,PRL2016Zurek,PRL2017MarisDetector,Zurek2017}. 

Landau's general formula, Eq. \ref{eq:rho_n}, yields when applied to the first term of the series expansions for the phonon and the roton dispersion relation (the so-called `simple Landau model' \cite{Landauroton,Landauroton2}) :
\noindent
$\rho_n^{(L)}=\rho_n^{Ph(L)}+\rho_n^{R(L)}$, with \\

\noindent
$\rho_n^{Ph(L)}=\frac{2\pi^2 {k_B}^4 T^4}{45 c^5 \hbar^3}$ \\
$\rho_n^{R(L)}= \frac{\hbar k_R^4 (\mu_R m_4)^{1/2}}{3 \sqrt{2} \pi^{3/2} (k_B T)^{1/2}} e^{-\frac{\Delta_R}{k_B T}}$\\

\begin{figure}[htbp]
	\centering
		\includegraphics[width=1.0\columnwidth]{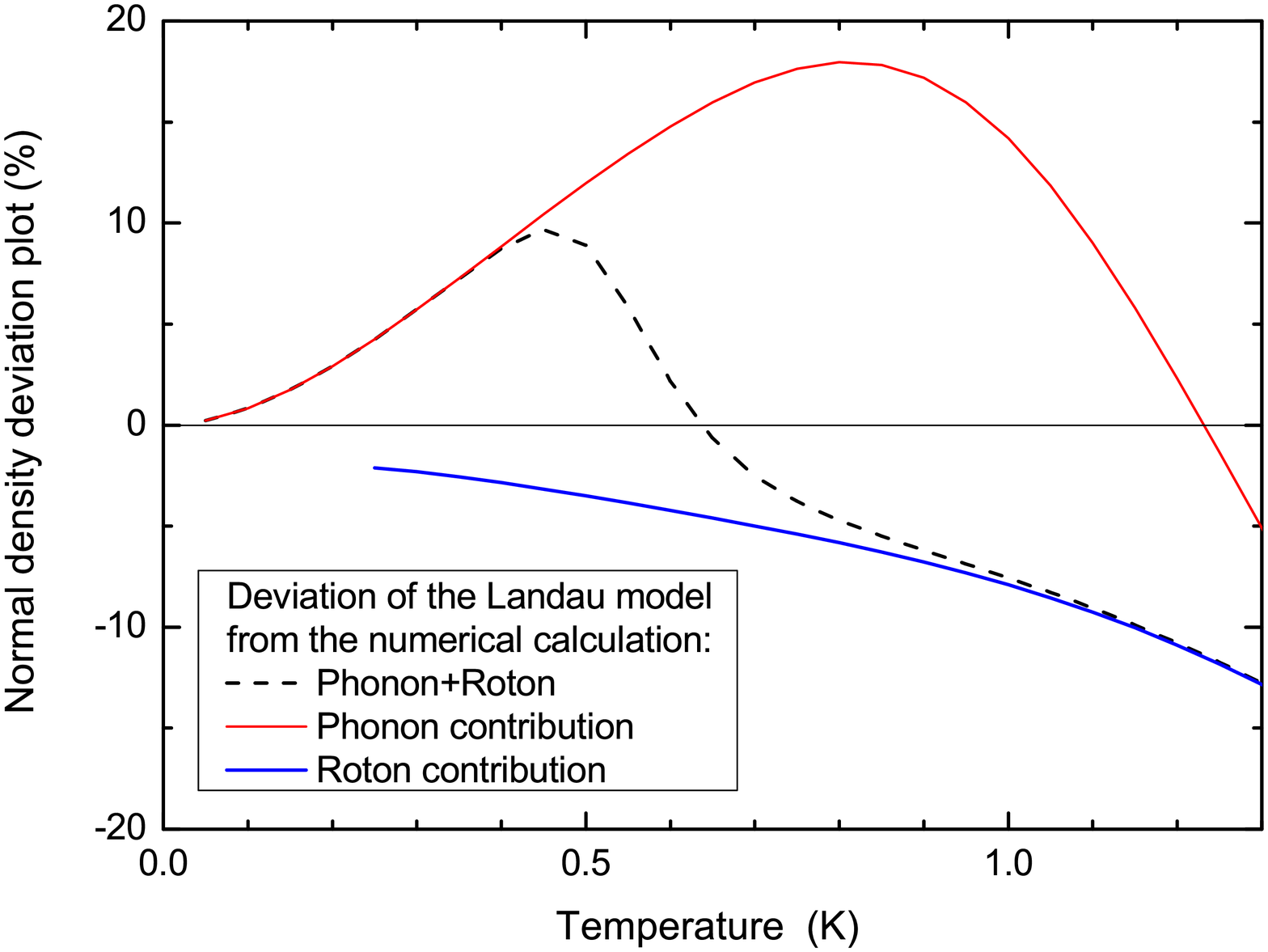}
	\caption{Normal fluid density deviation plot: percent difference between the `simple Landau model' normal densities \cite{Landauroton,Landauroton2,FetterWalecka}, and our numerical calculation with the full dispersion curve. Phonon, roton, and total normal density deviations are defined as 100\,($\rho_n^{Ph(L)}$-$\rho_n^{Ph}$)/$\rho_n^{Ph}$, 100\,($\rho_n^{R(L)}$-$\rho_n^{R}$)/$\rho_n^{R}$, and  100\,$((\rho_n^{Ph(L)}+\rho_n^{R(L)}$) - $\rho_n$)/$\rho_n$, respectively.  }
	\label{fig:DeviationLandau}
\end{figure}

Since these formulas are frequently used, it is interesting to check their validity range. A comparison between the expressions above and our complete numerical calculation with the measured dispersion relation, is shown in Fig. \ref{fig:deviationRho_n}. Deviations on the order of 10\% are found below 1\,K, even around 0.5\,K where this expression is often assumed to give a good approximation. The deviation is even larger if one considers the phonon and roton contributions separately, as shown in Fig. \ref{fig:DeviationLandau}. The superfluid densities are defined as $\rho_s=\rho-\rho_n$ (see Table \ref{tab:rho_n}).

\begingroup
\begin{table}[h]
\begin{ruledtabular}
\begin{tabular}{lllll}
$T$  & $\rho$  & $\rho_n$   & $\rho_n^{Ph}$  & $\rho_n^{R}$   \\ 
K    & g/cm$^3$& g/cm$^3$   &   g/cm$^3$     & g/cm$^3$   \\ \hline
0    & 0.14514 & --         & --             & --                 \\
0.05 & 0.14514 & 1.103E-10 & 1.103E-10     & 3.94E-74         \\
0.10 & 0.14514 & 1.754E-09 & 1.754E-09     & 6.65E-37         \\
0.15 & 0.14514 & 8.799E-09 & 8.799E-09     & 1.566E-24         \\
0.20 & 0.14514 & 2.749E-08 & 2.749E-08     & 2.305E-18         \\
0.25 & 0.14514 & 6.626E-08 & 6.626E-08     & 1.130E-14         \\
0.30 & 0.14514 & 1.355E-07 & 1.355E-07     & 3.213E-12         \\
0.35 & 0.14514 & 2.476E-07 & 2.474E-07     & 1.798E-10         \\
0.40 & 0.14514 & 4.196E-07 & 4.159E-07     & 3.649E-09         \\
0.45 & 0.14514 & 6.945E-07 & 6.567E-07     & 3.772E-08         \\
0.50 & 0.14514 & 1.230E-06 & 9.872E-07     & 2.432E-07         \\
0.55 & 0.14514 & 2.540E-06 & 1.427E-06     & 1.113E-06         \\
0.60 & 0.14514 & 5.939E-06 & 1.997E-06     & 3.943E-06         \\
0.65 & 0.14513 & 1.419E-05 & 2.722E-06     & 1.147E-05         \\
0.70 & 0.14513 & 3.220E-05 & 3.631E-06     & 2.857E-05         \\
0.75 & 0.14513 & 6.769E-05 & 4.757E-06     & 6.293E-05         \\
0.80 & 0.14513 & 1.316E-04 & 6.141E-06     & 1.254E-04         \\
0.85 & 0.14513 & 2.381E-04 & 7.835E-06     & 2.303E-04         \\
0.90 & 0.14512 & 4.048E-04 & 9.901E-06     & 3.95E-04         \\
0.95 & 0.14512 & 6.519E-04 & 1.242E-05     & 6.39E-04         \\
1.00 & 0.14512 & 0.001002  & 1.549E-05     & 9.87E-04         \\
1.05 & 0.14512 & 0.001480  & 1.922E-05     & 0.001461          \\
1.10 & 0.14511 & 0.002112  & 2.375E-05     & 0.00209          \\
1.15 & 0.14511 & 0.002925  & 2.923E-05     & 0.00289          \\
1.20 & 0.14512 & 0.003945  & 3.585E-05     & 0.00391          \\
1.25 & 0.14512 & 0.005201  & 4.378E-05     & 0.00516          \\
1.30 & 0.14512 & 0.006720  & 5.322E-05     & 0.00667          \\ 
\end{tabular}
\end{ruledtabular}
\caption{Total density $\rho$ (from Ref. \onlinecite{DonnellyBarenghi}) and normal fluid densities (total, phonon and roton contributions) as a function of temperature, calculated numerically using the measured dispersion relation (this work). See Supplemental Material at [URL will be inserted by publisher] for more detailed tables.}
\label{tab:rho_n}
\end{table}
\endgroup

\subsection{Number density of phonons and rotons} 
The number density of phonons $N_{Ph}$ and rotons $N_{R}$ as a function of temperature, of interest for instance in the ballistic regime, is given by the expressions:\\

$N_{Ph}=(2\pi^2)^{-1}\int_{0}^{k_{M}} (e^{\frac{\epsilon(k)}{k_B T}}-1)^{-1} k^2dk $  \\

$N_{R}=(2\pi^2)^{-1}\int_{k_{M}}^{k_{end}} (e^{\frac{\epsilon(k)}{k_B T}}-1)^{-1} k^2dk $ \\

The result is shown Fig. \ref{fig:NumberPerAtom} (see Supplemental Material at [URL will be inserted by publisher] for data tables.), converted to the number of phonons and rotons per helium atom, in order to emphasize the importance of the excitation density as the temperature exceeds T$\approx$1.3\,K. Near the lambda point, the roton density, an atomic-size excitation, approaches one per atom, as expected. 
 
 \begin{figure}[htbp]
	\centering
		\includegraphics[width=0.9\columnwidth]{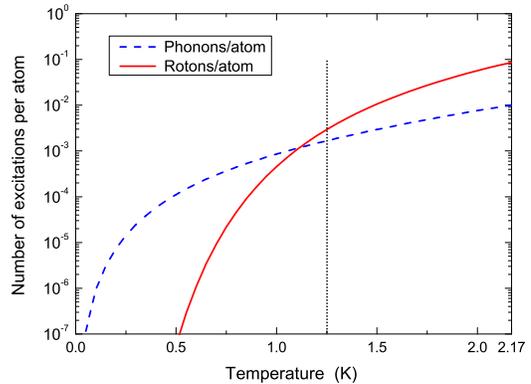}
	\caption{Number of phonon and roton excitations per atom. The numerical calculation performed with the measured low temperature dispersion curve, is accurate for T$<$1.3\,K.}
	\label{fig:NumberPerAtom}
\end{figure}

\section{Conclusions}
The main result of this work is the determination of the dispersion relation $\epsilon(k)$ of superfluid $^4$He in the whole wave-vector range at saturated vapor pressure, and in a large range for pressures up to 24 bar. Earlier neutron data  were scarce,  of low resolution, often contradictory. Roton and maxon parameters were affected by the instrumental resolution and the wave-vector range over which the dispersion curve was averaged out, leading to a substantial spread in the values found in the literature;  as the experimental techniques progressively improved, lower roton energies, wave-vectors and effective masses were obtained. 
Motivated by the lack of high accuracy data, Donnelly and coworkers \cite{BrooksDonnelly,DonnellyBarenghi} designed a `recommended dispersion relation', a spline curve making its way across the large error bars of some of the data available at that time. 

A \emph{measured} dispersion curve is now available. It is characterized by small error bars, and it agrees well with measurements performed by other techniques: ultrasound and compressibility at low wave-vectors, as well as  heat capacity, a technique providing  an important global test over a large wave-vector range.   

Our neutron measurements have been performed using time of flight (TOF) techniques with a 10$^5$ pixels detector matrix. Statistical errors are in general negligible in this work. Considerable effort has been devoted to understand the sources of systematic errors and the resulting corrections. This also allowed us to identify problems that affected earlier measurements.  
 
The energy scale of our spectrometer has been refined by a calibration at a single point, the roton energy $\Delta_R$=0.7418\,meV determined with an uncertainty of 1\,$\mu$eV by Stirling \cite{stirling-83,GlydeBook,StirlingExeter}. This is the main source of uncertainty of our results for $\epsilon$(k). It affects them in a global way, and they can therefore be corrected (the energies should be corrected proportionally, as indicated by Eq. \ref{Erefined}, and the wave-vectors recalculated using Eq. \ref{QAngle}) if a more accurate value of $\Delta_R$ becomes available. The remaining systematic errors originate from small defects in the IN5 spectrometer construction. Comparing data at two different neutron energies has allowed us to show that the corresponding effects are smaller than the global systematic error quoted above. They can be seen as oscillations in the curves, in particular around the roton minimum. No effort was done to suppress these by  averaging or smoothing, since they are a good indicator, for further applications of the present data, of residual systematic errors. 

The measured dispersion  curve has allowed us to calculate the thermodynamic properties of superfluid helium in the temperature range below 1.3\,K, where the non-interacting-excitations picture is valid. Our results provide the most accurate values for the heat capacity (direct measurements are affected by delicate issues of thermometry and temperature scales), the normal fluid density, and other thermodynamical properties. They also provide independently the  contributions from excitations in different wave-vector ranges, and in particular the phonon and the roton contributions. This information is of interest, for instance, for low temperature transport phenomena, the damping of nano-resonators in helium below 1\,K \cite{Guenault2019,2020-PRB-nano-osc}, particle detection \cite{PRL2016Zurek,PRL2017MarisDetector,Zurek2017}, and several other effects determined by ballistic phonons and roton quasiparticles.

In addition to these numerical calculations, we developed analytical expressions for the thermodynamical properties. We found that specific heat measurements were analyzed using inconsistent low temperature series expansions; no disagreement is observed between the neutron results and the specific heat data using our coherent expressions.   

The results on pure superfluid $^4$He have analogues in other systems. Rotons and maxons are studied in $^4$He in reduced dimensions, confined geometries, droplets spectrometry \cite{KroNavarroBook,bryan2018confined,2018-Nicolis,bossy2019,2019dropletsMoroshkin,2019droplets}, but also in $^3$He \cite{godfrin2012nature,pines2016emergent}, in cold atomic gases \cite{santos2003roton,chomaz2018observation,2020-maxon-roton}, classical liquids \cite{trachenko2015}, etc.: they are a general feature of many-body interacting systems.

Throughout this manuscript, we have considered `elementary excitations' (poles of the fully renormalized single-particle propagator) and `statistical quasi-particles' (elementary excitation energies defined as the functional derivative of the
total energy with respect to the distribution function), in the sense described by Balian and de Dominicis  
\cite{Balian1971,Carneiro1975,Carneiro1979}, as essentially identical concepts.
The description of the thermodynamic properties using a self-consistent temperature dependent dispersion relation has been explored by Donnelly and Roberts \cite{DonnellyRoberts}, it describes phenomenologically the thermodynamic properties. 
The other route, presently explored by different types of many-body microscopic calculations, seems more promising.  
The difference between the measured specific heat and that calculated for non-interacting quasi-particles using the dispersion curve has a simple behavior (see Fig. \ref{fig:Cv-fullRange}), and  provides information about roton-roton interactions \cite{Cohen1960,Fak2012}. Additional contributions have been predicted \cite{Carneiro1975,Carneiro1979} for the heat capacity, it would  be interesting to have a theoretical estimate of their magnitude and make a quantitative comparison with the present data. 

\section{Acknowledgements}
We are grateful to S. Triqueneaux and X. Tonon for their help with the experiments, to B. Gu\'erard and P. van Esch for helpful discussions on the IN5 detectors properties, and to H. Maris for pointing out Ref. \onlinecite{Rugar1984} to us. 
This work was supported by the European Microkelvin Platform. The research leading to these results has received funding from the European Union’s Horizon 2020 Research and Innovation Programme, under Grant Agreement no 824109.


\begin{thebibliography}{106}%
\makeatletter
\providecommand \@ifxundefined [1]{%
 \@ifx{#1\undefined}
}%
\providecommand \@ifnum [1]{%
 \ifnum #1\expandafter \@firstoftwo
 \else \expandafter \@secondoftwo
 \fi
}%
\providecommand \@ifx [1]{%
 \ifx #1\expandafter \@firstoftwo
 \else \expandafter \@secondoftwo
 \fi
}%
\providecommand \natexlab [1]{#1}%
\providecommand \enquote  [1]{``#1''}%
\providecommand \bibnamefont  [1]{#1}%
\providecommand \bibfnamefont [1]{#1}%
\providecommand \citenamefont [1]{#1}%
\providecommand \href@noop [0]{\@secondoftwo}%
\providecommand \href [0]{\begingroup \@sanitize@url \@href}%
\providecommand \@href[1]{\@@startlink{#1}\@@href}%
\providecommand \@@href[1]{\endgroup#1\@@endlink}%
\providecommand \@sanitize@url [0]{\catcode `\\12\catcode `\$12\catcode
  `\&12\catcode `\#12\catcode `\^12\catcode `\_12\catcode `\%12\relax}%
\providecommand \@@startlink[1]{}%
\providecommand \@@endlink[0]{}%
\providecommand \url  [0]{\begingroup\@sanitize@url \@url }%
\providecommand \@url [1]{\endgroup\@href {#1}{\urlprefix }}%
\providecommand \urlprefix  [0]{URL }%
\providecommand \Eprint [0]{\href }%
\providecommand \doibase [0]{http://dx.doi.org/}%
\providecommand \selectlanguage [0]{\@gobble}%
\providecommand \bibinfo  [0]{\@secondoftwo}%
\providecommand \bibfield  [0]{\@secondoftwo}%
\providecommand \translation [1]{[#1]}%
\providecommand \BibitemOpen [0]{}%
\providecommand \bibitemStop [0]{}%
\providecommand \bibitemNoStop [0]{.\EOS\space}%
\providecommand \EOS [0]{\spacefactor3000\relax}%
\providecommand \BibitemShut  [1]{\csname bibitem#1\endcsname}%
\let\auto@bib@innerbib\@empty
\bibitem [{\citenamefont {Nozi\`{e}res}\ and\ \citenamefont
  {Pines}(2018)}]{NozieresPines2018}%
  \BibitemOpen
  \bibfield  {author} {\bibinfo {author} {\bibfnamefont {P.}~\bibnamefont
  {Nozi\`{e}res}}\ and\ \bibinfo {author} {\bibfnamefont {D.}~\bibnamefont
  {Pines}},\ }\href@noop {} {\emph {\bibinfo {title} {The Theory of {Q}uantum
  Liquids}}},\ Advanced Book Classics\ (\bibinfo  {publisher} {CRC Press},\
  \bibinfo {year} {2018})\BibitemShut {NoStop}%
\bibitem [{\citenamefont {Fetter}\ and\ \citenamefont
  {Walecka}(2003)}]{FetterWalecka}%
  \BibitemOpen
  \bibfield  {author} {\bibinfo {author} {\bibfnamefont {A.~L.}\ \bibnamefont
  {Fetter}}\ and\ \bibinfo {author} {\bibfnamefont {J.~D.}\ \bibnamefont
  {Walecka}},\ }\href@noop {} {\emph {\bibinfo {title} {{Q}uantum Theory of
  Many-Particle Systems}}}\ (\bibinfo  {publisher} {Dover Publications},\
  \bibinfo {address} {Mineola, NY},\ \bibinfo {year} {2003})\BibitemShut
  {NoStop}%
\bibitem [{\citenamefont {Thouless}(2014)}]{ThoulessBook}%
  \BibitemOpen
  \bibfield  {author} {\bibinfo {author} {\bibfnamefont {D.~J.}\ \bibnamefont
  {Thouless}},\ }\href@noop {} {\emph {\bibinfo {title} {The {Q}uantum
  Mechanics of Many-Body Systems}}},\ \bibinfo {edition} {2nd}\ ed.\ (\bibinfo
  {publisher} {Dover Publications},\ \bibinfo {address} {Mineola, NY},\
  \bibinfo {year} {2014})\BibitemShut {NoStop}%
\bibitem [{\citenamefont {Landau}(1941)}]{Landauroton}%
  \BibitemOpen
  \bibfield  {author} {\bibinfo {author} {\bibfnamefont {L.}~\bibnamefont
  {Landau}},\ }\href@noop {} {\bibfield  {journal} {\bibinfo  {journal} {J.
  Phys. (Moscow)}\ }\textbf {\bibinfo {volume} {5}},\ \bibinfo {pages} {71}
  (\bibinfo {year} {1941})}\BibitemShut {NoStop}%
\bibitem [{\citenamefont {Landau}(1947)}]{Landauroton2}%
  \BibitemOpen
  \bibfield  {author} {\bibinfo {author} {\bibfnamefont {L.}~\bibnamefont
  {Landau}},\ }\href@noop {} {\bibfield  {journal} {\bibinfo  {journal} {J.
  Phys. (Moscow)}\ }\textbf {\bibinfo {volume} {11}},\ \bibinfo {pages} {91}
  (\bibinfo {year} {1947})}\BibitemShut {NoStop}%
\bibitem [{\citenamefont {Bogoliubov}(1947)}]{Bogoljubov}%
  \BibitemOpen
  \bibfield  {author} {\bibinfo {author} {\bibfnamefont {N.~N.}\ \bibnamefont
  {Bogoliubov}},\ }\href@noop {} {\bibfield  {journal} {\bibinfo  {journal} {J.
  Phys. U.S.S.R}\ }\textbf {\bibinfo {volume} {11}},\ \bibinfo {pages} {23}
  (\bibinfo {year} {1947})}\BibitemShut {NoStop}%
\bibitem [{\citenamefont {Jackson}\ \emph {et~al.}(1981)\citenamefont
  {Jackson}, \citenamefont {Jennings}, \citenamefont {Smith},\ and\
  \citenamefont {Lande}}]{SmithSolid}%
  \BibitemOpen
  \bibfield  {author} {\bibinfo {author} {\bibfnamefont {A.~D.}\ \bibnamefont
  {Jackson}}, \bibinfo {author} {\bibfnamefont {B.~K.}\ \bibnamefont
  {Jennings}}, \bibinfo {author} {\bibfnamefont {R.~A.}\ \bibnamefont {Smith}},
  \ and\ \bibinfo {author} {\bibfnamefont {A.}~\bibnamefont {Lande}},\
  }\href@noop {} {\bibfield  {journal} {\bibinfo  {journal} {Phys. Rev. B}\
  }\textbf {\bibinfo {volume} {24}},\ \bibinfo {pages} {105} (\bibinfo {year}
  {1981})}\BibitemShut {NoStop}%
\bibitem [{\citenamefont {Nozi\`eres}(2004)}]{NozSolid}%
  \BibitemOpen
  \bibfield  {author} {\bibinfo {author} {\bibfnamefont {P.}~\bibnamefont
  {Nozi\`eres}},\ }\href@noop {} {\bibfield  {journal} {\bibinfo  {journal} {J.
  Low Temp. Phys.}\ }\textbf {\bibinfo {volume} {137}},\ \bibinfo {pages} {45}
  (\bibinfo {year} {2004})}\BibitemShut {NoStop}%
\bibitem [{\citenamefont {Volovik}(2003)}]{volovik2003}%
  \BibitemOpen
  \bibfield  {author} {\bibinfo {author} {\bibfnamefont {G.~E.}\ \bibnamefont
  {Volovik}},\ }\href@noop {} {\emph {\bibinfo {title} {The Universe in a
  Helium Droplet}}},\ Vol.\ \bibinfo {volume} {117}\ (\bibinfo  {publisher}
  {Oxford University Press on Demand},\ \bibinfo {year} {2003})\BibitemShut
  {NoStop}%
\bibitem [{\citenamefont {Wen}(2004)}]{wen2004}%
  \BibitemOpen
  \bibfield  {author} {\bibinfo {author} {\bibfnamefont {X.-G.}\ \bibnamefont
  {Wen}},\ }\href@noop {} {\emph {\bibinfo {title} {Quantum Field Theory of
  Many-Body Systems: from the Origin of Sound to an Origin of Light and
  Electrons}}}\ (\bibinfo  {publisher} {Oxford University Press on Demand},\
  \bibinfo {year} {2004})\BibitemShut {NoStop}%
\bibitem [{\citenamefont {Glyde}(1994)}]{GlydeBook}%
  \BibitemOpen
  \bibfield  {author} {\bibinfo {author} {\bibfnamefont {H.~R.}\ \bibnamefont
  {Glyde}},\ }\href@noop {} {\emph {\bibinfo {title} {Excitations in Liquid and
  Solid Helium}}},\ Oxford Series on Neutron Scattering in Condensed Matter\
  (\bibinfo  {publisher} {Clarendon},\ \bibinfo {address} {Oxford},\ \bibinfo
  {year} {1994})\BibitemShut {NoStop}%
\bibitem [{\citenamefont {Glyde}(2017)}]{Glyde2017}%
  \BibitemOpen
  \bibfield  {author} {\bibinfo {author} {\bibfnamefont {H.~R.}\ \bibnamefont
  {Glyde}},\ }\href {\doibase 10.1088/1361-6633/aa7f90} {\bibfield  {journal}
  {\bibinfo  {journal} {Rep. Prog. Phys.}\ }\textbf {\bibinfo {volume} {81}},\
  \bibinfo {pages} {014501} (\bibinfo {year} {2017})}\BibitemShut {NoStop}%
\bibitem [{\citenamefont {Aldrich}\ and\ \citenamefont
  {Pines}(1976)}]{Aldrich}%
  \BibitemOpen
  \bibfield  {author} {\bibinfo {author} {\bibfnamefont {C.~H.}\ \bibnamefont
  {Aldrich}}\ and\ \bibinfo {author} {\bibfnamefont {D.}~\bibnamefont
  {Pines}},\ }\href@noop {} {\bibfield  {journal} {\bibinfo  {journal} {J. Low
  Temp. Phys.}\ }\textbf {\bibinfo {volume} {25}},\ \bibinfo {pages} {677}
  (\bibinfo {year} {1976})}\BibitemShut {NoStop}%
\bibitem [{\citenamefont {Lee}\ and\ \citenamefont {Lee}(1975)}]{LeeLee}%
  \BibitemOpen
  \bibfield  {author} {\bibinfo {author} {\bibfnamefont {D.~K.}\ \bibnamefont
  {Lee}}\ and\ \bibinfo {author} {\bibfnamefont {F.~J.}\ \bibnamefont {Lee}},\
  }\href@noop {} {\bibfield  {journal} {\bibinfo  {journal} {Phys. Rev. B}\
  }\textbf {\bibinfo {volume} {11}},\ \bibinfo {pages} {4318} (\bibinfo {year}
  {1975})}\BibitemShut {NoStop}%
\bibitem [{\citenamefont {Feenberg}(1969)}]{FeenbergBook}%
  \BibitemOpen
  \bibfield  {author} {\bibinfo {author} {\bibfnamefont {E.}~\bibnamefont
  {Feenberg}},\ }\href@noop {} {\emph {\bibinfo {title} {Theory of {Q}uantum
  Fluids}}}\ (\bibinfo  {publisher} {Academic},\ \bibinfo {address} {New
  York},\ \bibinfo {year} {1969})\BibitemShut {NoStop}%
\bibitem [{\citenamefont {Krotscheck}(1986)}]{EKthree}%
  \BibitemOpen
  \bibfield  {author} {\bibinfo {author} {\bibfnamefont {E.}~\bibnamefont
  {Krotscheck}},\ }\href@noop {} {\bibfield  {journal} {\bibinfo  {journal}
  {Phys. Rev. B}\ }\textbf {\bibinfo {volume} {33}},\ \bibinfo {pages} {3158}
  (\bibinfo {year} {1986})}\BibitemShut {NoStop}%
\bibitem [{\citenamefont {Boronat}\ and\ \citenamefont
  {Casulleras}(1994)}]{1994-BoronatQMC}%
  \BibitemOpen
  \bibfield  {author} {\bibinfo {author} {\bibfnamefont {J.}~\bibnamefont
  {Boronat}}\ and\ \bibinfo {author} {\bibfnamefont {J.}~\bibnamefont
  {Casulleras}},\ }\href {\doibase 10.1103/PhysRevB.49.8920} {\bibfield
  {journal} {\bibinfo  {journal} {Phys. Rev. B}\ }\textbf {\bibinfo {volume}
  {49}},\ \bibinfo {pages} {8920} (\bibinfo {year} {1994})}\BibitemShut
  {NoStop}%
\bibitem [{\citenamefont {Ceperley}(1995)}]{CeperleyRMP}%
  \BibitemOpen
  \bibfield  {author} {\bibinfo {author} {\bibfnamefont {D.~M.}\ \bibnamefont
  {Ceperley}},\ }\href@noop {} {\bibfield  {journal} {\bibinfo  {journal} {Rev.
  Mod. Phys.}\ }\textbf {\bibinfo {volume} {67}},\ \bibinfo {pages} {279}
  (\bibinfo {year} {1995})}\BibitemShut {NoStop}%
\bibitem [{\citenamefont {Boronat}\ and\ \citenamefont
  {Casulleras}(1997)}]{BoronatRoton}%
  \BibitemOpen
  \bibfield  {author} {\bibinfo {author} {\bibfnamefont {J.}~\bibnamefont
  {Boronat}}\ and\ \bibinfo {author} {\bibfnamefont {J.}~\bibnamefont
  {Casulleras}},\ }\href@noop {} {\bibfield  {journal} {\bibinfo  {journal}
  {Europhys. Lett.}\ }\textbf {\bibinfo {volume} {38}},\ \bibinfo {pages} {291}
  (\bibinfo {year} {1997})}\BibitemShut {NoStop}%
\bibitem [{\citenamefont {Pistolesi}(1998{\natexlab{a}})}]{Pistolesi}%
  \BibitemOpen
  \bibfield  {author} {\bibinfo {author} {\bibfnamefont {F.}~\bibnamefont
  {Pistolesi}},\ }\href@noop {} {\bibfield  {journal} {\bibinfo  {journal}
  {Phys. Rev. Lett.}\ }\textbf {\bibinfo {volume} {81}},\ \bibinfo {pages}
  {397} (\bibinfo {year} {1998}{\natexlab{a}})}\BibitemShut {NoStop}%
\bibitem [{\citenamefont {Fabrocini}\ \emph {et~al.}(2002)\citenamefont
  {Fabrocini}, \citenamefont {Fantoni},\ and\ \citenamefont
  {Krotscheck}}]{KroTriesteBook}%
  \BibitemOpen
  \bibfield  {author} {\bibinfo {author} {\bibfnamefont {A.}~\bibnamefont
  {Fabrocini}}, \bibinfo {author} {\bibfnamefont {S.}~\bibnamefont {Fantoni}},
  \ and\ \bibinfo {author} {\bibfnamefont {E.}~\bibnamefont {Krotscheck}},\
  }\href@noop {} {\emph {\bibinfo {title} {Introduction to Modern Methods of
  {Q}uantum Many--Body Theory and their Applications}}},\ \bibinfo {series}
  {Advances in {Q}uantum Many--Body Theory}, Vol.~\bibinfo {volume} {7}\
  (\bibinfo  {publisher} {World Scientific},\ \bibinfo {address} {Singapore},\
  \bibinfo {year} {2002})\BibitemShut {NoStop}%
\bibitem [{\citenamefont {Krotscheck}\ and\ \citenamefont
  {Navarro}(2002)}]{KroNavarroBook}%
  \BibitemOpen
  \bibfield  {author} {\bibinfo {author} {\bibfnamefont {E.}~\bibnamefont
  {Krotscheck}}\ and\ \bibinfo {author} {\bibfnamefont {J.}~\bibnamefont
  {Navarro}},\ }\href@noop {} {\emph {\bibinfo {title} {Microscopic Approaches
  to {Q}uantum Liquids in Confined Geometries}}},\ \bibinfo {series} {Advances
  in {Q}uantum Many--Body Theory}, Vol.~\bibinfo {volume} {4}\ (\bibinfo
  {publisher} {World Scientific},\ \bibinfo {address} {Singapore},\ \bibinfo
  {year} {2002})\BibitemShut {NoStop}%
\bibitem [{\citenamefont {Vitali}\ \emph {et~al.}(2010)\citenamefont {Vitali},
  \citenamefont {Rossi}, \citenamefont {Reatto},\ and\ \citenamefont
  {Galli}}]{Vitali2010}%
  \BibitemOpen
  \bibfield  {author} {\bibinfo {author} {\bibfnamefont {E.}~\bibnamefont
  {Vitali}}, \bibinfo {author} {\bibfnamefont {M.}~\bibnamefont {Rossi}},
  \bibinfo {author} {\bibfnamefont {L.}~\bibnamefont {Reatto}}, \ and\ \bibinfo
  {author} {\bibfnamefont {D.~E.}\ \bibnamefont {Galli}},\ }\href@noop {}
  {\bibfield  {journal} {\bibinfo  {journal} {Phys. Rev. B}\ }\textbf {\bibinfo
  {volume} {82}},\ \bibinfo {pages} {174510/1} (\bibinfo {year}
  {2010})}\BibitemShut {NoStop}%
\bibitem [{\citenamefont {Ferr\'e}\ and\ \citenamefont
  {Boronat}(2016)}]{2016FerreBoronat}%
  \BibitemOpen
  \bibfield  {author} {\bibinfo {author} {\bibfnamefont {G.}~\bibnamefont
  {Ferr\'e}}\ and\ \bibinfo {author} {\bibfnamefont {J.}~\bibnamefont
  {Boronat}},\ }\href {\doibase 10.1103/PhysRevB.93.104510} {\bibfield
  {journal} {\bibinfo  {journal} {Phys. Rev. B}\ }\textbf {\bibinfo {volume}
  {93}},\ \bibinfo {pages} {104510} (\bibinfo {year} {2016})}\BibitemShut
  {NoStop}%
\bibitem [{\citenamefont {Griffin}(1993)}]{GriffinBook}%
  \BibitemOpen
  \bibfield  {author} {\bibinfo {author} {\bibfnamefont {A.}~\bibnamefont
  {Griffin}},\ }\href@noop {} {\emph {\bibinfo {title} {Excitations in a
  Bose-condensed Liquid}}},\ Vol.~\bibinfo {volume} {4}\ (\bibinfo  {publisher}
  {Cambridge University Press},\ \bibinfo {year} {1993})\BibitemShut {NoStop}%
\bibitem [{\citenamefont {Gu\'enault}\ \emph {et~al.}(2019)\citenamefont
  {Gu\'enault}, \citenamefont {Guthrie}, \citenamefont {Haley}, \citenamefont
  {Kafanov}, \citenamefont {Pashkin}, \citenamefont {Pickett}, \citenamefont
  {Poole}, \citenamefont {Schanen}, \citenamefont {Tsepelin}, \citenamefont
  {Zmeev}, \citenamefont {Collin}, \citenamefont {Maillet},\ and\ \citenamefont
  {Gazizulin}}]{Guenault2019}%
  \BibitemOpen
  \bibfield  {author} {\bibinfo {author} {\bibfnamefont {A.~M.}\ \bibnamefont
  {Gu\'enault}}, \bibinfo {author} {\bibfnamefont {A.}~\bibnamefont {Guthrie}},
  \bibinfo {author} {\bibfnamefont {R.~P.}\ \bibnamefont {Haley}}, \bibinfo
  {author} {\bibfnamefont {S.}~\bibnamefont {Kafanov}}, \bibinfo {author}
  {\bibfnamefont {Y.~A.}\ \bibnamefont {Pashkin}}, \bibinfo {author}
  {\bibfnamefont {G.~R.}\ \bibnamefont {Pickett}}, \bibinfo {author}
  {\bibfnamefont {M.}~\bibnamefont {Poole}}, \bibinfo {author} {\bibfnamefont
  {R.}~\bibnamefont {Schanen}}, \bibinfo {author} {\bibfnamefont
  {V.}~\bibnamefont {Tsepelin}}, \bibinfo {author} {\bibfnamefont {D.~E.}\
  \bibnamefont {Zmeev}}, \bibinfo {author} {\bibfnamefont {E.}~\bibnamefont
  {Collin}}, \bibinfo {author} {\bibfnamefont {O.}~\bibnamefont {Maillet}}, \
  and\ \bibinfo {author} {\bibfnamefont {R.}~\bibnamefont {Gazizulin}},\ }\href
  {\doibase 10.1103/PhysRevB.100.020506} {\bibfield  {journal} {\bibinfo
  {journal} {Phys. Rev. B}\ }\textbf {\bibinfo {volume} {100}},\ \bibinfo
  {pages} {020506(R)} (\bibinfo {year} {2019})}\BibitemShut {NoStop}%
\bibitem [{\citenamefont {Gu\'enault}\ \emph {et~al.}(2020)\citenamefont
  {Gu\'enault}, \citenamefont {Guthrie}, \citenamefont {Haley}, \citenamefont
  {Kafanov}, \citenamefont {Pashkin}, \citenamefont {Pickett}, \citenamefont
  {Tsepelin}, \citenamefont {Zmeev}, \citenamefont {Collin}, \citenamefont
  {Gazizulin},\ and\ \citenamefont {Maillet}}]{2020-PRB-nano-osc}%
  \BibitemOpen
  \bibfield  {author} {\bibinfo {author} {\bibfnamefont {A.~M.}\ \bibnamefont
  {Gu\'enault}}, \bibinfo {author} {\bibfnamefont {A.}~\bibnamefont {Guthrie}},
  \bibinfo {author} {\bibfnamefont {R.~P.}\ \bibnamefont {Haley}}, \bibinfo
  {author} {\bibfnamefont {S.}~\bibnamefont {Kafanov}}, \bibinfo {author}
  {\bibfnamefont {Y.~A.}\ \bibnamefont {Pashkin}}, \bibinfo {author}
  {\bibfnamefont {G.~R.}\ \bibnamefont {Pickett}}, \bibinfo {author}
  {\bibfnamefont {V.}~\bibnamefont {Tsepelin}}, \bibinfo {author}
  {\bibfnamefont {D.~E.}\ \bibnamefont {Zmeev}}, \bibinfo {author}
  {\bibfnamefont {E.}~\bibnamefont {Collin}}, \bibinfo {author} {\bibfnamefont
  {R.}~\bibnamefont {Gazizulin}}, \ and\ \bibinfo {author} {\bibfnamefont
  {O.}~\bibnamefont {Maillet}},\ }\href {\doibase 10.1103/PhysRevB.101.060503}
  {\bibfield  {journal} {\bibinfo  {journal} {Phys. Rev. B}\ }\textbf {\bibinfo
  {volume} {101}},\ \bibinfo {pages} {060503(R)} (\bibinfo {year}
  {2020})}\BibitemShut {NoStop}%
\bibitem [{\citenamefont {Guo}\ and\ \citenamefont {McKinsey}(2013)}]{Guo2013}%
  \BibitemOpen
  \bibfield  {author} {\bibinfo {author} {\bibfnamefont {W.}~\bibnamefont
  {Guo}}\ and\ \bibinfo {author} {\bibfnamefont {D.~N.}\ \bibnamefont
  {McKinsey}},\ }\href {\doibase 10.1103/PhysRevD.87.115001} {\bibfield
  {journal} {\bibinfo  {journal} {Phys. Rev. D}\ }\textbf {\bibinfo {volume}
  {87}},\ \bibinfo {pages} {115001} (\bibinfo {year} {2013})}\BibitemShut
  {NoStop}%
\bibitem [{\citenamefont {Schutz}\ and\ \citenamefont
  {Zurek}(2016)}]{PRL2016Zurek}%
  \BibitemOpen
  \bibfield  {author} {\bibinfo {author} {\bibfnamefont {K.}~\bibnamefont
  {Schutz}}\ and\ \bibinfo {author} {\bibfnamefont {K.~M.}\ \bibnamefont
  {Zurek}},\ }\href {\doibase 10.1103/PhysRevLett.117.121302} {\bibfield
  {journal} {\bibinfo  {journal} {Phys. Rev. Lett.}\ }\textbf {\bibinfo
  {volume} {117}},\ \bibinfo {pages} {121302} (\bibinfo {year}
  {2016})}\BibitemShut {NoStop}%
\bibitem [{\citenamefont {Maris}\ \emph {et~al.}(2017)\citenamefont {Maris},
  \citenamefont {Seidel},\ and\ \citenamefont {Stein}}]{PRL2017MarisDetector}%
  \BibitemOpen
  \bibfield  {author} {\bibinfo {author} {\bibfnamefont {H.~J.}\ \bibnamefont
  {Maris}}, \bibinfo {author} {\bibfnamefont {G.~M.}\ \bibnamefont {Seidel}}, \
  and\ \bibinfo {author} {\bibfnamefont {D.}~\bibnamefont {Stein}},\ }\href
  {\doibase 10.1103/PhysRevLett.119.181303} {\bibfield  {journal} {\bibinfo
  {journal} {Phys. Rev. Lett.}\ }\textbf {\bibinfo {volume} {119}},\ \bibinfo
  {pages} {181303} (\bibinfo {year} {2017})}\BibitemShut {NoStop}%
\bibitem [{\citenamefont {Knapen}\ \emph {et~al.}(2017)\citenamefont {Knapen},
  \citenamefont {Lin},\ and\ \citenamefont {Zurek}}]{Zurek2017}%
  \BibitemOpen
  \bibfield  {author} {\bibinfo {author} {\bibfnamefont {S.}~\bibnamefont
  {Knapen}}, \bibinfo {author} {\bibfnamefont {T.}~\bibnamefont {Lin}}, \ and\
  \bibinfo {author} {\bibfnamefont {K.~M.}\ \bibnamefont {Zurek}},\ }\href
  {\doibase 10.1103/PhysRevD.95.056019} {\bibfield  {journal} {\bibinfo
  {journal} {Phys. Rev. D}\ }\textbf {\bibinfo {volume} {95}},\ \bibinfo
  {pages} {056019} (\bibinfo {year} {2017})}\BibitemShut {NoStop}%
\bibitem [{\citenamefont {Beauvois}\ \emph {et~al.}(2016)\citenamefont
  {Beauvois}, \citenamefont {Campbell}, \citenamefont {Dawidowski},
  \citenamefont {F{\aa}k}, \citenamefont {Godfrin}, \citenamefont {Krotscheck},
  \citenamefont {Lauter}, \citenamefont {Lichtenegger}, \citenamefont
  {Ollivier},\ and\ \citenamefont {Sultan}}]{Beauvois2016}%
  \BibitemOpen
  \bibfield  {author} {\bibinfo {author} {\bibfnamefont {K.}~\bibnamefont
  {Beauvois}}, \bibinfo {author} {\bibfnamefont {C.~E.}\ \bibnamefont
  {Campbell}}, \bibinfo {author} {\bibfnamefont {J.}~\bibnamefont
  {Dawidowski}}, \bibinfo {author} {\bibfnamefont {B.}~\bibnamefont {F{\aa}k}},
  \bibinfo {author} {\bibfnamefont {H.}~\bibnamefont {Godfrin}}, \bibinfo
  {author} {\bibfnamefont {E.}~\bibnamefont {Krotscheck}}, \bibinfo {author}
  {\bibfnamefont {H.-J.}\ \bibnamefont {Lauter}}, \bibinfo {author}
  {\bibfnamefont {T.}~\bibnamefont {Lichtenegger}}, \bibinfo {author}
  {\bibfnamefont {J.}~\bibnamefont {Ollivier}}, \ and\ \bibinfo {author}
  {\bibfnamefont {A.}~\bibnamefont {Sultan}},\ }\href@noop {} {\bibfield
  {journal} {\bibinfo  {journal} {Phys. Rev. B}\ }\textbf {\bibinfo {volume}
  {94}},\ \bibinfo {pages} {024504} (\bibinfo {year} {2016})}\BibitemShut
  {NoStop}%
\bibitem [{\citenamefont {Beauvois}\ \emph {et~al.}(2018)\citenamefont
  {Beauvois}, \citenamefont {Dawidowski}, \citenamefont {{F\aa k}},
  \citenamefont {Godfrin}, \citenamefont {Krotscheck}, \citenamefont
  {Ollivier},\ and\ \citenamefont {Sultan}}]{Beauvois2018}%
  \BibitemOpen
  \bibfield  {author} {\bibinfo {author} {\bibfnamefont {K.}~\bibnamefont
  {Beauvois}}, \bibinfo {author} {\bibfnamefont {J.}~\bibnamefont
  {Dawidowski}}, \bibinfo {author} {\bibfnamefont {B.}~\bibnamefont {{F\aa
  k}}}, \bibinfo {author} {\bibfnamefont {H.}~\bibnamefont {Godfrin}}, \bibinfo
  {author} {\bibfnamefont {E.}~\bibnamefont {Krotscheck}}, \bibinfo {author}
  {\bibfnamefont {J.}~\bibnamefont {Ollivier}}, \ and\ \bibinfo {author}
  {\bibfnamefont {A.}~\bibnamefont {Sultan}},\ }\href@noop {} {\bibfield
  {journal} {\bibinfo  {journal} {Phys. Rev. B}\ }\textbf {\bibinfo {volume}
  {97}},\ \bibinfo {pages} {184520} (\bibinfo {year} {2018})}\BibitemShut
  {NoStop}%
\bibitem [{\citenamefont {Campbell}\ \emph {et~al.}(2015)\citenamefont
  {Campbell}, \citenamefont {Krotscheck},\ and\ \citenamefont
  {Lichtenegger}}]{eomIII}%
  \BibitemOpen
  \bibfield  {author} {\bibinfo {author} {\bibfnamefont {C.~E.}\ \bibnamefont
  {Campbell}}, \bibinfo {author} {\bibfnamefont {E.}~\bibnamefont
  {Krotscheck}}, \ and\ \bibinfo {author} {\bibfnamefont {T.}~\bibnamefont
  {Lichtenegger}},\ }\href {\doibase 10.1103/PhysRevB.91.184510} {\bibfield
  {journal} {\bibinfo  {journal} {Phys. Rev. B}\ }\textbf {\bibinfo {volume}
  {91}},\ \bibinfo {pages} {184510} (\bibinfo {year} {2015})}\BibitemShut
  {NoStop}%
\bibitem [{\citenamefont {Beauvois}\ \emph {et~al.}(2019)\citenamefont
  {Beauvois}, \citenamefont {Godfrin}, \citenamefont {Krotscheck},\ and\
  \citenamefont {Zillich}}]{phonons}%
  \BibitemOpen
  \bibfield  {author} {\bibinfo {author} {\bibfnamefont {K.}~\bibnamefont
  {Beauvois}}, \bibinfo {author} {\bibfnamefont {H.}~\bibnamefont {Godfrin}},
  \bibinfo {author} {\bibfnamefont {E.}~\bibnamefont {Krotscheck}}, \ and\
  \bibinfo {author} {\bibfnamefont {R.}~\bibnamefont {Zillich}},\ }\href
  {\doibase 10.1007/s10909-019-02219-1} {\bibfield  {journal} {\bibinfo
  {journal} {J. Low Temp. Phys.}\ }\textbf {\bibinfo {volume} {197}},\ \bibinfo
  {pages} {113} (\bibinfo {year} {2019})}\BibitemShut {NoStop}%
\bibitem [{\citenamefont {Junker}\ and\ \citenamefont
  {Elbaum}(1977)}]{Junker1977}%
  \BibitemOpen
  \bibfield  {author} {\bibinfo {author} {\bibfnamefont {W.~R.}\ \bibnamefont
  {Junker}}\ and\ \bibinfo {author} {\bibfnamefont {C.}~\bibnamefont
  {Elbaum}},\ }\href {\doibase 10.1103/PhysRevB.15.162} {\bibfield  {journal}
  {\bibinfo  {journal} {Phys. Rev. B}\ }\textbf {\bibinfo {volume} {15}},\
  \bibinfo {pages} {162} (\bibinfo {year} {1977})}\BibitemShut {NoStop}%
\bibitem [{\citenamefont {Rugar}\ and\ \citenamefont
  {Foster}(1984)}]{Rugar1984}%
  \BibitemOpen
  \bibfield  {author} {\bibinfo {author} {\bibfnamefont {D.}~\bibnamefont
  {Rugar}}\ and\ \bibinfo {author} {\bibfnamefont {J.~S.}\ \bibnamefont
  {Foster}},\ }\href@noop {} {\bibfield  {journal} {\bibinfo  {journal} {Phys.
  Rev. B}\ }\textbf {\bibinfo {volume} {30}},\ \bibinfo {pages} {2595}
  (\bibinfo {year} {1984})}\BibitemShut {NoStop}%
\bibitem [{\citenamefont {Maris}\ and\ \citenamefont
  {Massey}(1970)}]{Maris1970}%
  \BibitemOpen
  \bibfield  {author} {\bibinfo {author} {\bibfnamefont {H.~J.}\ \bibnamefont
  {Maris}}\ and\ \bibinfo {author} {\bibfnamefont {W.~E.}\ \bibnamefont
  {Massey}},\ }\href {\doibase 10.1103/PhysRevLett.25.220} {\bibfield
  {journal} {\bibinfo  {journal} {Phys. Rev. Lett.}\ }\textbf {\bibinfo
  {volume} {25}},\ \bibinfo {pages} {220} (\bibinfo {year} {1970})}\BibitemShut
  {NoStop}%
\bibitem [{\citenamefont {Maris}(1977)}]{MarisRMP}%
  \BibitemOpen
  \bibfield  {author} {\bibinfo {author} {\bibfnamefont {H.~J.}\ \bibnamefont
  {Maris}},\ }\href@noop {} {\bibfield  {journal} {\bibinfo  {journal} {Rev.
  Mod. Phys.}\ }\textbf {\bibinfo {volume} {49}},\ \bibinfo {pages} {341}
  (\bibinfo {year} {1977})}\BibitemShut {NoStop}%
\bibitem [{\citenamefont {Sridhar}(1987)}]{Sridhar87}%
  \BibitemOpen
  \bibfield  {author} {\bibinfo {author} {\bibfnamefont {R.}~\bibnamefont
  {Sridhar}},\ }\href@noop {} {\bibfield  {journal} {\bibinfo  {journal}
  {Physics Reports}\ }\textbf {\bibinfo {volume} {146}},\ \bibinfo {pages}
  {259} (\bibinfo {year} {1987})}\BibitemShut {NoStop}%
\bibitem [{\citenamefont {Maris}(1973)}]{Maris1980}%
  \BibitemOpen
  \bibfield  {author} {\bibinfo {author} {\bibfnamefont {H.~J.}\ \bibnamefont
  {Maris}},\ }\href {\doibase 10.1103/PhysRevA.8.1980} {\bibfield  {journal}
  {\bibinfo  {journal} {Phys. Rev. A}\ }\textbf {\bibinfo {volume} {8}},\
  \bibinfo {pages} {1980} (\bibinfo {year} {1973})}\BibitemShut {NoStop}%
\bibitem [{\citenamefont {Kemoklidze}\ and\ \citenamefont
  {Pitaevskii}(1970)}]{Pitaevskii1970}%
  \BibitemOpen
  \bibfield  {author} {\bibinfo {author} {\bibfnamefont {M.~P.}\ \bibnamefont
  {Kemoklidze}}\ and\ \bibinfo {author} {\bibfnamefont {L.~P.}\ \bibnamefont
  {Pitaevskii}},\ }\href@noop {} {\bibfield  {journal} {\bibinfo  {journal}
  {Zh. Exsp. Teor. Fiz.}\ }\textbf {\bibinfo {volume} {59}},\ \bibinfo {pages}
  {2187} (\bibinfo {year} {1970})},\ \bibinfo {note} {[Sov. Phys. JETP 32, 1183
  (1971)]}\BibitemShut {NoStop}%
\bibitem [{\citenamefont {Feenberg}(1971)}]{Feenberg1971}%
  \BibitemOpen
  \bibfield  {author} {\bibinfo {author} {\bibfnamefont {E.}~\bibnamefont
  {Feenberg}},\ }\href@noop {} {\bibfield  {journal} {\bibinfo  {journal}
  {Phys. Rev. Lett.}\ }\textbf {\bibinfo {volume} {26}},\ \bibinfo {pages}
  {301} (\bibinfo {year} {1971})}\BibitemShut {NoStop}%
\bibitem [{\citenamefont {Aldrich}\ \emph {et~al.}(1976)\citenamefont
  {Aldrich}, \citenamefont {Pethick},\ and\ \citenamefont
  {Pines}}]{aldrich1976phonon}%
  \BibitemOpen
  \bibfield  {author} {\bibinfo {author} {\bibfnamefont {C.}~\bibnamefont
  {Aldrich}}, \bibinfo {author} {\bibfnamefont {C.}~\bibnamefont {Pethick}}, \
  and\ \bibinfo {author} {\bibfnamefont {D.}~\bibnamefont {Pines}},\
  }\href@noop {} {\bibfield  {journal} {\bibinfo  {journal} {J. Low Temp.
  Phys}\ }\textbf {\bibinfo {volume} {25}},\ \bibinfo {pages} {691} (\bibinfo
  {year} {1976})}\BibitemShut {NoStop}%
\bibitem [{\citenamefont {Abraham}\ \emph {et~al.}(1970)\citenamefont
  {Abraham}, \citenamefont {Eckstein}, \citenamefont {Ketterson}, \citenamefont
  {Kuchnir},\ and\ \citenamefont {Roach}}]{Abraham}%
  \BibitemOpen
  \bibfield  {author} {\bibinfo {author} {\bibfnamefont {B.~M.}\ \bibnamefont
  {Abraham}}, \bibinfo {author} {\bibfnamefont {Y.}~\bibnamefont {Eckstein}},
  \bibinfo {author} {\bibfnamefont {J.~B.}\ \bibnamefont {Ketterson}}, \bibinfo
  {author} {\bibfnamefont {M.}~\bibnamefont {Kuchnir}}, \ and\ \bibinfo
  {author} {\bibfnamefont {P.~R.}\ \bibnamefont {Roach}},\ }\href {\doibase
  10.1103/PhysRevA.1.250} {\bibfield  {journal} {\bibinfo  {journal} {Phys.
  Rev. A}\ }\textbf {\bibinfo {volume} {1}},\ \bibinfo {pages} {250} (\bibinfo
  {year} {1970})}\BibitemShut {NoStop}%
\bibitem [{\citenamefont {Donnelly}\ and\ \citenamefont
  {Barenghi}(1998)}]{DonnellyBarenghi}%
  \BibitemOpen
  \bibfield  {author} {\bibinfo {author} {\bibfnamefont {R.~J.}\ \bibnamefont
  {Donnelly}}\ and\ \bibinfo {author} {\bibfnamefont {C.~F.}\ \bibnamefont
  {Barenghi}},\ }\href {\doibase 10.1063/1.556028} {\bibfield  {journal}
  {\bibinfo  {journal} {J. Phys. Chem. Ref. Data}\ }\textbf {\bibinfo {volume}
  {27}},\ \bibinfo {pages} {1217} (\bibinfo {year} {1998})},\ \Eprint
  {http://arxiv.org/abs/https://doi.org/10.1063/1.556028}
  {https://doi.org/10.1063/1.556028} \BibitemShut {NoStop}%
\bibitem [{\citenamefont {Tanaka}\ \emph {et~al.}(2000)\citenamefont {Tanaka},
  \citenamefont {Hatakeyama}, \citenamefont {Noma},\ and\ \citenamefont
  {Satoh}}]{Tanaka2000}%
  \BibitemOpen
  \bibfield  {author} {\bibinfo {author} {\bibfnamefont {E.}~\bibnamefont
  {Tanaka}}, \bibinfo {author} {\bibfnamefont {K.}~\bibnamefont {Hatakeyama}},
  \bibinfo {author} {\bibfnamefont {S.}~\bibnamefont {Noma}}, \ and\ \bibinfo
  {author} {\bibfnamefont {T.}~\bibnamefont {Satoh}},\ }\href@noop {}
  {\bibfield  {journal} {\bibinfo  {journal} {Cryogenics}\ }\textbf {\bibinfo
  {volume} {40}},\ \bibinfo {pages} {365} (\bibinfo {year} {2000})}\BibitemShut
  {NoStop}%
\bibitem [{\citenamefont {Greywall}(1978)}]{Greywall78}%
  \BibitemOpen
  \bibfield  {author} {\bibinfo {author} {\bibfnamefont {D.~S.}\ \bibnamefont
  {Greywall}},\ }\href {\doibase 10.1103/PhysRevB.18.2127} {\bibfield
  {journal} {\bibinfo  {journal} {Phys. Rev. B}\ }\textbf {\bibinfo {volume}
  {18}},\ \bibinfo {pages} {2127} (\bibinfo {year} {1978})}\BibitemShut
  {NoStop}%
\bibitem [{\citenamefont {Phillips}\ \emph {et~al.}(1970)\citenamefont
  {Phillips}, \citenamefont {Waterfield},\ and\ \citenamefont
  {Hoffer}}]{Phillips1970}%
  \BibitemOpen
  \bibfield  {author} {\bibinfo {author} {\bibfnamefont {N.~E.}\ \bibnamefont
  {Phillips}}, \bibinfo {author} {\bibfnamefont {C.~G.}\ \bibnamefont
  {Waterfield}}, \ and\ \bibinfo {author} {\bibfnamefont {J.~K.}\ \bibnamefont
  {Hoffer}},\ }\href {\doibase 10.1103/PhysRevLett.25.1260} {\bibfield
  {journal} {\bibinfo  {journal} {Phys. Rev. Lett.}\ }\textbf {\bibinfo
  {volume} {25}},\ \bibinfo {pages} {1260} (\bibinfo {year}
  {1970})}\BibitemShut {NoStop}%
\bibitem [{\citenamefont {Greywall}(1980)}]{Greywall79ERRATUM}%
  \BibitemOpen
  \bibfield  {author} {\bibinfo {author} {\bibfnamefont {D.~S.}\ \bibnamefont
  {Greywall}},\ }\href {\doibase 10.1103/PhysRevB.21.1329} {\bibfield
  {journal} {\bibinfo  {journal} {Phys. Rev. B}\ }\textbf {\bibinfo {volume}
  {21}},\ \bibinfo {pages} {1329(E)} (\bibinfo {year} {1980})}\BibitemShut
  {NoStop}%
\bibitem [{\citenamefont {Brooks}\ and\ \citenamefont
  {Donnelly}(1978)}]{BrooksDonnelly}%
  \BibitemOpen
  \bibfield  {author} {\bibinfo {author} {\bibfnamefont {J.~S.}\ \bibnamefont
  {Brooks}}\ and\ \bibinfo {author} {\bibfnamefont {R.~J.}\ \bibnamefont
  {Donnelly}},\ }\href@noop {} {\bibfield  {journal} {\bibinfo  {journal} {J.
  Phys. Chem.}\ }\textbf {\bibinfo {volume} {6}},\ \bibinfo {pages} {51}
  (\bibinfo {year} {1978})}\BibitemShut {NoStop}%
\bibitem [{\citenamefont {Cowley}\ and\ \citenamefont
  {Woods}(1971)}]{CowleyWoods}%
  \BibitemOpen
  \bibfield  {author} {\bibinfo {author} {\bibfnamefont {R.~A.}\ \bibnamefont
  {Cowley}}\ and\ \bibinfo {author} {\bibfnamefont {A.~D.~B.}\ \bibnamefont
  {Woods}},\ }\href {\doibase 10.1139/p71-021} {\bibfield  {journal} {\bibinfo
  {journal} {Can. J. Phys.}\ }\textbf {\bibinfo {volume} {49}},\ \bibinfo
  {pages} {177} (\bibinfo {year} {1971})},\ \Eprint
  {http://arxiv.org/abs/http://dx.doi.org/10.1139/p71-021}
  {http://dx.doi.org/10.1139/p71-021} \BibitemShut {NoStop}%
\bibitem [{\citenamefont {Woods}\ \emph {et~al.}(1977)\citenamefont {Woods},
  \citenamefont {Hilton}, \citenamefont {Scherm},\ and\ \citenamefont
  {Stirling}}]{WoodsHilton}%
  \BibitemOpen
  \bibfield  {author} {\bibinfo {author} {\bibfnamefont {A.~D.~B.}\
  \bibnamefont {Woods}}, \bibinfo {author} {\bibfnamefont {P.~A.}\ \bibnamefont
  {Hilton}}, \bibinfo {author} {\bibfnamefont {R.}~\bibnamefont {Scherm}}, \
  and\ \bibinfo {author} {\bibfnamefont {W.}~\bibnamefont {Stirling}},\
  }\href@noop {} {\bibfield  {journal} {\bibinfo  {journal} {{J. Phys. C}}\
  }\textbf {\bibinfo {volume} {{10}}},\ \bibinfo {pages} {L45} (\bibinfo {year}
  {{1977}})}\BibitemShut {NoStop}%
\bibitem [{\citenamefont {Svensson}\ \emph {et~al.}(1975)\citenamefont
  {Svensson}, \citenamefont {Martel},\ and\ \citenamefont
  {Woods}}]{Svensson1975}%
  \BibitemOpen
  \bibfield  {author} {\bibinfo {author} {\bibfnamefont {E.}~\bibnamefont
  {Svensson}}, \bibinfo {author} {\bibfnamefont {P.}~\bibnamefont {Martel}}, \
  and\ \bibinfo {author} {\bibfnamefont {A.}~\bibnamefont {Woods}},\ }\href
  {\doibase https://doi.org/10.1016/0375-9601(75)90688-X} {\bibfield  {journal}
  {\bibinfo  {journal} {Phys. Lett.}\ }\textbf {\bibinfo {volume} {55}},\
  \bibinfo {pages} {151 } (\bibinfo {year} {1975})}\BibitemShut {NoStop}%
\bibitem [{\citenamefont {Svensson}\ \emph {et~al.}(1976)\citenamefont
  {Svensson}, \citenamefont {Martel}, \citenamefont {Sears},\ and\
  \citenamefont {Woods}}]{SvenssonMartel}%
  \BibitemOpen
  \bibfield  {author} {\bibinfo {author} {\bibfnamefont {E.}~\bibnamefont
  {Svensson}}, \bibinfo {author} {\bibfnamefont {P.}~\bibnamefont {Martel}},
  \bibinfo {author} {\bibfnamefont {V.}~\bibnamefont {Sears}}, \ and\ \bibinfo
  {author} {\bibfnamefont {A.}~\bibnamefont {Woods}},\ }\href@noop {}
  {\bibfield  {journal} {\bibinfo  {journal} {{Can. J. Phys.}}\ }\textbf
  {\bibinfo {volume} {{54}}},\ \bibinfo {pages} {2178} (\bibinfo {year}
  {{1976}})}\BibitemShut {NoStop}%
\bibitem [{\citenamefont {Svensson}\ \emph {et~al.}(1978)\citenamefont
  {Svensson}, \citenamefont {Scherm},\ and\ \citenamefont
  {Woods}}]{SvenssonScherm}%
  \BibitemOpen
  \bibfield  {author} {\bibinfo {author} {\bibfnamefont {E.~C.}\ \bibnamefont
  {Svensson}}, \bibinfo {author} {\bibfnamefont {R.}~\bibnamefont {Scherm}}, \
  and\ \bibinfo {author} {\bibfnamefont {A.~D.}\ \bibnamefont {Woods}},\ }\href
  {\doibase 10.1051/jphyscol:1978693} {\bibfield  {journal} {\bibinfo
  {journal} {J. Phys. (Paris) Colloques}\ }\textbf {\bibinfo {volume}
  {39-C6}},\ \bibinfo {pages} {211} (\bibinfo {year} {1978})}\BibitemShut
  {NoStop}%
\bibitem [{\citenamefont {Stirling}(1983)}]{stirling-83}%
  \BibitemOpen
  \bibfield  {author} {\bibinfo {author} {\bibfnamefont {W.~G.}\ \bibnamefont
  {Stirling}},\ }in\ \href@noop {} {\emph {\bibinfo {booktitle} {{$75^{\rm
  th}$} Jubilee Conference on Helium-4}}},\ \bibinfo {editor} {edited by\
  \bibinfo {editor} {\bibfnamefont {J.~G.~M.}\ \bibnamefont {Armitage}}}\
  (\bibinfo  {publisher} {World Scientific},\ \bibinfo {address} {Singapore},\
  \bibinfo {year} {1983})\BibitemShut {NoStop}%
\bibitem [{\citenamefont {Stirling}(1985)}]{StirlingPhonon}%
  \BibitemOpen
  \bibfield  {author} {\bibinfo {author} {\bibfnamefont {W.~G.}\ \bibnamefont
  {Stirling}},\ }in\ \href@noop {} {\emph {\bibinfo {booktitle} {Proceedings of
  the Second International Conference on Phonon Physics}}},\ \bibinfo {editor}
  {edited by\ \bibinfo {editor} {\bibfnamefont {J.}~\bibnamefont {Koll\`ar}},
  \bibinfo {editor} {\bibfnamefont {N.}~\bibnamefont {Kroo}}, \bibinfo {editor}
  {\bibfnamefont {M.}~\bibnamefont {Meynhard}}, \ and\ \bibinfo {editor}
  {\bibfnamefont {T.}~\bibnamefont {Siklos}}}\ (\bibinfo  {publisher} {World
  Scientific},\ \bibinfo {address} {Singapore},\ \bibinfo {year} {1985})\ p.\
  \bibinfo {pages} {829}\BibitemShut {NoStop}%
\bibitem [{\citenamefont {Stirling}(1991)}]{StirlingExeter}%
  \BibitemOpen
  \bibfield  {author} {\bibinfo {author} {\bibfnamefont {W.~G.}\ \bibnamefont
  {Stirling}},\ }in\ \href@noop {} {\emph {\bibinfo {booktitle} {Excitations in
  Two-Dimensional and Three-Dimensional {Q}uantum Fluids}}},\ \bibinfo {series}
  {NATO Advanced Study Institute, Series B: Physics}, Vol.\ \bibinfo {volume}
  {257},\ \bibinfo {editor} {edited by\ \bibinfo {editor} {\bibfnamefont
  {A.~F.~G.}\ \bibnamefont {Wyatt}}\ and\ \bibinfo {editor} {\bibfnamefont
  {H.~J.}\ \bibnamefont {Lauter}}}\ (\bibinfo  {publisher} {Plenum},\ \bibinfo
  {address} {New York},\ \bibinfo {year} {1991})\ pp.\ \bibinfo {pages}
  {25--46}\BibitemShut {NoStop}%
\bibitem [{\citenamefont {Talbot}\ \emph {et~al.}(1988)\citenamefont {Talbot},
  \citenamefont {Glyde}, \citenamefont {Stirling},\ and\ \citenamefont
  {Svensson}}]{Talbot-Glyde}%
  \BibitemOpen
  \bibfield  {author} {\bibinfo {author} {\bibfnamefont {E.~F.}\ \bibnamefont
  {Talbot}}, \bibinfo {author} {\bibfnamefont {H.~R.}\ \bibnamefont {Glyde}},
  \bibinfo {author} {\bibfnamefont {W.~G.}\ \bibnamefont {Stirling}}, \ and\
  \bibinfo {author} {\bibfnamefont {E.~C.}\ \bibnamefont {Svensson}},\
  }\href@noop {} {\bibfield  {journal} {\bibinfo  {journal} {Phys. Rev. B}\
  }\textbf {\bibinfo {volume} {{38}}},\ \bibinfo {pages} {11229} (\bibinfo
  {year} {{1988}})}\BibitemShut {NoStop}%
\bibitem [{\citenamefont {Dietrich}\ \emph {et~al.}(1972)\citenamefont
  {Dietrich}, \citenamefont {Graf}, \citenamefont {Huang},\ and\ \citenamefont
  {Passell}}]{Dietrich72}%
  \BibitemOpen
  \bibfield  {author} {\bibinfo {author} {\bibfnamefont {O.~W.}\ \bibnamefont
  {Dietrich}}, \bibinfo {author} {\bibfnamefont {E.~H.}\ \bibnamefont {Graf}},
  \bibinfo {author} {\bibfnamefont {C.~H.}\ \bibnamefont {Huang}}, \ and\
  \bibinfo {author} {\bibfnamefont {L.}~\bibnamefont {Passell}},\ }\href@noop
  {} {\bibfield  {journal} {\bibinfo  {journal} {Phys. Rev. A}\ }\textbf
  {\bibinfo {volume} {5}},\ \bibinfo {pages} {1377} (\bibinfo {year}
  {1972})}\BibitemShut {NoStop}%
\bibitem [{\citenamefont {{Stirling, W. G.}}\ \emph {et~al.}(1978)\citenamefont
  {{Stirling, W. G.}}, \citenamefont {{Copley, J. R. D.}},\ and\ \citenamefont
  {{Hilton, P. A.}}}]{StirlingCopley}%
  \BibitemOpen
  \bibfield  {author} {\bibinfo {author} {\bibnamefont {{Stirling, W. G.}}},
  \bibinfo {author} {\bibnamefont {{Copley, J. R. D.}}}, \ and\ \bibinfo
  {author} {\bibnamefont {{Hilton, P. A.}}},\ }in\ \href@noop {} {\emph
  {\bibinfo {booktitle} {IAEA Symposium, Neutron Inelastic Scattering,
  Vienna}}},\ Vol.\ \bibinfo {volume} {III}\ (\bibinfo {year} {1978})\ pp.\
  \bibinfo {pages} {45--52}\BibitemShut {NoStop}%
\bibitem [{\citenamefont {Andersen}(1991)}]{AndersenThesis}%
  \BibitemOpen
  \bibfield  {author} {\bibinfo {author} {\bibfnamefont {K.~H.}\ \bibnamefont
  {Andersen}},\ }\emph {\bibinfo {title} {A neutron scattering study of liquid
  $^4${H}e}},\ \href@noop {} {Ph.D. thesis},\ \bibinfo  {school} {Keele
  University} (\bibinfo {year} {1991})\BibitemShut {NoStop}%
\bibitem [{\citenamefont {F{\aa}k}\ and\ \citenamefont
  {Andersen}(1991)}]{Fak-Andersen-91}%
  \BibitemOpen
  \bibfield  {author} {\bibinfo {author} {\bibfnamefont {B.}~\bibnamefont
  {F{\aa}k}}\ and\ \bibinfo {author} {\bibfnamefont {K.~H.}\ \bibnamefont
  {Andersen}},\ }\href@noop {} {\bibfield  {journal} {\bibinfo  {journal}
  {Phys. Lett. A}\ }\textbf {\bibinfo {volume} {{160}}},\ \bibinfo {pages}
  {468} (\bibinfo {year} {{1991}})}\BibitemShut {NoStop}%
\bibitem [{\citenamefont {Andersen}\ \emph {et~al.}(1992)\citenamefont
  {Andersen}, \citenamefont {Stirling}, \citenamefont {Scherm}, \citenamefont
  {Stunault}, \citenamefont {F{\aa}k}, \citenamefont {Godfrin},\ and\
  \citenamefont {Dianoux}}]{Andersen92}%
  \BibitemOpen
  \bibfield  {author} {\bibinfo {author} {\bibfnamefont {K.~H.}\ \bibnamefont
  {Andersen}}, \bibinfo {author} {\bibfnamefont {W.~G.}\ \bibnamefont
  {Stirling}}, \bibinfo {author} {\bibfnamefont {R.}~\bibnamefont {Scherm}},
  \bibinfo {author} {\bibfnamefont {A.}~\bibnamefont {Stunault}}, \bibinfo
  {author} {\bibfnamefont {B.}~\bibnamefont {F{\aa}k}}, \bibinfo {author}
  {\bibfnamefont {H.}~\bibnamefont {Godfrin}}, \ and\ \bibinfo {author}
  {\bibfnamefont {A.}~\bibnamefont {Dianoux}},\ }\href@noop {} {\bibfield
  {journal} {\bibinfo  {journal} {Physica B: Condensed Matter}\ }\textbf
  {\bibinfo {volume} {180}},\ \bibinfo {pages} {851} (\bibinfo {year}
  {1992})}\BibitemShut {NoStop}%
\bibitem [{\citenamefont {Andersen}\ \emph {et~al.}(1994)\citenamefont
  {Andersen}, \citenamefont {Stirling}, \citenamefont {Scherm}, \citenamefont
  {Stunault}, \citenamefont {F{\aa}k}, \citenamefont {Godfrin},\ and\
  \citenamefont {Dianoux}}]{Andersen94a}%
  \BibitemOpen
  \bibfield  {author} {\bibinfo {author} {\bibfnamefont {K.~H.}\ \bibnamefont
  {Andersen}}, \bibinfo {author} {\bibfnamefont {W.~G.}\ \bibnamefont
  {Stirling}}, \bibinfo {author} {\bibfnamefont {R.}~\bibnamefont {Scherm}},
  \bibinfo {author} {\bibfnamefont {A.}~\bibnamefont {Stunault}}, \bibinfo
  {author} {\bibfnamefont {B.}~\bibnamefont {F{\aa}k}}, \bibinfo {author}
  {\bibfnamefont {H.}~\bibnamefont {Godfrin}}, \ and\ \bibinfo {author}
  {\bibfnamefont {A.~J.}\ \bibnamefont {Dianoux}},\ }\href@noop {} {\bibfield
  {journal} {\bibinfo  {journal} {J. Phys. Condens. Matter}\ }\textbf {\bibinfo
  {volume} {6}},\ \bibinfo {pages} {821} (\bibinfo {year} {1994})}\BibitemShut
  {NoStop}%
\bibitem [{\citenamefont {Andersen}\ and\ \citenamefont
  {Stirling}(1994)}]{Andersen94b}%
  \BibitemOpen
  \bibfield  {author} {\bibinfo {author} {\bibfnamefont {K.~H.}\ \bibnamefont
  {Andersen}}\ and\ \bibinfo {author} {\bibfnamefont {W.~G.}\ \bibnamefont
  {Stirling}},\ }\href@noop {} {\bibfield  {journal} {\bibinfo  {journal} {J.
  Phys. Cond. Matter}\ }\textbf {\bibinfo {volume} {6}},\ \bibinfo {pages}
  {5805} (\bibinfo {year} {1994})}\BibitemShut {NoStop}%
\bibitem [{\citenamefont {Gibbs}(1996)}]{GibbsThesis}%
  \BibitemOpen
  \bibfield  {author} {\bibinfo {author} {\bibfnamefont {M.~R.}\ \bibnamefont
  {Gibbs}},\ }\emph {\bibinfo {title} {The collective excitations of superfluid
  $^4${H}e: the dependence on pressure and the effect of restricted
  geometry}},\ \href@noop {} {Ph.D. thesis},\ \bibinfo  {school} {Keele
  University} (\bibinfo {year} {1996})\BibitemShut {NoStop}%
\bibitem [{\citenamefont {Gibbs}\ \emph {et~al.}(1999)\citenamefont {Gibbs},
  \citenamefont {Andersen}, \citenamefont {Stirling},\ and\ \citenamefont
  {Schober}}]{AndersenRoton}%
  \BibitemOpen
  \bibfield  {author} {\bibinfo {author} {\bibfnamefont {M.~R.}\ \bibnamefont
  {Gibbs}}, \bibinfo {author} {\bibfnamefont {K.~H.}\ \bibnamefont {Andersen}},
  \bibinfo {author} {\bibfnamefont {W.~G.}\ \bibnamefont {Stirling}}, \ and\
  \bibinfo {author} {\bibfnamefont {H.}~\bibnamefont {Schober}},\ }\href@noop
  {} {\bibfield  {journal} {\bibinfo  {journal} {J. Phys.: Condens. Matter}\
  }\textbf {\bibinfo {volume} {11}},\ \bibinfo {pages} {603} (\bibinfo {year}
  {1999})}\BibitemShut {NoStop}%
\bibitem [{\citenamefont {Glyde}\ \emph {et~al.}(1998)\citenamefont {Glyde},
  \citenamefont {Gibbs}, \citenamefont {Stirling},\ and\ \citenamefont
  {Adams}}]{Glyde-Gibbs-98}%
  \BibitemOpen
  \bibfield  {author} {\bibinfo {author} {\bibfnamefont {H.~R.}\ \bibnamefont
  {Glyde}}, \bibinfo {author} {\bibfnamefont {M.~R.}\ \bibnamefont {Gibbs}},
  \bibinfo {author} {\bibfnamefont {W.~G.}\ \bibnamefont {Stirling}}, \ and\
  \bibinfo {author} {\bibfnamefont {M.~A.}\ \bibnamefont {Adams}},\ }\href@noop
  {} {\bibfield  {journal} {\bibinfo  {journal} {{Europhysics Letters}}\
  }\textbf {\bibinfo {volume} {{43}}},\ \bibinfo {pages} {422} (\bibinfo {year}
  {{1998}})}\BibitemShut {NoStop}%
\bibitem [{\citenamefont {Pearce}\ \emph
  {et~al.}(2001{\natexlab{a}})\citenamefont {Pearce}, \citenamefont {Azuah},
  \citenamefont {Stirling}, \citenamefont {Dimeo}, \citenamefont {Sokol},\ and\
  \citenamefont {Adams}}]{Pearce-Azuah-Stirling}%
  \BibitemOpen
  \bibfield  {author} {\bibinfo {author} {\bibfnamefont {J.}~\bibnamefont
  {Pearce}}, \bibinfo {author} {\bibfnamefont {R.}~\bibnamefont {Azuah}},
  \bibinfo {author} {\bibfnamefont {W.}~\bibnamefont {Stirling}}, \bibinfo
  {author} {\bibfnamefont {R.}~\bibnamefont {Dimeo}}, \bibinfo {author}
  {\bibfnamefont {P.}~\bibnamefont {Sokol}}, \ and\ \bibinfo {author}
  {\bibfnamefont {M.}~\bibnamefont {Adams}},\ }\href@noop {} {\bibfield
  {journal} {\bibinfo  {journal} {J. Low Temp. Phys.}\ }\textbf {\bibinfo
  {volume} {{124}}},\ \bibinfo {pages} {573} (\bibinfo {year}
  {{2001}}{\natexlab{a}})}\BibitemShut {NoStop}%
\bibitem [{\citenamefont {Andersen}\ \emph {et~al.}(1996)\citenamefont
  {Andersen}, \citenamefont {Bossy}, \citenamefont {Cook}, \citenamefont
  {Randl},\ and\ \citenamefont {Ragazzoni}}]{GrenobleRotonsT}%
  \BibitemOpen
  \bibfield  {author} {\bibinfo {author} {\bibfnamefont {K.~H.}\ \bibnamefont
  {Andersen}}, \bibinfo {author} {\bibfnamefont {J.}~\bibnamefont {Bossy}},
  \bibinfo {author} {\bibfnamefont {J.~C.}\ \bibnamefont {Cook}}, \bibinfo
  {author} {\bibfnamefont {O.~G.}\ \bibnamefont {Randl}}, \ and\ \bibinfo
  {author} {\bibfnamefont {J.-L.}\ \bibnamefont {Ragazzoni}},\ }\href@noop {}
  {\bibfield  {journal} {\bibinfo  {journal} {Phys. Rev. Lett.}\ }\textbf
  {\bibinfo {volume} {77}},\ \bibinfo {pages} {4043} (\bibinfo {year}
  {1996})}\BibitemShut {NoStop}%
\bibitem [{\citenamefont {Lovesey}(1984)}]{Lovesey}%
  \BibitemOpen
  \bibfield  {author} {\bibinfo {author} {\bibfnamefont {S.~W.}\ \bibnamefont
  {Lovesey}},\ }\href@noop {} {\emph {\bibinfo {title} {Theory of Neutron
  Scattering from Condensed Matter}}}\ (\bibinfo  {publisher} {Clarendon
  Press},\ \bibinfo {address} {Oxford},\ \bibinfo {year} {1984})\BibitemShut
  {NoStop}%
\bibitem [{\citenamefont {Schober}(2014)}]{schober2014introduction}%
  \BibitemOpen
  \bibfield  {author} {\bibinfo {author} {\bibfnamefont {H.}~\bibnamefont
  {Schober}},\ }\href@noop {} {\bibfield  {journal} {\bibinfo  {journal}
  {Journal of Neutron Research}\ }\textbf {\bibinfo {volume} {17}},\ \bibinfo
  {pages} {109} (\bibinfo {year} {2014})}\BibitemShut {NoStop}%
\bibitem [{\citenamefont {Ollivier}\ \emph {et~al.}(2010)\citenamefont
  {Ollivier}, \citenamefont {Mutka},\ and\ \citenamefont {Didier}}]{IN5_2010}%
  \BibitemOpen
  \bibfield  {author} {\bibinfo {author} {\bibfnamefont {J.}~\bibnamefont
  {Ollivier}}, \bibinfo {author} {\bibfnamefont {H.}~\bibnamefont {Mutka}}, \
  and\ \bibinfo {author} {\bibfnamefont {L.}~\bibnamefont {Didier}},\ }\href
  {\doibase 10.1080/10448631003757573} {\bibfield  {journal} {\bibinfo
  {journal} {Neutron News}\ }\textbf {\bibinfo {volume} {21}},\ \bibinfo
  {pages} {22} (\bibinfo {year} {2010})},\ \Eprint
  {http://arxiv.org/abs/https://doi.org/10.1080/10448631003757573}
  {https://doi.org/10.1080/10448631003757573} \BibitemShut {NoStop}%
\bibitem [{\citenamefont {Ollivier}\ and\ \citenamefont
  {Mutka}(2011)}]{IN5_2011}%
  \BibitemOpen
  \bibfield  {author} {\bibinfo {author} {\bibfnamefont {J.}~\bibnamefont
  {Ollivier}}\ and\ \bibinfo {author} {\bibfnamefont {H.}~\bibnamefont
  {Mutka}},\ }\href {\doibase 10.1143/JPSJS.80SB.SB003} {\bibfield  {journal}
  {\bibinfo  {journal} {J. Phys. Soc. Jpn}\ }\textbf {\bibinfo {volume} {80}},\
  \bibinfo {pages} {SB003} (\bibinfo {year} {2011})},\ \Eprint
  {http://arxiv.org/abs/https://doi.org/10.1143/JPSJS.80SB.SB003}
  {https://doi.org/10.1143/JPSJS.80SB.SB003} \BibitemShut {NoStop}%
\bibitem [{lam()}]{lamp}%
  \BibitemOpen
  \href {http://www.ill.eu/data\_treat/lamp} {}\bibinfo {note} {{\sf LAMP}
  (Large Array Manipulation Program) {
  http://www.ill.eu/data\_treat/lamp}}\BibitemShut {NoStop}%
\bibitem [{\citenamefont {Davison}(1966)}]{Davison_1966}%
  \BibitemOpen
  \bibfield  {author} {\bibinfo {author} {\bibfnamefont {W.~D.}\ \bibnamefont
  {Davison}},\ }\href {\doibase 10.1088/0370-1328/87/1/314} {\bibfield
  {journal} {\bibinfo  {journal} {Proc. Phys. Soc. London}\ }\textbf {\bibinfo
  {volume} {87}},\ \bibinfo {pages} {133} (\bibinfo {year} {1966})}\BibitemShut
  {NoStop}%
\bibitem [{\citenamefont {Boronat}(2002)}]{JordiQFSBook}%
  \BibitemOpen
  \bibfield  {author} {\bibinfo {author} {\bibfnamefont {J.}~\bibnamefont
  {Boronat}},\ }in\ \href@noop {} {\emph {\bibinfo {booktitle} {Microscopic
  Approaches to {Q}uantum Liquids in Confined Geometries}}},\ \bibinfo {editor}
  {edited by\ \bibinfo {editor} {\bibfnamefont {E.}~\bibnamefont {Krotscheck}}\
  and\ \bibinfo {editor} {\bibfnamefont {J.}~\bibnamefont {Navarro}}}\
  (\bibinfo  {publisher} {World Scientific},\ \bibinfo {address} {Singapore},\
  \bibinfo {year} {2002})\ pp.\ \bibinfo {pages} {21--90}\BibitemShut {NoStop}%
\bibitem [{\citenamefont {Feynman}(1954)}]{Feynman1954}%
  \BibitemOpen
  \bibfield  {author} {\bibinfo {author} {\bibfnamefont {R.~P.}\ \bibnamefont
  {Feynman}},\ }\href {\doibase 10.1103/PhysRev.94.262} {\bibfield  {journal}
  {\bibinfo  {journal} {Phys. Rev.}\ }\textbf {\bibinfo {volume} {94}},\
  \bibinfo {pages} {262} (\bibinfo {year} {1954})}\BibitemShut {NoStop}%
\bibitem [{\citenamefont {Boninsegni}\ and\ \citenamefont
  {Ceperley}(1996)}]{CeperleyExcitationPIMC}%
  \BibitemOpen
  \bibfield  {author} {\bibinfo {author} {\bibfnamefont {M.}~\bibnamefont
  {Boninsegni}}\ and\ \bibinfo {author} {\bibfnamefont {D.~M.}\ \bibnamefont
  {Ceperley}},\ }\href@noop {} {\bibfield  {journal} {\bibinfo  {journal} {J.
  Low Temp. Phys.}\ }\textbf {\bibinfo {volume} {104}},\ \bibinfo {pages} {339}
  (\bibinfo {year} {1996})}\BibitemShut {NoStop}%
\bibitem [{\citenamefont {Boronat}\ and\ \citenamefont
  {Casulleras}(1998)}]{BoronatCasulleras}%
  \BibitemOpen
  \bibfield  {author} {\bibinfo {author} {\bibfnamefont {J.}~\bibnamefont
  {Boronat}}\ and\ \bibinfo {author} {\bibfnamefont {J.}~\bibnamefont
  {Casulleras}},\ }\href@noop {} {\bibfield  {journal} {\bibinfo  {journal} {J.
  Low Temp. Phys.}\ }\textbf {\bibinfo {volume} {110}},\ \bibinfo {pages} {443}
  (\bibinfo {year} {1998})}\BibitemShut {NoStop}%
\bibitem [{\citenamefont {F{\aa}k}\ and\ \citenamefont
  {Bossy}(1998{\natexlab{a}})}]{Fak98b}%
  \BibitemOpen
  \bibfield  {author} {\bibinfo {author} {\bibfnamefont {B.}~\bibnamefont
  {F{\aa}k}}\ and\ \bibinfo {author} {\bibfnamefont {J.}~\bibnamefont
  {Bossy}},\ }\href@noop {} {\bibfield  {journal} {\bibinfo  {journal} {J. Low
  Temp. Phys.}\ }\textbf {\bibinfo {volume} {112}},\ \bibinfo {pages} {1}
  (\bibinfo {year} {1998}{\natexlab{a}})}\BibitemShut {NoStop}%
\bibitem [{\citenamefont {F{\aa}k}\ and\ \citenamefont
  {Bossy}(1998{\natexlab{b}})}]{faak1998excitations}%
  \BibitemOpen
  \bibfield  {author} {\bibinfo {author} {\bibfnamefont {B.}~\bibnamefont
  {F{\aa}k}}\ and\ \bibinfo {author} {\bibfnamefont {J.}~\bibnamefont
  {Bossy}},\ }\href@noop {} {\bibfield  {journal} {\bibinfo  {journal} {J. Low
  Temp. Phys.}\ }\textbf {\bibinfo {volume} {113}},\ \bibinfo {pages} {531}
  (\bibinfo {year} {1998}{\natexlab{b}})}\BibitemShut {NoStop}%
\bibitem [{\citenamefont {Pistolesi}(1998{\natexlab{b}})}]{PistolesiJLTP}%
  \BibitemOpen
  \bibfield  {author} {\bibinfo {author} {\bibfnamefont {F.}~\bibnamefont
  {Pistolesi}},\ }\href@noop {} {\bibfield  {journal} {\bibinfo  {journal} {J.
  Low Temp. Phys.}\ }\textbf {\bibinfo {volume} {113}},\ \bibinfo {pages} {597}
  (\bibinfo {year} {1998}{\natexlab{b}})}\BibitemShut {NoStop}%
\bibitem [{\citenamefont {Pearce}\ \emph
  {et~al.}(2001{\natexlab{b}})\citenamefont {Pearce}, \citenamefont {Azuah},
  \citenamefont {F{\aa}k}, \citenamefont {Sakhel}, \citenamefont {Glyde},\ and\
  \citenamefont {Stirling}}]{PearceAzuah}%
  \BibitemOpen
  \bibfield  {author} {\bibinfo {author} {\bibfnamefont {J.~V.}\ \bibnamefont
  {Pearce}}, \bibinfo {author} {\bibfnamefont {R.~T.}\ \bibnamefont {Azuah}},
  \bibinfo {author} {\bibfnamefont {B.}~\bibnamefont {F{\aa}k}}, \bibinfo
  {author} {\bibfnamefont {A.~R.}\ \bibnamefont {Sakhel}}, \bibinfo {author}
  {\bibfnamefont {H.~R.}\ \bibnamefont {Glyde}}, \ and\ \bibinfo {author}
  {\bibfnamefont {W.~G.}\ \bibnamefont {Stirling}},\ }\href@noop {} {\bibfield
  {journal} {\bibinfo  {journal} {J. Phys.: Condens. Matter}\ }\textbf
  {\bibinfo {volume} {13}},\ \bibinfo {pages} {4421} (\bibinfo {year}
  {2001}{\natexlab{b}})}\BibitemShut {NoStop}%
\bibitem [{\citenamefont {Donnelly}\ \emph {et~al.}(1981)\citenamefont
  {Donnelly}, \citenamefont {Donnelly},\ and\ \citenamefont
  {Hills}}]{DonnellyDonnellyHills}%
  \BibitemOpen
  \bibfield  {author} {\bibinfo {author} {\bibfnamefont {R.~J.}\ \bibnamefont
  {Donnelly}}, \bibinfo {author} {\bibfnamefont {J.~A.}\ \bibnamefont
  {Donnelly}}, \ and\ \bibinfo {author} {\bibfnamefont {R.~N.}\ \bibnamefont
  {Hills}},\ }\href@noop {} {\bibfield  {journal} {\bibinfo  {journal} {J. Low
  Temp. Phys.}\ }\textbf {\bibinfo {volume} {44}},\ \bibinfo {pages} {471}
  (\bibinfo {year} {1981})}\BibitemShut {NoStop}%
\bibitem [{\citenamefont {Donnelly}\ and\ \citenamefont
  {Roberts}(1977)}]{DonnellyRoberts}%
  \BibitemOpen
  \bibfield  {author} {\bibinfo {author} {\bibfnamefont {R.~J.}\ \bibnamefont
  {Donnelly}}\ and\ \bibinfo {author} {\bibfnamefont {P.~H.}\ \bibnamefont
  {Roberts}},\ }\href@noop {} {\bibfield  {journal} {\bibinfo  {journal} {J.
  Low Temp. Phys}\ }\textbf {\bibinfo {volume} {27}},\ \bibinfo {pages} {687}
  (\bibinfo {year} {1977})}\BibitemShut {NoStop}%
\bibitem [{\citenamefont {Wiebes}(1969)}]{WiebesThesis}%
  \BibitemOpen
  \bibfield  {author} {\bibinfo {author} {\bibfnamefont {J.}~\bibnamefont
  {Wiebes}},\ }\href@noop {} {Ph.D. thesis},\ \bibinfo  {school} {Kammerlingh
  Onnes Laboratory} (\bibinfo {year} {1969})\BibitemShut {NoStop}%
\bibitem [{\citenamefont {Morishita}\ \emph {et~al.}(1989)\citenamefont
  {Morishita}, \citenamefont {Kuroda}, \citenamefont {Sawada},\ and\
  \citenamefont {Satoh}}]{morishita1989mean}%
  \BibitemOpen
  \bibfield  {author} {\bibinfo {author} {\bibfnamefont {M.}~\bibnamefont
  {Morishita}}, \bibinfo {author} {\bibfnamefont {T.}~\bibnamefont {Kuroda}},
  \bibinfo {author} {\bibfnamefont {A.}~\bibnamefont {Sawada}}, \ and\ \bibinfo
  {author} {\bibfnamefont {T.}~\bibnamefont {Satoh}},\ }\href@noop {}
  {\bibfield  {journal} {\bibinfo  {journal} {J. Low Temp. Phys.}\ }\textbf
  {\bibinfo {volume} {76}},\ \bibinfo {pages} {387} (\bibinfo {year}
  {1989})}\BibitemShut {NoStop}%
\bibitem [{\citenamefont {Bryan}\ and\ \citenamefont
  {Sokol}(2018)}]{bryan2018confined}%
  \BibitemOpen
  \bibfield  {author} {\bibinfo {author} {\bibfnamefont {M.~S.}\ \bibnamefont
  {Bryan}}\ and\ \bibinfo {author} {\bibfnamefont {P.~E.}\ \bibnamefont
  {Sokol}},\ }\href@noop {} {\bibfield  {journal} {\bibinfo  {journal} {Phys.
  Rev. B}\ }\textbf {\bibinfo {volume} {97}},\ \bibinfo {pages} {184511}
  (\bibinfo {year} {2018})}\BibitemShut {NoStop}%
\bibitem [{\citenamefont {Nicolis}\ and\ \citenamefont
  {Penco}(2018)}]{2018-Nicolis}%
  \BibitemOpen
  \bibfield  {author} {\bibinfo {author} {\bibfnamefont {A.}~\bibnamefont
  {Nicolis}}\ and\ \bibinfo {author} {\bibfnamefont {R.}~\bibnamefont
  {Penco}},\ }\href {\doibase 10.1103/PhysRevB.97.134516} {\bibfield  {journal}
  {\bibinfo  {journal} {Phys. Rev. B}\ }\textbf {\bibinfo {volume} {97}},\
  \bibinfo {pages} {134516} (\bibinfo {year} {2018})}\BibitemShut {NoStop}%
\bibitem [{\citenamefont {Bossy}\ \emph {et~al.}(2019)\citenamefont {Bossy},
  \citenamefont {Ollivier},\ and\ \citenamefont {Glyde}}]{bossy2019}%
  \BibitemOpen
  \bibfield  {author} {\bibinfo {author} {\bibfnamefont {J.}~\bibnamefont
  {Bossy}}, \bibinfo {author} {\bibfnamefont {J.}~\bibnamefont {Ollivier}}, \
  and\ \bibinfo {author} {\bibfnamefont {H.~R.}\ \bibnamefont {Glyde}},\
  }\href@noop {} {\bibfield  {journal} {\bibinfo  {journal} {Phys. Rev. B}\
  }\textbf {\bibinfo {volume} {99}},\ \bibinfo {pages} {165425} (\bibinfo
  {year} {2019})}\BibitemShut {NoStop}%
\bibitem [{\citenamefont {Moroshkin}\ and\ \citenamefont
  {Kono}(2019)}]{2019dropletsMoroshkin}%
  \BibitemOpen
  \bibfield  {author} {\bibinfo {author} {\bibfnamefont {P.}~\bibnamefont
  {Moroshkin}}\ and\ \bibinfo {author} {\bibfnamefont {K.}~\bibnamefont
  {Kono}},\ }\href {\doibase 10.1103/PhysRevB.99.104512} {\bibfield  {journal}
  {\bibinfo  {journal} {Phys. Rev. B}\ }\textbf {\bibinfo {volume} {99}},\
  \bibinfo {pages} {104512} (\bibinfo {year} {2019})}\BibitemShut {NoStop}%
\bibitem [{\citenamefont {Escart\'{\i}n}\ \emph {et~al.}(2019)\citenamefont
  {Escart\'{\i}n}, \citenamefont {Ancilotto}, \citenamefont {Barranco},\ and\
  \citenamefont {Pi}}]{2019droplets}%
  \BibitemOpen
  \bibfield  {author} {\bibinfo {author} {\bibfnamefont {J.~M.}\ \bibnamefont
  {Escart\'{\i}n}}, \bibinfo {author} {\bibfnamefont {F.}~\bibnamefont
  {Ancilotto}}, \bibinfo {author} {\bibfnamefont {M.}~\bibnamefont {Barranco}},
  \ and\ \bibinfo {author} {\bibfnamefont {M.}~\bibnamefont {Pi}},\ }\href
  {\doibase 10.1103/PhysRevB.99.140505} {\bibfield  {journal} {\bibinfo
  {journal} {Phys. Rev. B}\ }\textbf {\bibinfo {volume} {99}},\ \bibinfo
  {pages} {140505(R)} (\bibinfo {year} {2019})}\BibitemShut {NoStop}%
\bibitem [{\citenamefont {Godfrin}\ \emph {et~al.}(2012)\citenamefont
  {Godfrin}, \citenamefont {Meschke}, \citenamefont {Lauter}, \citenamefont
  {Sultan}, \citenamefont {B{\"o}hm}, \citenamefont {Krotscheck},\ and\
  \citenamefont {Panholzer}}]{godfrin2012nature}%
  \BibitemOpen
  \bibfield  {author} {\bibinfo {author} {\bibfnamefont {H.}~\bibnamefont
  {Godfrin}}, \bibinfo {author} {\bibfnamefont {M.}~\bibnamefont {Meschke}},
  \bibinfo {author} {\bibfnamefont {H.-J.}\ \bibnamefont {Lauter}}, \bibinfo
  {author} {\bibfnamefont {A.}~\bibnamefont {Sultan}}, \bibinfo {author}
  {\bibfnamefont {H.~M.}\ \bibnamefont {B{\"o}hm}}, \bibinfo {author}
  {\bibfnamefont {E.}~\bibnamefont {Krotscheck}}, \ and\ \bibinfo {author}
  {\bibfnamefont {M.}~\bibnamefont {Panholzer}},\ }\href@noop {} {\bibfield
  {journal} {\bibinfo  {journal} {Nature}\ }\textbf {\bibinfo {volume} {483}},\
  \bibinfo {pages} {576} (\bibinfo {year} {2012})}\BibitemShut {NoStop}%
\bibitem [{\citenamefont {Pines}(2016)}]{pines2016emergent}%
  \BibitemOpen
  \bibfield  {author} {\bibinfo {author} {\bibfnamefont {D.}~\bibnamefont
  {Pines}},\ }\href@noop {} {\bibfield  {journal} {\bibinfo  {journal} {Rep.
  Prog. Phys.}\ }\textbf {\bibinfo {volume} {79}},\ \bibinfo {pages} {092501}
  (\bibinfo {year} {2016})}\BibitemShut {NoStop}%
\bibitem [{\citenamefont {Santos}\ \emph {et~al.}(2003)\citenamefont {Santos},
  \citenamefont {Shlyapnikov},\ and\ \citenamefont
  {Lewenstein}}]{santos2003roton}%
  \BibitemOpen
  \bibfield  {author} {\bibinfo {author} {\bibfnamefont {L.}~\bibnamefont
  {Santos}}, \bibinfo {author} {\bibfnamefont {G.~V.}\ \bibnamefont
  {Shlyapnikov}}, \ and\ \bibinfo {author} {\bibfnamefont {M.}~\bibnamefont
  {Lewenstein}},\ }\href@noop {} {\bibfield  {journal} {\bibinfo  {journal}
  {Phys. Rev. Lett.}\ }\textbf {\bibinfo {volume} {90}},\ \bibinfo {pages}
  {250403} (\bibinfo {year} {2003})}\BibitemShut {NoStop}%
\bibitem [{\citenamefont {Chomaz}\ \emph {et~al.}(2018)\citenamefont {Chomaz},
  \citenamefont {van Bijnen}, \citenamefont {Petter}, \citenamefont {Faraoni},
  \citenamefont {Baier}, \citenamefont {Becher}, \citenamefont {Mark},
  \citenamefont {Waechtler}, \citenamefont {Santos},\ and\ \citenamefont
  {Ferlaino}}]{chomaz2018observation}%
  \BibitemOpen
  \bibfield  {author} {\bibinfo {author} {\bibfnamefont {L.}~\bibnamefont
  {Chomaz}}, \bibinfo {author} {\bibfnamefont {R.~M.}\ \bibnamefont {van
  Bijnen}}, \bibinfo {author} {\bibfnamefont {D.}~\bibnamefont {Petter}},
  \bibinfo {author} {\bibfnamefont {G.}~\bibnamefont {Faraoni}}, \bibinfo
  {author} {\bibfnamefont {S.}~\bibnamefont {Baier}}, \bibinfo {author}
  {\bibfnamefont {J.~H.}\ \bibnamefont {Becher}}, \bibinfo {author}
  {\bibfnamefont {M.~J.}\ \bibnamefont {Mark}}, \bibinfo {author}
  {\bibfnamefont {F.}~\bibnamefont {Waechtler}}, \bibinfo {author}
  {\bibfnamefont {L.}~\bibnamefont {Santos}}, \ and\ \bibinfo {author}
  {\bibfnamefont {F.}~\bibnamefont {Ferlaino}},\ }\href@noop {} {\bibfield
  {journal} {\bibinfo  {journal} {Nature Physics}\ }\textbf {\bibinfo {volume}
  {14}},\ \bibinfo {pages} {442} (\bibinfo {year} {2018})}\BibitemShut
  {NoStop}%
\bibitem [{\citenamefont {McCormack}\ \emph {et~al.}(2020)\citenamefont
  {McCormack}, \citenamefont {Nath},\ and\ \citenamefont
  {Li}}]{2020-maxon-roton}%
  \BibitemOpen
  \bibfield  {author} {\bibinfo {author} {\bibfnamefont {G.}~\bibnamefont
  {McCormack}}, \bibinfo {author} {\bibfnamefont {R.}~\bibnamefont {Nath}}, \
  and\ \bibinfo {author} {\bibfnamefont {W.}~\bibnamefont {Li}},\ }\href
  {\doibase 10.1103/PhysRevA.102.023319} {\bibfield  {journal} {\bibinfo
  {journal} {Phys. Rev. A}\ }\textbf {\bibinfo {volume} {102}},\ \bibinfo
  {pages} {023319} (\bibinfo {year} {2020})}\BibitemShut {NoStop}%
\bibitem [{\citenamefont {Trachenko}\ and\ \citenamefont
  {Brazhkin}(2015)}]{trachenko2015}%
  \BibitemOpen
  \bibfield  {author} {\bibinfo {author} {\bibfnamefont {K.}~\bibnamefont
  {Trachenko}}\ and\ \bibinfo {author} {\bibfnamefont {V.}~\bibnamefont
  {Brazhkin}},\ }\href@noop {} {\bibfield  {journal} {\bibinfo  {journal} {Rep.
  Prog. Phys.}\ }\textbf {\bibinfo {volume} {79}},\ \bibinfo {pages} {016502}
  (\bibinfo {year} {2015})}\BibitemShut {NoStop}%
\bibitem [{\citenamefont {Balian}\ and\ \citenamefont
  {de~Dominicis}(1971)}]{Balian1971}%
  \BibitemOpen
  \bibfield  {author} {\bibinfo {author} {\bibfnamefont {R.}~\bibnamefont
  {Balian}}\ and\ \bibinfo {author} {\bibfnamefont {C.}~\bibnamefont
  {de~Dominicis}},\ }\href@noop {} {\bibfield  {journal} {\bibinfo  {journal}
  {Annals of Physics}\ }\textbf {\bibinfo {volume} {62}},\ \bibinfo {pages}
  {229} (\bibinfo {year} {1971})}\BibitemShut {NoStop}%
\bibitem [{\citenamefont {Carneiro}\ and\ \citenamefont
  {Pethick}(1975)}]{Carneiro1975}%
  \BibitemOpen
  \bibfield  {author} {\bibinfo {author} {\bibfnamefont {G.~M.}\ \bibnamefont
  {Carneiro}}\ and\ \bibinfo {author} {\bibfnamefont {C.~J.}\ \bibnamefont
  {Pethick}},\ }\href {\doibase 10.1103/PhysRevB.11.1106} {\bibfield  {journal}
  {\bibinfo  {journal} {Phys. Rev. B}\ }\textbf {\bibinfo {volume} {11}},\
  \bibinfo {pages} {1106} (\bibinfo {year} {1975})}\BibitemShut {NoStop}%
\bibitem [{\citenamefont {Carneiro}(1979)}]{Carneiro1979}%
  \BibitemOpen
  \bibfield  {author} {\bibinfo {author} {\bibfnamefont {G.~M.}\ \bibnamefont
  {Carneiro}},\ }\href {\doibase 10.1103/PhysRevB.19.215} {\bibfield  {journal}
  {\bibinfo  {journal} {Phys. Rev. B}\ }\textbf {\bibinfo {volume} {19}},\
  \bibinfo {pages} {215} (\bibinfo {year} {1979})}\BibitemShut {NoStop}%
\bibitem [{\citenamefont {Cohen}(1960)}]{Cohen1960}%
  \BibitemOpen
  \bibfield  {author} {\bibinfo {author} {\bibfnamefont {M.}~\bibnamefont
  {Cohen}},\ }\href {\doibase 10.1103/PhysRev.118.27} {\bibfield  {journal}
  {\bibinfo  {journal} {Phys. Rev.}\ }\textbf {\bibinfo {volume} {118}},\
  \bibinfo {pages} {27} (\bibinfo {year} {1960})}\BibitemShut {NoStop}%
\bibitem [{\citenamefont {F\aa{}k}\ \emph {et~al.}(2012)\citenamefont
  {F\aa{}k}, \citenamefont {Keller}, \citenamefont {Zhitomirsky},\ and\
  \citenamefont {Chernyshev}}]{Fak2012}%
  \BibitemOpen
  \bibfield  {author} {\bibinfo {author} {\bibfnamefont {B.}~\bibnamefont
  {F\aa{}k}}, \bibinfo {author} {\bibfnamefont {T.}~\bibnamefont {Keller}},
  \bibinfo {author} {\bibfnamefont {M.~E.}\ \bibnamefont {Zhitomirsky}}, \ and\
  \bibinfo {author} {\bibfnamefont {A.~L.}\ \bibnamefont {Chernyshev}},\ }\href
  {\doibase 10.1103/PhysRevLett.109.155305} {\bibfield  {journal} {\bibinfo
  {journal} {Phys. Rev. Lett.}\ }\textbf {\bibinfo {volume} {109}},\ \bibinfo
  {pages} {155305} (\bibinfo {year} {2012})}\BibitemShut {NoStop}%
\end{thebibliography}
\end{document}